\pdfoutput=1
\documentclass[12pt,a4paper]{article}
\usepackage{a4wide,hyperref}
\usepackage{subcaption}
\usepackage[english]{babel}
\usepackage{amsmath}   
\usepackage{graphicx}  
\usepackage{dcolumn}   
\usepackage{bm}        
\usepackage{amssymb}   
\usepackage{epsfig}
\usepackage{epstopdf}
\usepackage[utf8]{inputenc}
\usepackage{mdframed}
\usepackage{amsmath}
\usepackage{wasysym}
\usepackage{tabu}
\usepackage[toc]{appendix}
\usepackage{color,soul} 
\usepackage{datetime}
\usepackage{IEEEtrantools}
\usepackage{tikz}
\usetikzlibrary{calc, 3d}
\usepackage{physics}
\usepackage{enumerate}
\usepackage{asymptote}
\usepackage{comment}

\newcommand{\be}{\begin{equation}}
\newcommand{\ee}{\end{equation}}
\newcommand{\bea}{\begin{eqnarray}}
\newcommand{\eea}{\end{eqnarray}}
\newcommand{\bean}{\begin{eqnarray*}}
\newcommand{\eean}{\end{eqnarray*}}

\begin{document}
\begin{titlepage}
    \numberwithin{equation}{section}
    \begin{flushright}
        \small

        \normalsize
    \end{flushright}
    \vspace{0.8 cm}

    \begin{center}

        \hspace*{-1cm}\mbox{\LARGE \textbf{Real-Time Holography and Hybrid WKB}}\\ \vspace{0.5cm}
        \mbox{\LARGE \textbf{ for BTZ Wormholes}}\\
        \vspace{0.5cm}

        \medskip

        \vspace{1.2 cm} {\large Vasil Dimitrov$^1$, Daniel R. Mayerson$^{1,2}$, Vincent Min$^1$}\\

        \vspace{1cm} {$^1$ Institute for Theoretical Physics, KU Leuven,\\
            Celestijnenlaan 200D, B-3001 Leuven, Belgium}

        \vspace{0.5cm} {$^2$ Universit\'e Paris-Saclay, CNRS, CEA,\\ Institut de physique th\'eorique,\\ 91191, Gif-sur-Yvette, France.}

        \vspace{0.5cm}

        \vspace{.5cm}
        vasko.dimitrov, daniel.mayerson, \\
        vincent.min @ kuleuven.be\\

        \vspace{2cm}

        \textbf{Abstract}
    \end{center}


 \noindent   We study probe scalar correlation functions in a Solodukhin wormhole corresponding to the non-rotating BTZ black hole, as a toy model for microstate geometries thereof. Using real-time holography, we obtain the retarded scalar correlator in the wormhole geometry and quantitatively compare it to the result of the hybrid WKB method for the same correlator. We also calculate an off-diagonal correlator $\sim\langle H L L H' \rangle$ involving two different (heavy) wormhole states.

\end{titlepage}

\newpage
\setcounter{tocdepth}{2}
\tableofcontents

\section{Introduction}
The quantum nature of the black hole microstructure has remained elusive, most notably due to the information paradox \cite{Hawking:1976ra,Mathur:2009hf,Harlow:2014yka,Raju:2020smc}.
The fuzzball proposal in string theory \cite{Mathur:2005zp,Mathur:2009hf,Bena:2013dka,Bena:2007kg,Shigemori:2020yuo} aims to resolve this paradox by replacing the black hole and its horizon entirely by quantum, stringy, horizon-scale microstructure. The examples of such microstructure that can be studied in supergravity are called microstate geometries, and are smooth, horizonless solutions of supergravity.
Perhaps the best studied examples of such microstate geometries are the superstrata geometries \cite{Bena:2014qxa,Bena:2015bea,Bena:2016agb,Bena:2016ypk,Bena:2017xbt,Heidmann:2019xrd,Mayerson:2020tcl,Shigemori:2020yuo,Mayerson:2020acj}, which are microstates of the BMPV black hole \cite{Breckenridge:1996is} in five dimensions (which uplifts to a six-dimensional black string). The BMPV black hole admits a near-horizon $\text{BTZ} \times S^3$  decoupling limit metric; the superstrata geometries effectively replace this by a metric that interpolates between $\text{BTZ} \times S^3$ in the UV and a smooth, geometrical cap in the IR.
The possible shapes of the cap, determined by the fluxes in supergravity, represent the different microstates. 
Using the AdS/CFT correspondence, one can study these microstates from the point of view of the dual CFT \cite{Giusto:2015dfa,Giusto:2019qig}. 
The microstates in supergravity correspond to semi-classical, coherent states in the CFT, created by acting with different heavy operators on the D1/D5 ground state.\footnote{Heavy operators are ones with scaling dimension $\Delta \sim c$ and light operators are one with scaling dimension $\Delta \sim O(1)$, where $c$ is the central charge of the CFT.} 

One can probe these microstate geometries with a light field (that does not backreact) such as a scalar; a quantity of interest is thus the two-point function of such a light field in a particular supergravity background. This corresponds in the dual CFT to a four-point correlation function $\sim \langle H LL H\rangle$ of two light operators $L$ (for the light probe) and two heavy operators $H$ (for the background state), and can, in some cases, be studied using CFT methods \cite{Bombini:2017sge,Giusto:2018ovt,Ceplak:2021wak,Ceplak:2021wzz}.

To calculate this correlation function in the supergravity solution, one has to solve the wave equation for the light field propagating in the (heavy) background, and study the asymptotic behavior of that solution to obtain the two-point function in this background following the standard rules of the AdS/CFT dictionary. In practice, this can be technically challenging as the wave equation is not always separable.
Even when it is separable, separating out the growing and decaying modes near the boundary unambiguously (which is necessary to extract the correlator) seems to require an exact solution \cite{Bena:2019azk}.
This issue was addressed in \cite{Bena:2019azk} (and developed further for asymptotically flat geometries in \cite{Bena:2020yii}) using the so-called \textit{hybrid WKB} method. There, they used an exact solution near the UV boundary (allowing the extraction of the correlator) which was matched to the approximate WKB solution in the IR through an intermediate point. For the superstrata (in particular, the $(1,0,n)$ family),
this calculation revealed that the scalar correlator reproduces the initial black hole correlator fall off. Only after a large amount of time (proportional to the throat length), do we see echoes due to the reflection of the scalar wave off the smooth cap at the interior.

The appearance of echoes is a standard feature of many models that introduce structure at the horizon scale, and are a focus for potential experimental explorations of black hole microstates and other horizon-scale compact objects (often called ECOs) \cite{Maggio:2021ans,Mayerson:2020tpn,Cardoso:2019rvt,Barack:2018yly}. Note, however, that at the moment there has been no conclusive evidence of echoes appearing in current observations \cite{Ashton:2016xff,Abedi:2017isz,Westerweck:2017hus,Abedi:2018pst,Lo:2018sep,Nielsen:2018lkf,Testa:2018bzd,Tsang:2018uie,Abedi:2018npz,Abedi:2020sgg,Uchikata:2019frs,LongoMicchi:2020cwm,Maggio:2021ans}. It is also unclear whether one would expect echoes to appear in a typical state; as argued in our previous work \cite{Dimitrov:2020txx}, a typical superposition state of many coherent heavy states would presumably lead to the exponential suppression of any echo structure in the correlator. Echoes in supersymmetric microstate geometries have been analyzed in \cite{Ikeda:2021uvc}; these results suggest that as the microstructure becomes more compact (making the microstate more typical), the echoes become more suppressed and washed out (as also argued in \cite{Bacchini:2021fig}).
See also \cite{Martinec:2020cml,Ceplak:2021kgl} for a discussion of how additional stringy tidal effects can excite heavy string modes to delay and smear echoes significantly, even in a coherent, atypical microstate. 

We have previously argued that a Solodukhin wormhole \cite{Solodukhin:2005qy} (see also \cite{Damour:2007ap}) can be used as a toy model of the more complicated superstrata microstate geometries, as it captures some of its essential features \cite{Dimitrov:2020txx}. Such a wormhole connects two asymptotically $\text{AdS}_3$ regions through a wormhole throat (which sits at the would-be horizon scale) and behaves roughly as a ``cap'' where the boundary conditions on the second $\text{AdS}_3$ boundary reflect back the information. Using this toy model, we obtained a correlator that is qualitatively very similar to the $(1,0,n)$ superstratum correlator of \cite{Bena:2019azk}, containing an initial fall-off identical to that of the black hole, and with echoes at times of order the wormhole throat length.

In this paper, we investigate such scalar correlators in the Solodukhin wormhole toy models in more detail. In particular, we use the Skenderis-van Rees formalism of real-time holography \cite{Skenderis:2008dh,Skenderis:2008dg} to calculate this correlator exactly. This allows us to compare the exact answer to the hybrid WKB answer, which is the first time such an explicit comparison is made. It also allows us to calculate all possible two-point correlators (Feynman, Wightman, retarded, advanced), as opposed to only the retarded one, as one calculates using the hybrid WKB method.

Note also that \cite{Bena:2019azk,Bena:2020yii,Dimitrov:2020txx} (mainly) consider microstates of a supersymmetric black hole, which has an extremal BTZ throat. 
In this paper, we use the non-rotating (non-extremal) BTZ black hole and its corresponding Solodukhin wormhole as the starting point for our calculations.
As such, one can view our toy model as a playground to explore basic features of correlators of non-extremal microstate geometries.


The real-time holography formalism has the additional benefit that it in principle allows us to compute correlators in a background that describes \emph{off-diagonal} correlators in the CFT such as $\langle H L L H'\rangle$, where the initial and final heavy states are \emph{different}.\footnote{Note that such a scenario can only be of physical relevance if the timescale of observation is much larger that timescale of such a transition, and moreover that the probability of transition between neighboring heavy states is sufficiently high. We will be considering the case where the transition happens between infinitesimally close states.} These no longer correspond to a scalar correlator in a single geometry, but rather (using the real-time formalism) can be seen to correspond to a \emph{transition} between different (Euclidean) geometries. Such correlators are the first step towards understanding the correlators in a mixed-state given by a density matrix constructed from several heavy states $H_i$ --- one would hope that the black hole correlator will be arbitrarily well approximated by such a density matrix correlator using sufficiently many microstate geometry heavy states \cite{Bena:2019azk}.
We model this transition in the Solodukhin wormhole toy model and compute this off-diagonal correlator for an infinitesimal change in heavy state. To our knowledge, this is the first ever such calculation that explores off-diagonal correlators, albeit in a toy model for microstate geometries. Our results suggest that the sharp echoes (of diagonal correlators) will get smoothed out in off-diagonal correlators.

\medskip 

This paper is organized as follows: In section \ref{sec:BTZandWH} we introduce the BTZ black hole, its corresponding Solodukhin wormhole modification, as well as discuss the scalar wave equations in these backgrounds. In section \ref{sec:SvR_calculations}, we calculate the scalar correlator using the real-time holography formalism to successively more complicated geometries: the BTZ black hole, its Solodukhin wormhole, and the (infinitesimal) transition between two such wormholes. 
In section \ref{sec:AdS3WKB}, we apply hybrid WKB to obtain the (approximate) retarded correlator of the wormhole, and compare it to the exact result  obtained in section \ref{sec:SvRWH}. In section \ref{sec:posspaceprop}, we explicitly calculate the appropriate contour integrals on the momentum-space correlators obtained in section \ref{sec:SvR_calculations} to obtain the position-space correlators of BTZ, the wormhole and the transitioning wormhole. The appendices \ref{sec:appendix_SK} and \ref{sec:appendix_SvR} contain a brief overview of the necessary details of the Schwinger-Keldysh real-time correlator formalism in quantum field theory and Skenderis-van Rees real-time holography.

\section{BTZ, Wormhole and Wave Equation}\label{sec:BTZandWH}
The rest of this paper will use the non-rotating BTZ black hole metric and its corresponding Solodukhin-type wormhole \cite{Solodukhin:2005qy}. In this section, we will review the BTZ black hole and introduce the wormhole geometry as well as a few of its basic properties. We will also discuss the wave equation of a minimally coupled scalar in these geometries. Note that most of this introductory and review material was also presented in \cite{Dimitrov:2020txx}, which also included the expressions for the extremal BTZ black hole and corresponding Solodukhin wormhole.

\subsection{The Non-Rotating BTZ Black Hole}
The non-rotating BTZ black hole \cite{Banados:1992wn,Banados:1992gq} has metric:
\begin{align}
    \label{eq:BTZmetric}
    \begin{aligned}
        \dd{s}^2 & = -X(r) \dd{t}^2 + \frac{ \dd{r}^2 }{Y(r)}+ r^2 \dd{\varphi}^2 \,,          \\
        X(r)     & =Y(r)= \frac{(r^2 - r^2_+)}{R^2} \,, 
    \end{aligned} \qquad
    \begin{cases}
        \begin{aligned}
            r       & \in [r_+,\infty) \,,     \\
            t       & \in (-\infty,\infty) \,, \\
            \varphi & \sim \varphi + 2 \pi \,,
        \end{aligned}
    \end{cases}
\end{align}
where $R$ is the AdS radius. These coordinates cover the patch outside the horizon ($r>r_+$). 
The asymptotic mass of the black hole, in units where $8G_3=1$, reads
\begin{align}
    M = \frac{r^2_+}{R^2} \,,
\end{align}
while the temperature and the entropy of the black hole are given by
\begin{align}
    T=\frac{r_+}{2 \pi R^2} \,,  \qquad S=4\pi r_+ \,.
    \label{eq: temp and entropy of BTZ}
\end{align}
In the holographically dual $\text{CFT}_2$, the black hole corresponds to a thermal state with the same temperature.
\begin{figure}[ht]
    \begin{subfigure}[h]{0.5\linewidth}
        \includegraphics[width=\linewidth]{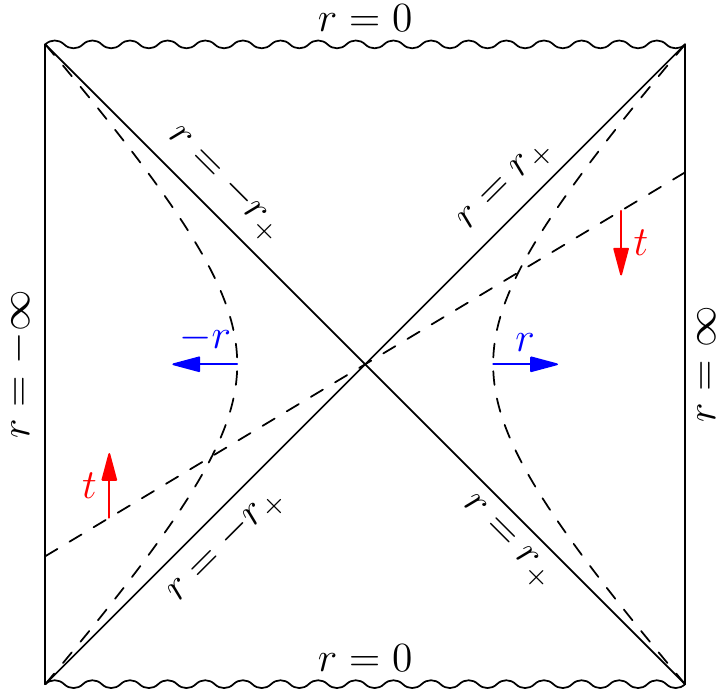}
    \end{subfigure}
    \hfill
    \begin{subfigure}[h]{0.44\linewidth}
        \includegraphics[width=\linewidth]{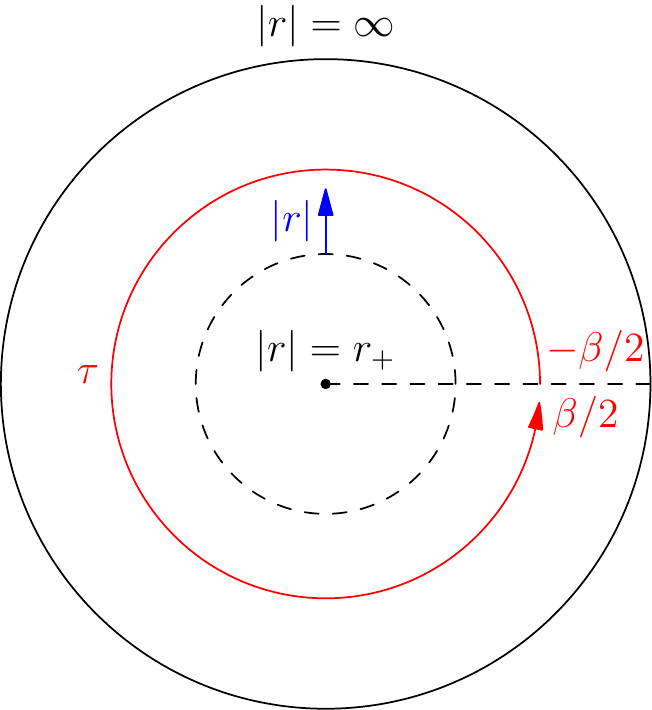}
    \end{subfigure}
    \caption{The non-rotating BTZ black hole: Lorentzian signature (left) and Euclidean signature (right). The Euclidean black hole has periodic time $\tau \sim \tau + \beta$, where $\beta = 1/T$.}
    \label{fig:black_hole_conformal_diagrams}
\end{figure}

\subsection{The Non-Rotating Wormhole}
To modify the non-rotating BTZ metric (\ref{eq:BTZmetric}) into a wormhole in the style of Solodukhin \cite{Solodukhin:2005qy} (see also \cite{Damour:2007ap}), we simply alter by hand one of the metric components, giving:
\begin{align}
    \begin{aligned}
        \label{eq:WH3metric} \dd{s}^2 & = -X(r) \dd{t}^2 + \frac{\dd{r}^2}{Y(r)} + r^2 \dd{\varphi}^2 \,,                                                            \\
        X(r)                          & =\frac{r^2 -r_\lambda^2}{R^2} \,, \quad    Y(r) = \frac{r^2 -r_+^2}{R^2} \,, \quad r_\lambda^2 \equiv r_+^2(1-\lambda^2) \,, \\
    \end{aligned} \qquad
    \begin{cases}
        \begin{aligned}
            r       & \in (-\infty,-r_+]\cup [r_+,\infty) \,, \\
            t       & \in (-\infty,\infty) \,,                \\
            \varphi & \sim \varphi + 2 \pi \,.
        \end{aligned}
    \end{cases}
\end{align}
For each value of the (small) dimensionless parameter $\lambda>0$, this gives us a wormhole geometry. For $r\gg r_+$, the smallness of $\lambda$ ensures that the metric looks almost like that of the original black hole. However, the global structure of spacetime is drastically altered, as there is no longer a horizon in the spacetime. Instead, we have essentially glued a second copy of this asymptotic flat spacetime (with $r<-r_+$) to the old horizon position, thus creating a traversable wormhole.

In \cite{Dimitrov:2020txx}, we discussed the stress energy tensor of this geometry and show that it requires the presence of exotic matter, localized roughly where the two throats meet. There, we also compute the holographic stress energy tensor and show that it differs from the one of the corresponding black hole at $\mathcal{O}(\lambda^2)$.

\begin{figure}[ht]
    \begin{center}
        \begin{subfigure}[c]{0.5\linewidth}
            \includegraphics[width=\linewidth]{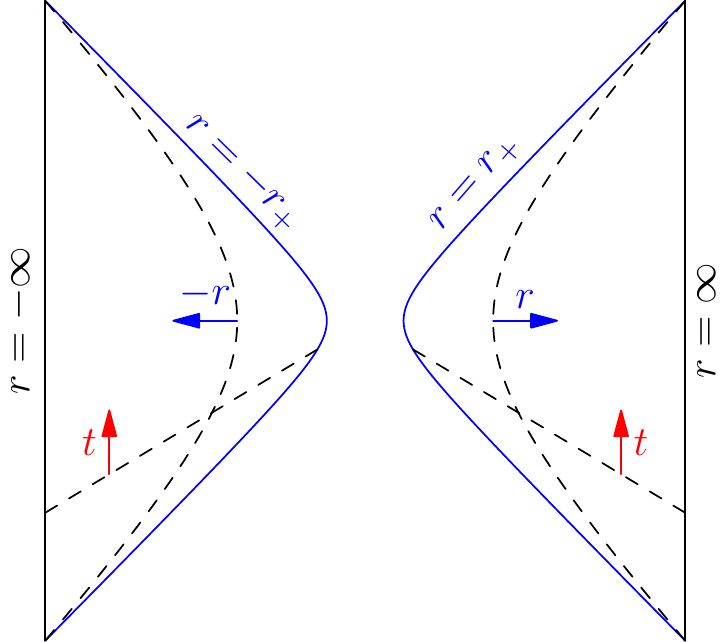}
        \end{subfigure}
        \ \ \ \ \ \ \ \ \ \ \ \
        \begin{subfigure}[c]{0.3\linewidth}
            \includegraphics[width=\linewidth]{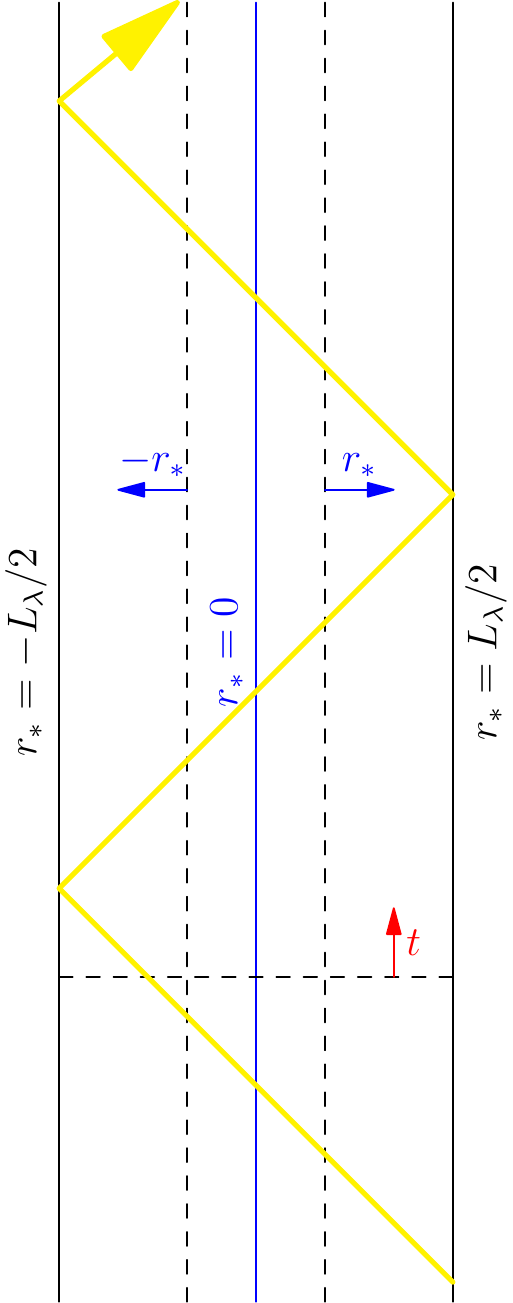}
        \end{subfigure}
    \end{center}
    \caption{Conformal diagrams of the Lorentzian non-rotating wormhole in terms of $r$ (left) and $r_*$ (right) coordinates. In the $r$ coordinate we see that the wormhole is created by a left and a right BTZ wedge glued together (the blue solid line) slightly outside of the BTZ horizon. The $r_*$ coordinate, which ranges continuously in $[-L_\lambda /2,L_\lambda /2]$, makes the causal structure more apparent: the glued surfaces $r=\pm r_+$ (on the left) both coincide with the surface $r_*=0$ (on the right). The gluing is smooth and a radial light-ray (yellow) can explore the entire wormhole by bouncing infinitely many times between the two boundaries. The diagrams are the same as in \cite{Dimitrov:2020txx}.}
    \label{fig:non_rot_wh}
\end{figure}

Further, \cite{Dimitrov:2020txx} calculated explicitly the tortoise coordinate of the extended, two-sided geometry, showing that it has a finite range.
  This allows us to define the effective length $L_\lambda$ of the wormhole as \cite{Dimitrov:2020txx}:
\be \label{eq:WH3defthroatlength} L_\lambda  \equiv \frac{R^2}{r_+} \log \frac{16}{\lambda^2} \,. \ee
For completeness, we present figure \ref{fig:non_rot_wh}, also appearing in \cite{Dimitrov:2020txx}, that shows that causal structure of the space time in $r$ and $r_*$ coordinates.

\subsection{The Scalar Wave Equation}\label{sec:scalarwave}
A minimally coupled scalar field $\Phi$ on the BTZ metric background (\ref{eq:BTZmetric}) or the wormhole metric (\ref{eq:WH3metric}) has the general equation of motion:
\be \nabla^2 \Phi = m^2 \Phi \,. \ee
We can solve this by a separation of variables:
\be \label{eq:scalar3Dsep} \Phi(t,r,\varphi) = e^{-i\omega t + i k \varphi} \Phi_r(r) \,, \ee
where the radial function $\Phi_r(r)$ must satisfy the differential equation:
\be \label{eq:scalar3Dradial} \Phi_r'' + \left( \frac{1}{r} + \frac{X'}{2X} + \frac{Y'}{2Y}\right) \Phi_r' + \left( \frac{\omega^2}{X Y} - \frac{k^2}{r^2 Y}-\frac{m^2}{Y}\right) \Phi_r = 0 \,. \ee
We are interested in finding the scalar wave solutions to (\ref{eq:scalar3Dradial}) in both the black hole and wormhole metrics. Note that for the BTZ black hole, the solutions to (\ref{eq:scalar3Dradial}) are explicitly known (see (\ref{eq:BTZ_modes})). 

To find the solutions to (\ref{eq:scalar3Dradial}) for the wormhole, note that we can introduce appropriate tortoise coordinates $r_*$ in both the black hole and wormhole geometries such that for $\phi = \Phi_r \sqrt{r/R}$, the equation (\ref{eq:scalar3Dradial}) becomes:
\be (\partial_{r_*}^2 - V(r_*))\phi(r_*) = 0.\ee
Further, using the relation between the respective black hole and wormhole tortoise coordinates, one finds:
\be V^\text{WH}(r_*) = \theta(r_*) V^\text{BH}(r_*) + \left(r_*\rightarrow -r_*\right) + \mathcal{O}(\lambda^2),\ee
so that the wormhole scalar potential is the same as the black hole potential on each side of the wormhole, up to $\mathcal{O}(\lambda^2)$ corrections. This means that the solutions on each side of the wormhole can be taken to be the black hole ones, $\phi^\text{WH}(r) = \phi^\text{BH}(r)$ and  $\phi^\text{WH}(-r) = \phi^\text{BH}(r)$. This must then be supplemented with a continuity condition at the center of the wormhole throat (at $r_*=0$), which gives
\begin{equation} \label{eq:rtmatchingcond}
    \Phi_r(r_t) = \Phi_r(-r_t) \,, \qquad \partial_r \Phi_r(r_t) = \partial_r \Phi_r (-r_t) \,,
\end{equation}
where we have defined the ``gluing point'' radius:
\be \label{eq:rstarnonrotWH} r_t = r_+\left(1 + \frac{\lambda^2}{8}\right) + \mathcal{O}(\lambda^4) \,. \ee
For more details on this matching, see \cite{Dimitrov:2020txx}.

\section{Correlators via Real Time Holography}\label{sec:SvR_calculations}
In this section, we will use the Skenderis-van Rees formalism of real-time holography \cite{Skenderis:2008dg,Skenderis:2008dh} to calculate various two-point correlation functions (Feynman, Wightman, and retarded) for a minimally coupled scalar field in various geometries of interest: the (non-rotating) BTZ black hole, the (non-rotating) Solodukhin wormhole, and a time-dependent geometry that transitions between two Solodukhin wormholes with different $\lambda$.

A brief overview as well as the relevant formulas of real-time holography and its preliminary Schwinger-Keldysh formalism are assembled in appendices \ref{sec:appendix_SK} and \ref{sec:appendix_SvR}.

\subsection{Correlators of the BTZ Black Hole}\label{sec:SvRBTZ}
Here, we use the real-time formalism to obtain the two-point functions of a minimally coupled scalar field in a non-rotating BTZ background. This was first done in the context of real-time holography in \cite{Skenderis:2008dg} and later expanded and generalized in \cite{vanRees:2009rw,Botta-Cantcheff:2018brv,Botta-Cantcheff:2019apr}; earlier works include \cite{Son:2002sd,Herzog:2002pc}. The calculation can also be generalized to consider minimally coupled gauge fields in this background \cite{deBoer:2018qqm}.

\begin{figure}[ht]
    \centering
    \begin{subfigure}[c]{0.48\textwidth}
        \includegraphics{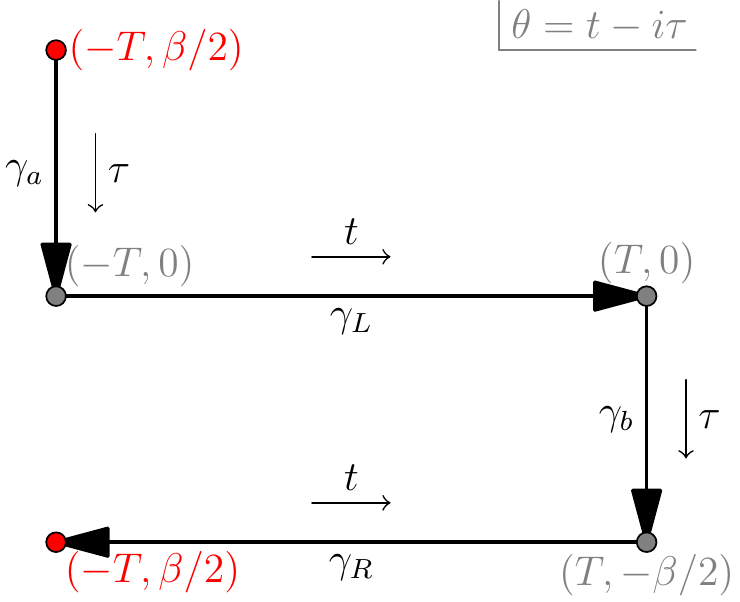}
    \end{subfigure}
    \begin{subfigure}[c]{0.48\textwidth}
        \includegraphics{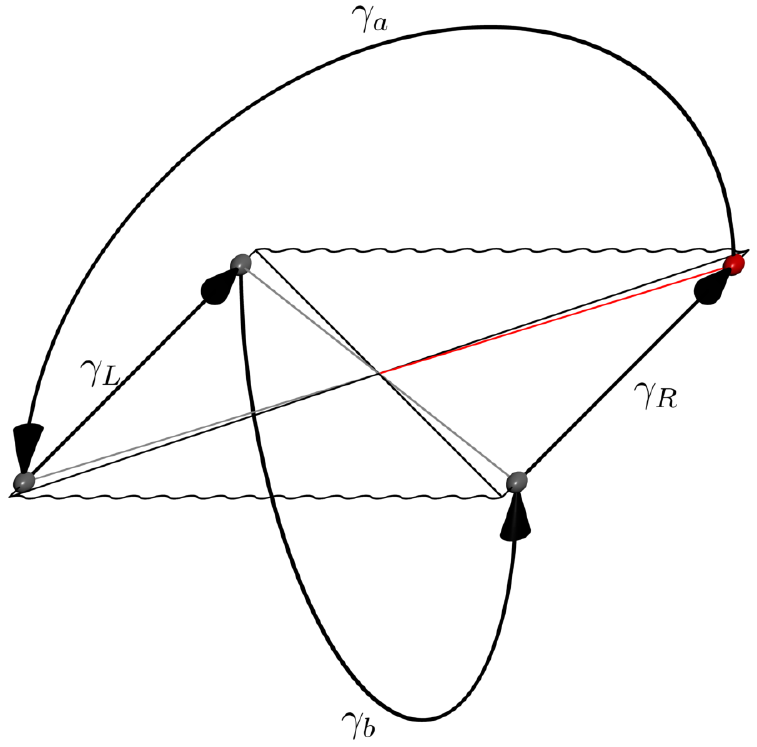}
    \end{subfigure}
    \caption{The thermal Schwinger-Keldysh contour in $\text{CFT}_2$ (left) and the mixed-signature bulk spacetime, corresponding to it (right). On the boundary contour (left), the grey dots signify gluing between an Euclidean and a Lorentzian segment and the red dots at $(-T,\pm\beta/2)$ are identified and also glued. On the bulk contour (right) the fields are glued at fixed time slices extending all the way into the bulk until the horizon at $r=r_+$ is reached.}
    \label{fig:bh_contours}
\end{figure}

\paragraph{Schwinger-Keldysh contour}
To find the BTZ two-point correlation function of a minimally coupled scalar field, we start with a thermal Schwinger-Keldysh contour (\ref{eq:general_SK_contour}) with $\sigma_a = \sigma_{b} = \sigma_c = \beta/2$ and identified endpoints \cite{Keldysh:1964ud,Skenderis:2008dg}.  Then the forward (resp. backward) running Lorentzian segments are interpreted as segments on the left (resp. right) boundary: $\gamma_1 \equiv \gamma_L$ and $\gamma_{2} \equiv \gamma_R$; the Euclidean segments are connected to both $L$ and $R$ akin to a Hartle-Hawking state \cite{Maldacena:2001kr,Israel:1976ur}. Thus, the ranges of the complex Keldysh time $\theta = t -i\tau$ are as follows; see figure \ref{fig:bh_contours} (left):
\begin{align}
    \begin{aligned}
        \gamma_a   & : \quad \theta \in \qty[-T+i\beta/2,-T] \,, \\
        \gamma_{b} & : \quad \theta \in \qty[T,T-i\beta/2] \,,
    \end{aligned} \qquad
    \begin{aligned}
        \gamma_L   & : \quad \theta \in \qty[-T,T]     \,,               \\
        \gamma_{R} & : \quad \theta \in \qty[-T-i\beta/2,T-i\beta/2] \,.
    \end{aligned}
    \label{eq:bh_SK_contour}
\end{align}
To construct the bulk dual to this field theory contour, we take the upper half of the thermal circle in Euclidean time, whose time ranges in $\tau \in [-\beta/2,0]$, and we glue it to the line $t=-T$ in the Lorentzian non-rotating BTZ (see figure \ref{fig:black_hole_conformal_diagrams}). Similarly, we glue the lower half of the thermal circle, whose time ranges in $\tau\in [0,\beta/2]$, to the line $t=T$; note that $\tau=\pm\beta/2$ are identified. The resulting bulk contour is depicted in figure \ref{fig:bh_contours} (right).
The bulk gluings dictate the following matching conditions that the piecewise bulk fields must obey:
\begin{align}
    \begin{aligned}
        \Phi_{a}(0,\varphi,\abs{r})       & = \Phi_{L}(-T,\varphi,-r) \,,            & \partial_{\tau} \Phi_{a}(0,\varphi,\abs{r})       & = -i\partial_{t}\Phi_{L}(-T,\varphi,-r)         \,,     \\
        \Phi_{L}(T,\varphi,-r)            & = \Phi_{b}(0,\varphi,\abs{r}) \,,        & -i\partial_{t} \Phi_{L}(T,\varphi,-r)             & = \partial_{\tau}\Phi_{b}(0,\varphi,\abs{r})       \,,  \\
        \Phi_{b}(\beta/2,\varphi,\abs{r}) & = \Phi_{R}(T,\varphi,r) \,,              & \partial_{\tau} \Phi_{b}(\beta/2,\varphi,\abs{r}) & = -i\partial_{t}\Phi_{R}(T,\varphi,r)             \,,   \\
        \Phi_{R}(-T,\varphi,r)            & = \Phi_{a}(-\beta/2,\varphi,\abs{r}) \,, & -i\partial_{t} \Phi_{R}(-T,\varphi,r)             & = \partial_{\tau}\Phi_{a}(-\beta/2,\varphi,\abs{r}) \,.
    \end{aligned}
    \label{eq:bh_matching_conditions}
\end{align}

\paragraph{Scalar wave solutions}
We can use separation of variables as in (\ref{eq:scalar3Dsep}) to solve the equations of motion for the minimally coupled scalar on the different pieces of the Schwinger-Keldysh contour:
\begin{align}
    \Phi_i= e^{-i \omega t} e^{ik\varphi} \tilde{s}_i(\omega,k) f(\omega,k,r) \,, \qquad \Phi_I = e^{-\omega \tau} e^{ik\varphi} \tilde{s}_I(\omega,k) f(\omega,k,r) \,,
\end{align}
with $i=\{L,R\}$ for the Lorentzian segments and $I=\{a,b\}$ for the Euclidean segments. The radial function $f(\omega,k,r)$ is a solution to the radial wave equation (\ref{eq:scalar3Dradial}) with the metric functions given in (\ref{eq:BTZmetric}) with $r_-=0$; the general solution is a linear combination of $f^{\pm}$ with:
\begin{align}
    f^{\pm}(\omega, k, r) & = \mathcal{N}^\pm_{\omega k}\left(1-\frac{r_+^2}{r^2}\right)^{\pm i r_+ \overline{\omega}/2}  \nonumber                                                                        \\
                          & \qquad \times \phantom{a}_2F_1\qty({i\over 2}\qty(\pm r_+ \overline{\omega} -\overline{k}),{i\over 2}\qty(\pm r_+ \overline{\omega} +\overline{k});1\pm i r_+ \overline{\omega} ;1-\frac{r_+^2}{r^2}),
    \label{eq:BTZ_modes}
\end{align}
where we have introduced the rescaled quantities $\overline{k} = R k/r_+$ and $\overline{\omega} = R^2 \omega/r_+^2$. At the horizon, these functions behave as:
\begin{align}
    f^{\pm}(\omega,k,r) \sim (r-r_+)^{\pm{i r_+\overline{\omega} \over 2}} \sim e^{\pm i \omega r_*},
\end{align}
where $r_*$ is the standard black hole tortoise coordinate, so that $f^+$ (resp. $f^-$) is identified with outgoing (resp. ingoing) modes.\footnote{We do not impose ingoing boundary conditions at the horizon so that we are in principle able to compute all correlators; imposing ingoing boundary conditions would mean we are computing the retarded correlator \cite{vanRees:2009rw,Son:2002sd,Herzog:2002pc}.}
We choose the normalization constants:
\begin{align}
    \mathcal{N}^\pm_{\omega k} = {\Gamma\qty(1\pm {i \over 2}(r_+ \overline{\omega} - \overline{k}))\Gamma\qty(1\pm {i \over 2}(r_+ \overline{\omega} + \overline{k})) \over \Gamma(1\pm ir_+ \overline{\omega})} \,,
    \label{eq:normalization_bh_modes}
\end{align}
such that at the UV boundary $r\rightarrow\infty$, we have $f^{\pm}\rightarrow 1$; the asymptotic expansion is then:
\begin{align}
    f^{\pm}(\omega,k,r)   & = 1+{r_+^2 \over r^2} \alpha(\omega,k) \qty( \beta^{\pm}(\omega,k)+\log{r_+^2 \over r^2}) + \dots \,,  \label{eq:UV_exp_non_rot_BTZ}                                                                                                       \\
    \alpha(\omega,k)      & = -{1 \over 4}\qty(r_+^2 \overline{\omega}^2 - \overline{k}^2 ) \nonumber                            \,,                                                                                                                                   \\
    \beta^{\pm}(\omega,k) & = \psi \qty(1+{i \over 2}\qty(\pm r_+ \overline{\omega} -\overline{k}) )+ \psi\qty(1+{i \over 2}\qty(\pm r_+ \overline{\omega} +\overline{k})) \pm {2ir_+\overline{\omega} \over r^2_+ \overline{\omega}^2 -\overline{k}^2} \,,  \nonumber
\end{align}
where $\psi(\cdot)$ is the digamma function and we have given $\beta^\pm$ up to constants which do not affect the poles of $\alpha \beta^{\pm}$. These poles of $f^\pm$ can be obtained from the poles of the normalization factors (or equivalently from the poles of $\alpha \beta^\pm$) and are located at
\begin{align}\label{eq:BTZQNMs}
    R \omega^{\pm}_{nk+} =  k \pm i \frac{2r_+}{R} n \,, \quad R\omega^{\pm}_{nk-} = - k \pm i \frac{2r_+}{R} n \,,   \qquad n \in \{1,2,3,\dots \} \,,  \quad k \in \mathbb{Z} \,.
\end{align}
The $\omega^-_{nk\pm}$ frequencies with negative imaginary part are the poles corresponding to the ingoing mode $f^-$ and thus are the quasinormal modes of the non-rotating BTZ black hole \cite{Cardoso:2001hn} which are exponentially damped at late times (corresponding to the scalar field leaking into the black hole horizon). Note that the poles of $f^\pm$ precisely correspond to the poles of the normalization factors $\mathcal{N}^\pm_{\omega k}$ in (\ref{eq:normalization_bh_modes}).

Finally, we define a more convenient basis of modes as
\begin{align}\label{eq:gdef}
    g^\pm = {1\over 2} (f^+ \pm f^-) \,,
\end{align}
where $g^+$ (resp. $g^-$) is the non-normalizable mode (resp. normalizable mode) as it asymptotes to $1$ (resp. $0$) in the UV ($r\rightarrow \infty$).

\begin{figure}[ht!]
    \centering
    \includegraphics{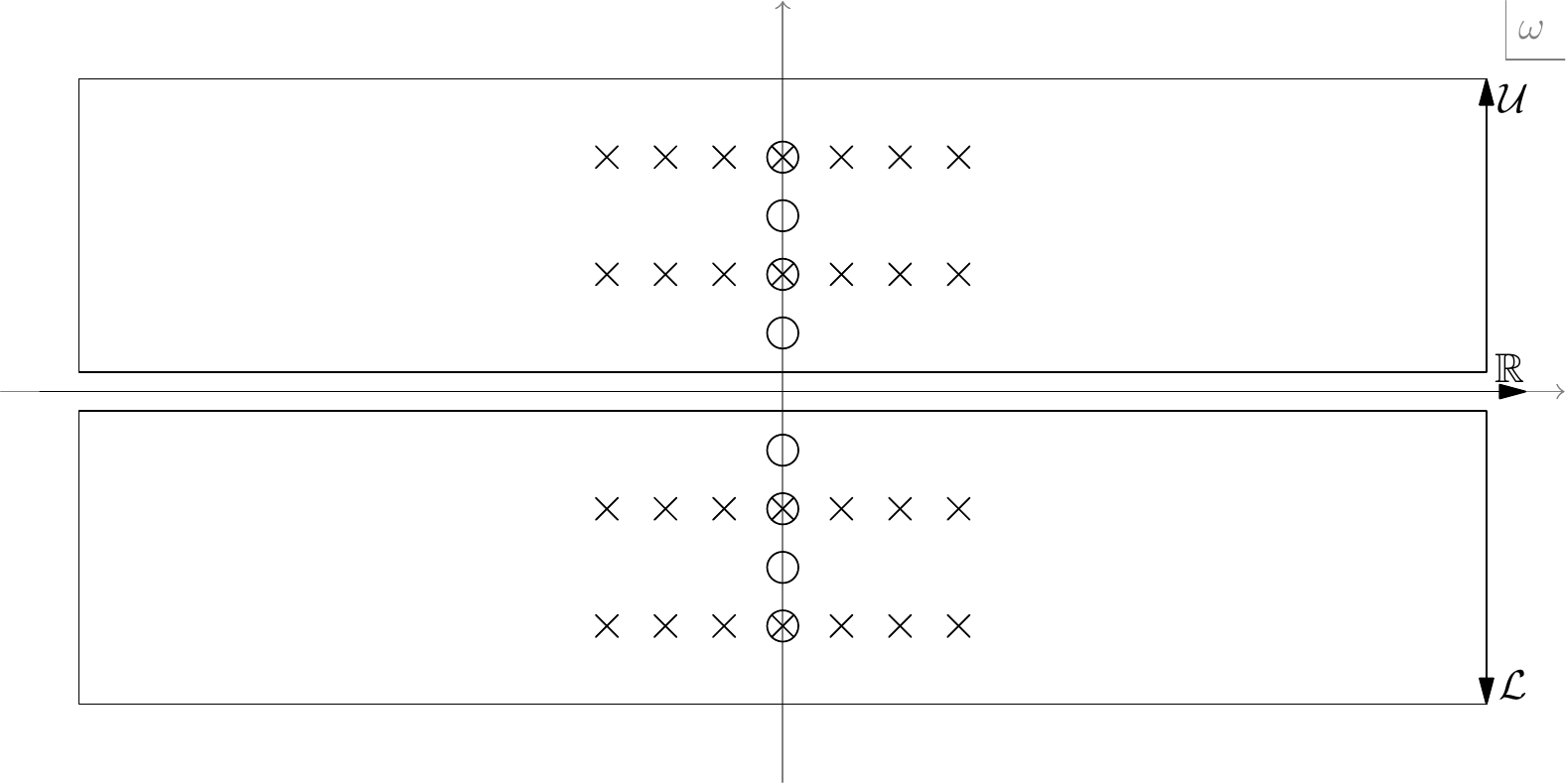}
    \caption{The contours $\mathcal{U},\mathcal{L},\mathbb{R}$ in the complex $\omega$-plane. The crosses in the upper (resp. lower) half plane are the poles of $f^+$ (resp. $f^-$); each vertical line and its reflection across the $y$-axis corresponds to the poles of $f^\pm$ at fixed $k$. The circles are the poles of $A_{ij}$ (i.e. the poles of $n(\omega)$), which we discuss in section \ref{sec:bh_pos_space_correlators}.}
    \label{fig:complexcontours}
\end{figure}

\paragraph{Constructing bulk fields piecewise}
Our next task is to construct the piecewise bulk fields. A generic Lorentzian field is expanded as
\begin{align}
    \begin{aligned}
        \Phi_i & = \sum_k e^{i k\varphi} \int_{\mathcal{C}_i}\dd{\omega} e^{-i\omega t} \qty(g^+\delta_{ij} + g^-A_{ij})\tilde{s}_j \,,
    \end{aligned}
    \label{eq:bh_lorentzian_field_ansatz}
\end{align}
where $A_{ij}$ are arbitrary meromorphic functions of $\omega$ and $\mathcal{C}_i$ is an arbitrary contour that avoids the poles of $A_{ij}$ and $g^\pm$.  Note that (\ref{eq:bh_lorentzian_field_ansatz}) indeed corresponds to introducing a source $s_i$ on the boundary:
\begin{align}\label{eq:FTofsBH}
    \lim_{r \rightarrow \pm \infty} \Phi_i = \sum_k e^{ik\varphi} \int_{\mathcal{C}_i}\dd{\omega} e^{-i\omega t} \delta_{ij}\tilde{s}_j \equiv s_i \,.
\end{align}
The real time formalism fixes both $\mathcal{C}_i$ and $A_{ij}$ by the matching conditions. We parametrize the freedom in $\mathcal{C}_i$ as:
\begin{align}
    \begin{aligned}
        \Phi_i & = \sum_k e^{ik\varphi}\left[ \int_{\mathbb{R}}\dd{\omega} e^{-i\omega t} \qty(g^+\delta_{ij} + g^-A_{ij})\tilde{s}_j \right.                                            \\
               & \qquad \left.    + \int_{\mathcal{U}} \dd{\omega} e^{-i\omega t} g^- U_{ij}\tilde{s}_j +  \int_{\mathcal{L}} \dd{\omega} e^{-i\omega t}g^- L^-_{ij}\tilde{s}_j \right].
    \end{aligned}\label{eq:BHlorentzfieldsUL}
\end{align}
Here $\mathcal{U}$ is a positively oriented contour around the upper-half-plane and $\mathcal{L}$ is a negatively oriented contour around the lower-half-plane, see figure \ref{fig:complexcontours}. The coefficients $U_{ij}, L_{ij}$ thus pick up additional normalizable mode contributions, related to the choice of contour in (\ref{eq:bh_lorentzian_field_ansatz}).

The Euclidean fields do not have sources as we are interested in the thermal state without extra excitations (see \cite{Botta-Cantcheff:2019apr} for a more general treatment), so they are composed solely of normalizable modes:
\begin{align}
    \begin{aligned}
        \Phi_I & = \sum_k e^{ik\varphi}\left[ \int_{\mathcal{U}} \dd{\omega} e^{-\omega \tau} g^- U_{Ij}\tilde{s}_j +  \int_{\mathcal{L}} \dd{\omega} e^{-\omega \tau}  g^-L_{Ij}\tilde{s}_j \right].
    \end{aligned}
\end{align}

\paragraph{Applying the matching conditions}
At early times $t\sim -T$ (resp. late times $t\sim T$), the sources all lie in the future (resp. past) so we must close the Feynman integral in the upper (resp. lower) half plane (see also the discussion in appendix \ref{sec:app:emptyAdS} and especially around (\ref{eq:ads_lorentzian_fields_at_early_and_late_times})), so that:
\begin{align}
    \begin{aligned}
        \Phi_i(t\sim -T) & = \sum_k e^{ik\varphi}\left[ \int_{\mathcal{U}} \dd{\omega} e^{-i\omega t} g^- \qty(\delta_{ij}+A_{ij}+U_{ij})\tilde{s}_j +  \int_{\mathcal{L}} \dd{\omega} e^{-i\omega t} g^- L_{ij}\tilde{s}_j \right],  \\
        \Phi_i(t\sim T)  & = \sum_k e^{ik\varphi}\left[ \int_{\mathcal{U}} \dd{\omega} e^{-i\omega t} g^- U_{ij}\tilde{s}_j +  \int_{\mathcal{L}} \dd{\omega} e^{-i\omega t} g^- \qty(-\delta_{ij}+A_{ij}+L_{ij})\tilde{s}_j \right],
    \end{aligned}
    \label{eq:bh_lor_field_at_late_and_early_times}
\end{align}
where we have used that:
\begin{align}
    \begin{aligned}
        \int_\mathcal{U}\dd{\omega} e^{-i\omega t} g^+\delta_{ij}  \tilde{s}_j & = \int_\mathcal{U}\dd{\omega} e^{-i\omega t} {1\over 2}f^+\delta_{ij} \tilde{s}_j = \int_\mathcal{U}\dd{\omega} e^{-i\omega t} g^-\delta_{ij} \tilde{s}_j \,,    \\
        \int_\mathcal{L}\dd{\omega} e^{-i\omega t} g^+ \delta_{ij} \tilde{s}_j & = \int_\mathcal{L}\dd{\omega} e^{-i\omega t} {1\over 2}f^-\delta_{ij} \tilde{s}_j = \int_\mathcal{L}\dd{\omega} e^{-i\omega t}(-g^-) \delta_{ij} \tilde{s}_j \,.
    \end{aligned}
\end{align}
Since each of the contours picks up independent pole residues (i.e. normalizable modes), the two contour integrals  $\mathcal{U}$, $\mathcal{L}$ must be independently matched, giving:
\begin{align}
     & \begin{aligned}
        \mathcal{U}: &  & U_{aj}e^{-i\omega T} & = \delta_{Lj}+A_{Lj} +U_{Lj}                 \,,           \\
                     &  & U_{bj}e^{i\omega T}  & = U_{Lj}                                        \,,        \\
                     &  & U_{bj}e^{i\omega T}  & = U_{Rj}e^{\beta \omega /2}                        \,,     \\
                     &  & U_{aj}e^{-i\omega T} & = \qty(\delta_{Rj}+A_{Rj} +U_{Rj})e^{-\beta \omega /2} \,,
    \end{aligned} &
     & \begin{aligned}
        \mathcal{L}: &  & L_{aj}e^{-i\omega T} & = L_{Lj}                               \,,                 \\
                     &  & L_{bj}e^{i\omega T}  & = -\delta_{Lj}+A_{Lj} +L_{Lj}             \,,              \\
                     &  & L_{bj}e^{i\omega T}  & = \qty(-\delta_{Rj}+A_{Rj} +L_{Rj})e^{\beta \omega /2} \,, \\
                     &  & L_{aj}e^{-i\omega T} & = L_{Rj}e^{-\beta \omega /2} \,.
    \end{aligned}
    \label{eq:BH_uncostrained_matching_equs}
\end{align}
First, we solve for the Lorentzian functions $U_{ij},L_{ij}$ in terms of $A_{ij}$:
\begin{align}
    \begin{aligned}
        U_{Lj} & = U_{Rj}e^{\beta \omega/2} = - (\delta_{Lj}+A_{Lj}){e^{\beta \omega} \over e^{\beta \omega} -1} + (\delta_{Rj}+A_{Rj}){e^{\beta \omega/2} \over e^{\beta \omega} -1} \,, \\
        L_{Lj} & = L_{Rj}e^{-\beta \omega/2} = - (\delta_{Lj}-A_{Lj}){1 \over e^{\beta \omega} -1} + (\delta_{Rj}-A_{Rj}){e^{\beta \omega/2} \over e^{\beta \omega} -1} \,.
    \end{aligned}\label{eq:BHULinfuncA}
\end{align}
Note that no factors of $e^{\pm i\omega T}$ appear in (\ref{eq:BHULinfuncA}) (as opposed to (\ref{eq:BH_uncostrained_matching_equs})). Now, regularity at late times of the $\mathcal{U}$ contour integral in (\ref{eq:bh_lor_field_at_late_and_early_times}) for $\omega\rightarrow i\infty$ forces us to set $U_{ij}=0$; similarly, regularity of the $\mathcal{L}$ integral at early times imposes $L_{ij}=0$. These regularity conditions, together with (\ref{eq:BH_uncostrained_matching_equs}), now fix all unknown functions uniquely:
\begin{align}
    \begin{aligned}
        A_{ij}               & =
        2\mqty(
        -(n+1/2)             & \sqrt{n(n+1)}                                                      \\
        -\sqrt{n(n+1)}       & (n+1/2)
        )_{ij}          \,,       & n(\omega)     & = {1 \over e^{\beta \omega} -1}    \,,             \\
        U_{aj}               & = 2\mqty(
        -n                   & \sqrt{n(n+1)}
        )_je^{i\omega T} \,, & U_{bj}        & = 0     \,,                                        \\
        L_{aj}               & =0 \,,        & L_{bj}                                 & = 2\mqty(
        -(n+1)               & \sqrt{n(n+1)}
        )_je^{-i\omega T} \,.
    \end{aligned}
    \label{eq: bh A func}
\end{align}
Note that $U_{aj}$ and $L_{bj}$ indeed have appropriate exponential tails in $\mathbb{R}$, coming from $n(\omega)$, that ensure regularity of the Euclidean fields when $\omega \rightarrow \pm \infty$; moreover, the factors of $e^{\pm i\omega T}$ ensure regularity (at all times) for $\omega \rightarrow \pm i\infty$.

\paragraph{Extracting the correlators}
Finally, having solved the bulk fields entirely, we can extract the correlation functions from them. For this, we extract the $r_+^2 / r^2$ term in the expansion of the Lorentzian fields (see appendix \ref{sec:appendix_SvR}):
\begin{align}
    \phi^{(2)}_i = {1\over 2}\sum_k e^{ik\varphi}\int_{\mathbb{R}} \dd{\omega} e^{-i\omega t} \qty[ \alpha\beta^+\qty(\delta_{ij}+A_{ij}) \tilde{s}_j + \alpha\beta^-\qty(\delta_{ij}-A_{ij}) \tilde{s}_j] \,.
\end{align}
Then, we use the Fourier transform $s(t,\varphi)$ of $\tilde{s}_j(\omega,k)$ defined in (\ref{eq:FTofsBH}) and perform functional derivatives with respect to $s_j(t',\varphi')$ to obtain the black hole correlators:
\begin{align}
    i G^{ij}_{\text{BH}}(x,x') & = {2(-1)^{\delta_{ij}-\delta_{iR}+1} \over i} \eval{ { \delta \phi^{(2)}_i(x)  \over \delta s_j(x')} }_{s_i=0}
    \nonumber                                                                                                                                                                                                                                                      \\
                               & = {(-1)^{\delta_{ij}-\delta_{iR}+1} \over 4 \pi^2 i} \sum_k e^{ik (\varphi-\varphi')}\int_{\mathbb{R}} \dd{\omega} e^{-i\omega (t-t')} \qty[ \qty(\delta_{ij}+A_{ij}) \alpha\beta^+ + \qty(\delta_{ij}-A_{ij}) \alpha\beta^-] \,.
\end{align}
Explicitly, we can write:
\begin{align}
    \begin{aligned}
        iG^{LL}_{\text{BH}}(x,x') & = {1 \over 2 \pi^2 i} \sum_k e^{ik (\varphi-\varphi')}\int_{\mathbb{R}} \dd{\omega} e^{-i\omega (t-t')} \qty[ -n \alpha\beta^+ + \qty(n+1) \alpha\beta^-] \,,                 \\
        iG^{LR}_{\text{BH}}(x,x') & = {1 \over 2 \pi^2 i} \sum_k e^{ik (\varphi-\varphi')}\int_{\mathbb{R}} \dd{\omega} e^{-i\omega (t-t')} \qty[ -\sqrt{n(n+1)} \alpha\beta^+ + \sqrt{n(n+1)} \alpha\beta^-] \,, \\
        iG^{RL}_{\text{BH}}(x,x') & = {1 \over 2 \pi^2 i} \sum_k e^{ik (\varphi-\varphi')}\int_{\mathbb{R}} \dd{\omega} e^{-i\omega (t-t')} \qty[ -\sqrt{n(n+1)} \alpha\beta^+ + \sqrt{n(n+1)} \alpha\beta^-] \,, \\
        iG^{RR}_{\text{BH}}(x,x') & = {1 \over 2 \pi^2 i} \sum_k e^{ik (\varphi-\varphi')}\int_{\mathbb{R}} \dd{\omega} e^{-i\omega (t-t')} \qty[ -(n+1) \alpha\beta^+ + n \alpha\beta^-] \,.
    \end{aligned}
    \label{eq:bh_correlators}
\end{align}
These correlators require a careful interpretation. First, we identify the Feynman correlator as $iG^F_{\text{BH}}\equiv iG^{LL}_{\text{BH}} $ and the anti-Feynman correlator as $iG^{\overline{F}}_{\text{BH}}\equiv iG^{RR}_{\text{BH}} = \qty[iG^{LL}_\text{BH}(x,x')]^*$. The interpretation of this is that the $R$-region Feynman correlator behaves as a $L$-region anti-Feynman correlator, in accordance with the notion that the time in $R$ in the thermofield-double state runs oppositely to the time in $L$, see figure \ref{fig:black_hole_conformal_diagrams}. Next, we note that $iG^{LR}_{\text{BH}} = iG^{RL}_{\text{BH}}$; they are not conjugate and should not be identified with Wightman functions. Instead, these functions measure correlations between the $L$ and $R$ regions as ``propagated through'' the non-traversable Einstein-Rosen bridge that connects them. As for the Wightman functions, our contour (\ref{eq:bh_SK_contour}) cannot produce them explicitly; rather, one would first need to introduce a `doubled' contour containing also a reverse-$L$ and a reverse-$R$ segments. However, a simple shortcut to find the Wightman functions without this calculation is to use (\ref{eq:wightman_correlators_in_terms_of_Feynmans}), which gives them in terms of the Feynman correlators.

We are most interested in the retarded propagator (see section \ref{sec:appendix_SK} and in particular (\ref{eq:GRphysical})), which according to (\ref{eq:retarded_and_advenced_correlators_in_terms_of_Feynmans}) is given by
\begin{align}
    \label{eq:BHretardedinfuncLLRR} iG^{R}_{\text{BH}}(x,x') & = \theta(t-t')[iG^{LL}_{\text{BH}}(x,x') - iG^{RR}_{\text{BH}}(x,x')] \nonumber                                                                                   \\
                                                             & = {\theta(t-t') \over 2 \pi^2 i} \sum_k e^{ik (\varphi-\varphi')}\int_{\mathbb{R}} \dd{\omega} e^{-i\omega (t-t')} \qty[ \alpha\beta^+ + \alpha\beta^-] \nonumber \\
                                                             & = {\theta(t-t') \over 2 \pi^2 i} \sum_k e^{ik (\varphi-\varphi')}\int_{\mathcal{L}} \dd{\omega} e^{-i\omega (t-t')} \alpha\beta^- \,,
\end{align}
where we used the pole structure of $\alpha\beta^{\pm}$ to simplify the integral; note that the contour $\mathcal{L}$ simply picks up all of the poles in the complex lower half plane (see figure \ref{fig:complexcontours}). We will find the explicit expression for this position-space propagator in section \ref{sec:bh_pos_space_correlators}.

\subsection{Correlators of the Solodukhin Wormhole}\label{sec:SvRWH}
Here, we will apply the real-time holography formalism to the non-rotating Solodukhin wormhole with metric (\ref{eq:WH3metric}).

\begin{figure}[ht]
    \centering
    \begin{subfigure}[c]{0.48\textwidth}
        \includegraphics{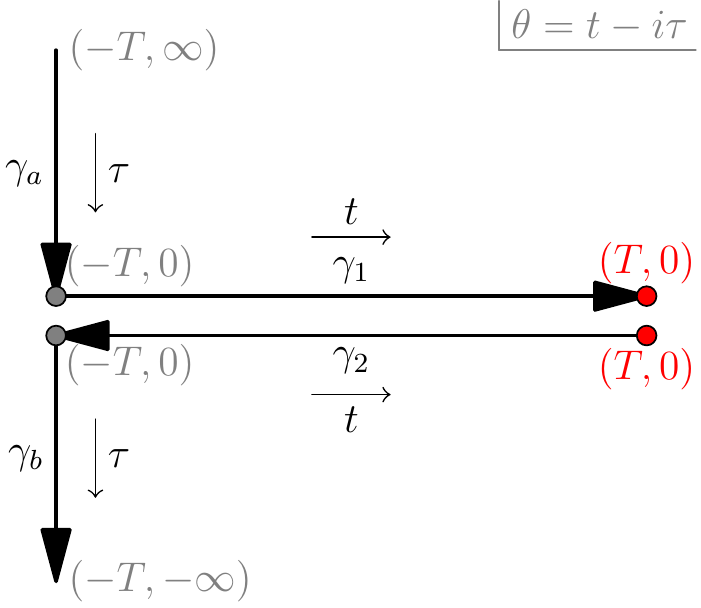}
    \end{subfigure}
    \begin{subfigure}[c]{0.48\textwidth}
        \includegraphics{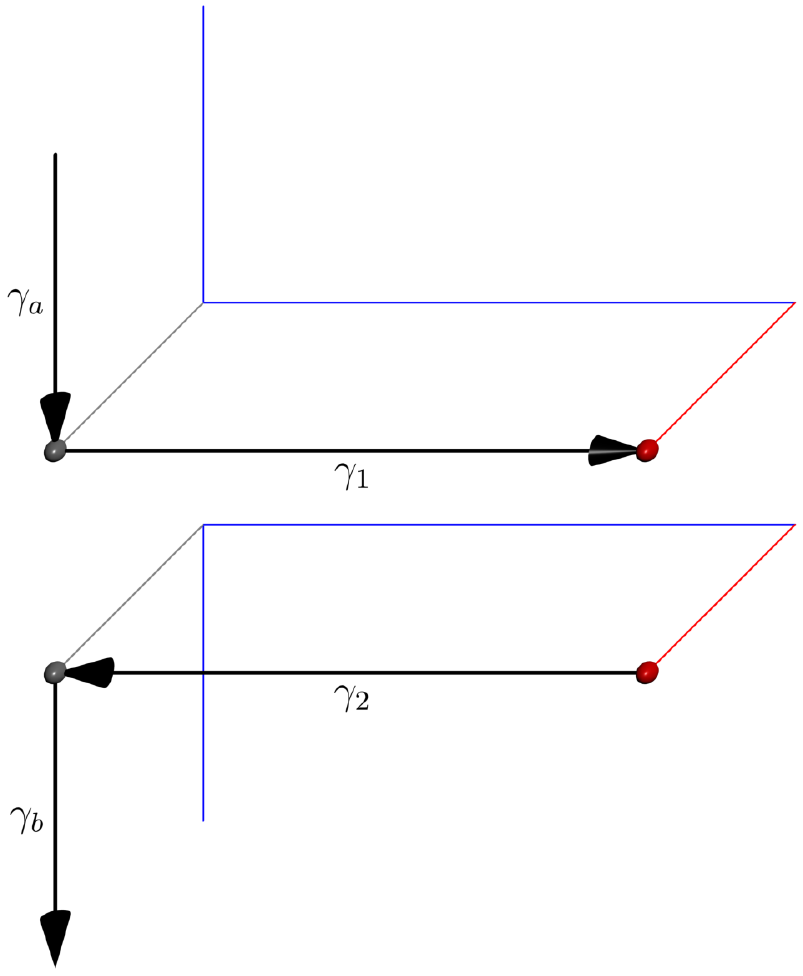}
    \end{subfigure}
    \caption{The pure-state Schwinger-Keldysh contour in $\text{CFT}_2$ (left) and the mixed-signature bulk spacetime, corresponding to it (right). On the boundary contour (left) the grey dots signify gluing between an Euclidean and a Lorentzian segment and the red dots are identified and also glued. On the bulk contour (right) the fields are glued on a fixed time slices extending all the way into the bulk. The blue line is the suface $r_*=0$. On this mixed-signature diagram we have included only the right side of the wormhole, but the reader should envision that the left side is glued to the blue line as in figure \ref{fig:non_rot_wh}.}
    \label{fig:pure_state_SK_contours}
\end{figure}

\paragraph{Schwinger-Keldysh contour} Since the wormhole is a pure state, we take a pure state (zero temperature) Schwinger-Keldysh contour, similar to that in empty AdS in appendix \ref{sec:appendix_SK}. This consists of two Euclidean ($\gamma_I$, $I=a,b$) and two Lorentzian ($\gamma_i$, $i=1,2$) segments, whose Keldysh time $\theta = t -i\tau$ runs as
\begin{align}
    \begin{aligned}
        \gamma_a   & : \quad \theta \in \qty[-T+i\infty,-T] \,, \\
        \gamma_{b} & : \quad \theta \in \qty[-T,-T-i\infty] \,,
    \end{aligned} \qquad
    \begin{aligned}
        \gamma_{1} & : \quad \theta \in \qty[-T,T] \,, \\
        \gamma_{2} & : \quad \theta \in \qty[-T,T] \,.
    \end{aligned}
    \label{eq:wh_SK_contour}
\end{align}
To construct the bulk dual to this contour, we follow the same strategy as for empty AdS (see appendix \ref{sec:app:emptyAdS}). We have two Lorentzian pieces and one Euclidean piece, which is split in two and glued to the the ends of the Lorentzian pieces --- see figure \ref{fig:pure_state_SK_contours}. These gluings dictate the matching conditions that the piecewise bulk fields obey:
\begin{align}
    \begin{aligned}
        \Phi_a(0,\varphi,r) & = \Phi_1(-T,\varphi,r) \,, & \partial_{\tau} \Phi_a(0,\varphi,r) & = -i \partial_{t}\Phi_1(-T,\varphi,r) \,, \\
        \Phi_1(T,\varphi,r) & = \Phi_2(T,\varphi,r)  \,, & \partial_{t} \Phi_1(T,\varphi,r)    & =  \partial_{t}\Phi_2(T,\varphi,r) \,,    \\
        \Phi_b(0,\varphi,r) & = \Phi_2(-T,\varphi,r) \,, & \partial_{\tau}\Phi_b(0,\varphi,r)  & = -i\partial_{t} \Phi_2(-T,\varphi,r) \,.
    \end{aligned}\label{eq:SvRWHmatchingconds}
\end{align}

\paragraph{Scalar wave solutions}
As discussed in section \ref{sec:scalarwave}, the solutions of the scalar wave equation to $\mathcal{O}(\lambda^2)$ on each side of the wormhole are those of the BTZ black hole itself, i.e. the modes $f^\pm$ in (\ref{eq:BTZ_modes}). These need to be supplemented with the continuity condition (\ref{eq:rtmatchingcond}) at $r_t = r_+(1+\lambda^2/8)$. Note that at the gluing point $\pm r_t$, the modes $f^{\pm}$ behave as:
\begin{align}
    \begin{aligned}
        f^+(\omega,k,\pm r_t)            & = \mathcal{N}^+_{\omega k} \left(\frac{\lambda}{2}\right)^{i r_+ \overline{\omega}}                       \,,                       & f^-(\omega,k,\pm r_t)            & = \mathcal{N}^-_{\omega k} \left(\frac{\lambda}{2}\right)^{- i r_+\overline{\omega}}                  \,,                           \\
        \partial_r f^+(\omega,k,\pm r_t) & = \pm {4i \overline{\omega} \over \lambda^2} \mathcal{N}^+_{\omega k} \left(\frac{\lambda}{2}\right)^{ i r_+ \overline{\omega}} \,, & \partial_r f^-(\omega,k,\pm r_t) & = \mp {4i \overline{\omega} \over \lambda^2} \mathcal{N}^-_{\omega k} \left(\frac{\lambda}{2}\right)^{-i r_+ \overline{\omega}} \,,
    \end{aligned}
    \label{eq:bh_modes_at_gluing_point}
\end{align}
where we remember that the normalization factors $\mathcal{N}^\pm_{\omega k}$ are given by (\ref{eq:normalization_bh_modes}).

We are interested in the holographic correlators for the CFT that sits on the boundary on one side --- say, the right, i.e. $r>r_t$ --- of the wormhole. Thus, the holographic sources live here, at $r\rightarrow \infty$. It is then natural to impose that there are no sources present on the \emph{other}, i.e. left, boundary at $r\rightarrow -\infty$. This means that the only allowed mode must be proportional to $f^+-f^-$ for $r\leq -r_t$, which, using (\ref{eq:bh_modes_at_gluing_point}), means the allowed mode for $r>r_t$ is:\footnote{The expression for $f^{\lambda}$ on the left side of the wormhole can be inferred from (\ref{eq:deftildef}), (\ref{eq:bh_modes_at_gluing_point}), and the matching conditions (\ref{eq:rtmatchingcond}). Note that specifying the wavefunction on (only) the right side uniquely determines the wavefunction on the left side (and vice versa). In particular, the mode (\ref{eq:deftildef}) is well-defined and well-behaved on the right side, and from symmetry it will then also be so on the left side.}
\be \label{eq:deftildef} f^{\lambda} \equiv \frac{1}{1-\kappa_{\lambda}^2} (f^+ - \kappa_{\lambda}^2 f^-) \,,  \qquad \kappa_{\lambda}(\omega,k) \equiv \frac{\mathcal{N}_{\omega k}^+}{\mathcal{N}_{\omega k}^-}\left(\frac{\lambda}{2}\right)^{i\, 2 r_+\overline{\omega}} \,, \ee
which has been normalized so that $\lim_{r\rightarrow\infty} f^{\lambda} = 1$. Note that imposing the condition of no sources on the left boundary is the analogue of the IR regularity condition in empty AdS (see appendix \ref{sec:app:emptyAdS}) that also eliminates one linear combination of the scalar wave solutions.

In principle, we are free to choose different, fixed boundary conditions on the left boundary (including a non-normalizable piece). Alternatively, we could have performed a holographic analysis on the direct product of the CFTs on the right \emph{and} left boundaries, introducing sources on both boundaries. This allows us to also compute the ``mixed'' left-right correlation functions (similar to the $LR$ and $RL$ correlators of the eternal black hole in (\ref{eq:bh_correlators})); the correlators we are calculating here are then simply the ``right-right'' correlators (where functional derivatives are taken with respect to the right sources and the left sources are set to zero).

While the BTZ black hole modes $f^+$ (resp. $f^-$) have poles in the upper (resp. lower) complex $\omega$ plane, a careful analysis shows that $f^{\lambda}$ does \emph{not} have these (or other) poles with non-zero imaginary part. Instead, $f^{\lambda}$ only has poles on the real line, at locations where $\kappa_{\lambda} = \pm 1$. Explicitly, these normal modes are the solutions to the transcendental equation:
\begin{align}
    {\Gamma\qty(1 + {i \over 2}(r_+ \overline{\omega} - \overline{k}))\Gamma\qty(1 + {i \over 2}(r_+ \overline{\omega} + \overline{k}))\Gamma(1 - ir_+ \overline{\omega}) \over \Gamma\qty(1 - {i \over 2}(r_+ \overline{\omega} - \overline{k}))\Gamma\qty(1 - {i \over 2}(r_+ \overline{\omega} + \overline{k}))\Gamma(1+ ir_+ \overline{\omega})} \left(\frac{\lambda}{2}\right)^{i\, 2r_+\overline{\omega}}=\pm 1 \,.
    \label{eq:wh_normal_modes}
\end{align}
This equation cannot be solved analytically in general; however, there are two frequency regimes where the (approximate) solution is simple. First of all, in the ``low-frequency'' regime:
\be \label{eq:SvRWHlowfreqregime} \omega \sim \frac{1}{L_\lambda} \,, \ee
where we used the wormhole throat length definition (\ref{eq:WH3defthroatlength}); note that $L_\lambda =R^2/r_+ \log (16\lambda^{-2})\gg R,r_+$ (i.e. it is the largest scale in the solution). Then, we have the solutions:
\be \label{eq:SvRWHlowfreqNMs} \omega_n \approx n\frac{\pi}{L_\lambda} \,,  \qquad n\in \mathbb{Z} \,. \ee
Second, in the ``high-frequency'' regime:\footnote{In this limit, we will also always assume $r_+\overline{\omega} \gg \overline{k}$.}
\be \label{eq:SvRWHhighfreqregime} \omega \gg \frac{r_+}{R^2} \,, \ee
or equivalently $\overline{\omega}\gg r_+^{-1}$; note that then automatically $\omega\gg L_\lambda^{-1}$.
In this limit, we find:
\be \label{eq:SvRWHhighfreqNMs} \omega_n \approx \left( n \pm \frac12\right)\frac{\pi}{L_\lambda} \,,  \qquad n\in \mathbb{Z} \,, \ee
Curiously, these differ only from the low frequency normal modes (\ref{eq:SvRWHlowfreqNMs}) by a shift of $\pi/(2L_\lambda)$. The intermediate frequency regime must then interpolate between the solutions (\ref{eq:SvRWHlowfreqNMs}) and (\ref{eq:SvRWHhighfreqNMs}). In any case, we can conclude that the modes are (approximately) equally spaced with a spacing inversely proportional to the wormhole length. Note that the wormhole thus only has \emph{normal} modes and no quasinormal modes, as appropriate for a pure state in AdS,\footnote{Without a horizon and in AdS asymptotics, there is nowhere that the scalar field can ``leak'' out to, so the amplitude of an eigenfunction cannot decrease. This is in contrast to Damour-Solodukhin wormholes in flat space, which have quasinormal modes (with a very small imaginary part) since the scalar wavefunction can ``leak'' out to flat infinity \cite{Bueno:2017hyj}.} such as also empty AdS as discussed in appendix \ref{sec:app:emptyAdS}.

\begin{figure}[ht!]
    \centering
    \includegraphics{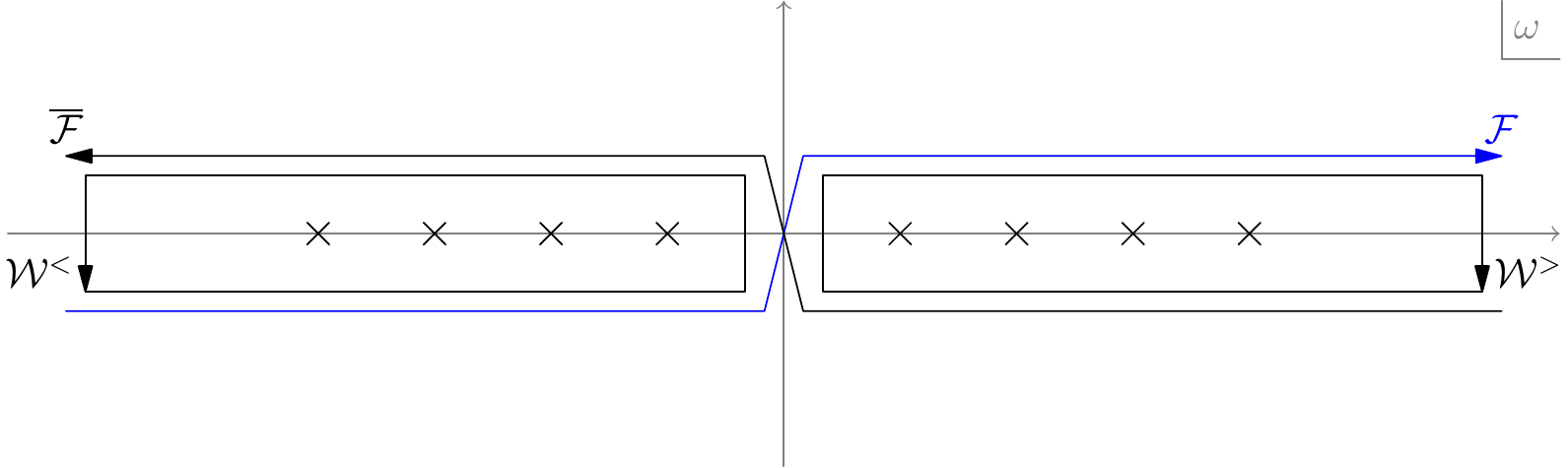}
    \caption{The contours $\mathcal{F}, \overline{\mathcal{F}},\mathcal{W}^>,\mathcal{W}^<$ in the complex $\omega$-plane. The crosses on the real line are the poles of $\tilde{f}$.}
    \label{fig:omega_plane_contours_AdS/WH}
\end{figure}

\paragraph{Constructing bulk fields piecewise}
Our next task is to construct the piecewise bulk fields; this construction is analogous to the construction of the fields in empty AdS in appendix \ref{sec:app:emptyAdS}. The Lorentzian fields are expanded as
\begin{align}
    \begin{aligned}
        \Phi_i & = \sum_k e^{i k\varphi} \int_{\mathcal{C}_i}\dd{\omega} e^{-i\omega t} f^{\lambda} \delta_{ij} \tilde{s}_{j} \,,
    \end{aligned}
    \label{eq:wh_lorentzian_field_ansatz}
\end{align}
where in this notation $i,j=\qty{1,2}$ specify whether we are on the forward or the backward Lorentzian bulk piece. This ansatz has the correct UV asymptotics since:
\begin{align}\label{eq:FTofsWH}
    \begin{aligned}
        \lim_{r \rightarrow \infty} \Phi_i   & = \sum_k e^{ik\varphi} \int_{\mathcal{C}_i}\dd{\omega} e^{-i\omega t} f^{\lambda}\delta_{ij}\tilde{s}_{j} \equiv s_{i} \,, \\
        \lim_{r \rightarrow - \infty} \Phi_i & = \sum_k e^{ik\varphi} \int_{\mathcal{C}_i}\dd{\omega} e^{-i\omega t} f^{\lambda}\delta_{ij}\tilde{s}_{j} \equiv 0 \,.
    \end{aligned}
\end{align}
We parametrize the freedom in choosing the contour $\mathcal{C}_i$ by writing (again, precisely analogous to empty AdS in appendix \ref{sec:app:emptyAdS}):
\begin{align}
    \begin{aligned}
        \Phi_i & = \sum_k e^{ik\varphi}\left[ \int_{\mathcal{F}}\dd{\omega} e^{-i\omega t} f^{\lambda}\delta_{ij} \tilde{s}_{j} + \int_{\mathcal{W}^<}\dd{\omega} e^{-i\omega t} f^{\lambda} V_{ij} \tilde{s}_{j}  +  \int_{\mathcal{W}^>}\dd{\omega} e^{-i\omega t}f^{\lambda} W_{ij} \tilde{s}_{j} \right].
    \end{aligned}
    \label{eq:wh_lor_fields}
\end{align}
Here, $V_{ij}$ and $W_{ij}$ are arbitrary analytic functions of $\omega$, $\mathcal{F}$ is a Feynman contour, $\mathcal{W}^>$ is a negatively oriented contour around $\mathbb{R}_{>0}$ and $\mathcal{W}^<$ is a positively oriented contour around $\mathbb{R}_{<0}$, see figure \ref{fig:omega_plane_contours_AdS/WH}.
Note that, as opposed to the black hole field ansatz (\ref{eq:BHlorentzfieldsUL}), we do not include a term with an upper or lower complex plane contour $\mathcal{U}$ or $\mathcal{L}$ as $f^{\lambda}$ (and $V_{ij}$ and $W_{ij}$) have no poles there.
The Euclidean fields, since there are no Euclidean sources, are decomposed as
\begin{align}
    \begin{aligned}
        \Phi_I & = \sum_k e^{ik\varphi}\left[ \int_{\mathcal{W}^<}\dd{\omega} e^{-\omega \tau}f^{\lambda} V_{Ij}\tilde{s}_{j} +  \int_{\mathcal{W}^>}\dd{\omega} e^{-\omega \tau}f^{\lambda} W_{Ij} \tilde{s}_{j} \right].
    \end{aligned}\label{eq:SvRWHEfields}
\end{align}
Regularity on $\gamma_a$ ($\tau<0$) as $\omega\rightarrow +\infty$ further fixes $W_{aj}=0$, and regularity on $\gamma_b$ ($\tau>0$)  as $\omega\rightarrow -\infty$ fixes $V_{bj} = 0$.

\paragraph{Applying matching conditions}
The Schwinger-Keldysh contour (\ref{eq:wh_SK_contour}), the piecewise bulk fields (\ref{eq:wh_lor_fields}) and (\ref{eq:SvRWHEfields}), and the matching conditions (\ref{eq:SvRWHmatchingconds}) are precisely the same as those of empty AdS discussed in appendix \ref{sec:app:emptyAdS} after substituting the AdS scalar mode $f$ with the wormhole mode $f^{\lambda}$ used here. Thus, applying and solving the matching conditions proceeds precisely the same way as in appendix \ref{sec:app:emptyAdS}, and the results are:
\begin{align}
    \begin{aligned}
        V_{ij} & = \mqty(0 & -1                                                                 \\ 0 & -1) \,, & W_{ij} &= \mqty(0 & 0 \\ 1 & -1) \,, \\
        V_{aj} & = \mqty(1 & -1) e^{i\omega T} \,, & W_{bj} & = \mqty(1 & -1) e^{i\omega T} \,.
    \end{aligned}
\end{align}
The Euclidean and Lorentzian fields are thus completely determined in terms of the sources $\tilde{s}_i$.

\paragraph{Extracting the correlators}
Extracting the correlators of the wormhole happens entirely analogous to empty AdS in appendix \ref{sec:app:emptyAdS} (again, with the substitution of the AdS mode $f$ by $f^{\lambda}$). The result for the retarded correlator is then:
\begin{align}
    \nonumber iG^{R}_{\text{WH}}(x,x') & = {\theta(t-t') \over 2 \pi^2 i} \sum_k e^{ik (\varphi-\varphi')} \int_{\mathcal{F}-\overline{\mathcal{F}}}\dd{\omega} e^{-i\omega (t-t')} \alpha \beta^{\lambda}  \,,                     \\
    \nonumber                          & = {\theta(t-t') \over 2 \pi^2 i} \sum_k e^{ik (\varphi-\varphi')} \int_{\mathbb{R}+i\epsilon}\dd{\omega} e^{-i\omega (t-t')} \alpha \beta^{\lambda} \,,                                    \\
                                       & = {\theta(t-t') \over 2 \pi^2 i} \sum_k e^{ik (\varphi-\varphi')} \int_{\mathcal{W}^>-\mathcal{W}^<}\dd{\omega} e^{-i\omega (t-t')} \alpha \beta^{\lambda} \,,     \label{eq: wh retarded}
\end{align}
where in the second line we have used the relation $\mathcal{F}-\overline{\mathcal{F}} = \mathbb{R}+i\epsilon$, with $\epsilon>0$ a small, real, and positive parameter, and in the third line we have used that $\theta(t-t')$ forces the contour of $\mathbb{R}+i\epsilon$ to close in the lower half plane, see figure \ref{fig:omega_plane_contours_AdS/WH}. Note that
\begin{equation}
    \beta^{\lambda} = \frac{1}{1-\kappa_{\lambda}^2} \qty(\beta^+ - \kappa_{\lambda}^2 \beta^-) \,,
\end{equation}
and $\alpha$ is the same as for the BTZ black hole, see (\ref{eq:UV_exp_non_rot_BTZ}).

\subsection{Feynman Correlator of a Wormhole Transition}\label{sec:wormhole_transition}
The real-time holography formalism can also be used to calculate ``off-diagonal'' correlators such as $\langle \lambda' | \mathcal{O}\mathcal{O} | \lambda \rangle$, where we denote the light scalar operator dual to the probe scalar field $\Phi$ by $\mathcal{O}$. To calculate such a correlator, one must prepare Euclidean states with wormhole parameter $\lambda$, resp. $\lambda'$ on $\gamma_a$, resp. $\gamma_b$ (using the notation of (\ref{eq:wh_SK_contour})). The Lorentzian piece on $\gamma_{1,2}$ is then in principle completely determined by the boundary conditions obtained by ``gluing'' the two Euclidean sections to the Lorentzian piece appropriately.

In this way, the Lorentzian piece \emph{must} become time-dependent, as it interpolates between a geometry with wormhole parameter $\lambda$ at (Lorentzian) time $t=-T$ and a geometry with parameter $\lambda'$ at $t=+T$. We will model this time-dependence as a ``sudden'' transition at time $t=t_0$ between the two wormhole states $\ket{\lambda}, \ket{\lambda'}$. We will demand that $|\lambda-\lambda'|\ll \lambda$ so that this procedure can provide a meaningful approximation.\footnote{Note that the Solodukhin wormhole does not satisfy the (vacuum) Einstein equations in any case, and that we have never specified explicitly what matter would be necessary to support it. As long as $|\lambda-\lambda'|\ll \lambda$, then the time-dependent Lorentzian ``transition'' geometry will actually have the same approximate energy-momentum tensor supporting it as the original (time-independent) wormhole.}

Having set up the problem in this simplified way, we will now show that it is relatively straightforward to obtain the explicit results for the ``off-diagonal'' momentum space correlator to $\mathcal{O}(\lambda-\lambda')$. We will consider (only) the Feynman correlator, as this is the natural one to consider when the (heavy) initial and final states are not the same.

\begin{figure}[ht!]
    \centering
    \includegraphics{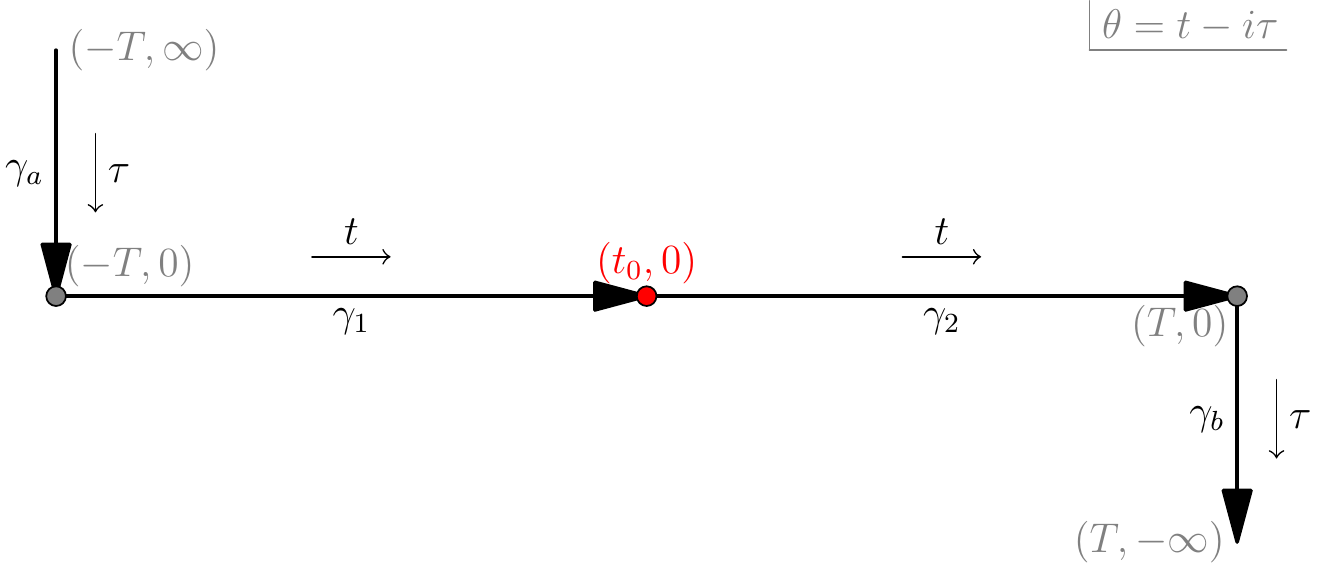}
    \caption{The in-out Schwinger-Keldysh contour appropriate for a transition between wormhole states. On the segments $\gamma_a$ and $\gamma_1$ the state is $\ket{\lambda}$. Then at Lorentzian time $t=t_0$ a ``heavy'' operator is suddenly applied (red bullet on the diagram) that changes the state to $\ket{\lambda'}$. This new state then forms that background on the segments $\gamma_2$ and $\gamma_b$.}
    \label{fig:wh_trans_SK}
\end{figure}

\paragraph{Schwinger-Keldysh contour} We take an in-out Schwinger-Keldysh contour consisting of two Euclidean ($\gamma_I, \   I=a,b$) and two Lorentzian ($\gamma_i, \ i=1,2$) segments, whose Keldysh time $\theta = t-i\tau$ runs as
\begin{align}
    \begin{aligned}
        \gamma_a   & : \quad \theta \in \qty[-T+i\infty,-T] \,, \\
        \gamma_{b} & : \quad \theta \in \qty[T,T-i\infty] \,,
    \end{aligned} \qquad
    \begin{aligned}
        \gamma_{1} & : \quad \theta \in \qty[-T,t_0] \,, \\
        \gamma_{2} & : \quad \theta \in \qty[t_0,T] \,.
    \end{aligned}
    \label{eq:wh_transition_SK_contour}
\end{align}
As mentioned above, we imagine that at time $t=t_0$, there is an instantaneous change in the wormhole state from $\ket{\lambda}$ to $\ket{\lambda'}$. We depict this contour in figure \ref{fig:wh_trans_SK}; note that this contour is set up to calculate the Feynman correlator as mentioned above.
The bulk version of this Schwinger-Keldysh contour follows analogously to the bulk versions that we previously considered and essentially fills in the bulk in such a way that the bulk fields obey smoothness conditions (in $r$) at the two different throat positions. Thus, we have the (temporal) matching conditions:
\begin{align}
    \begin{aligned}
        \Phi_a(0,\varphi,r)   & = \Phi_1(-T,\varphi,r) \,,   & \partial_{\tau} \Phi_a(0,\varphi,r) & = -i \partial_{t}\Phi_1(-T,\varphi,r) \,, \\
        \Phi_1(t_0,\varphi,r) & = \Phi_2(t_0,\varphi,r)  \,, & \partial_{t} \Phi_1(t_0,\varphi,r)  & =  \partial_{t}\Phi_2(t_0,\varphi,r) \,,  \\
        \Phi_b(0,\varphi,r)   & = \Phi_2(T,\varphi,r) \,,    & \partial_{\tau}\Phi_b(0,\varphi,r)  & = -i\partial_{t} \Phi_2(T,\varphi,r) \,.
    \end{aligned}\label{eq:wh_trans_mathching_cond}
\end{align}
These matching equations are valid at all $r$ from the throat position at $r_t$ to infinity; however, note that the throat position is now dependent on the time (i.e. whether $t<t_0$ or $t>t_0$). For $t=t_0$ we can take the lower bound on $r$ as $\min( r_{t,\lambda},r_{t,\lambda'})$.

\paragraph{Scalar wave solutions}
We follow the same reasoning as in section \ref{sec:SvRWH}. In particular, we choose to only put sources on the right side of the wormhole. Since for $t<t_0$, resp. $t>t_0$, the wormhole geometry is given by the time-independent geometry with wormhole parameter $\lambda$, resp. $\lambda'$, it follows that the scalar solutions are simply $f^\lambda$, resp. $f^{\lambda'}$, given by (\ref{eq:deftildef}). Note that these functions have normal mode poles specified by (\ref{eq:wh_normal_modes}) with $\lambda$, resp. $\lambda'$.

\paragraph{Constructing bulk fields piecewise} We express the Lorentzian fields as
\begin{align}
    \begin{aligned}
        \Phi_1 & = \sum_k e^{ik\varphi}\left[ \int_{\mathcal{F}}\dd{\omega} e^{-i\omega t} f^{\lambda}\delta_{1j} \tilde{s}_{j} + \int_{\mathcal{W}^<}\dd{\omega} e^{-i\omega t} f^{\lambda} V_{1j} \tilde{s}_{j}  +  \int_{\mathcal{W}^>}\dd{\omega} e^{-i\omega t}f^{\lambda} W_{1j} \tilde{s}_{j} \right],    \\
        \Phi_2 & = \sum_k e^{ik\varphi}\left[ \int_{\mathcal{F}}\dd{\omega} e^{-i\omega t} f^{\lambda'}\delta_{2j} \tilde{s}_{j} + \int_{\mathcal{W}^<}\dd{\omega} e^{-i\omega t} f^{\lambda'} V_{2j} \tilde{s}_{j}  +  \int_{\mathcal{W}^>}\dd{\omega} e^{-i\omega t}f^{\lambda'} W_{2j} \tilde{s}_{j} \right],
    \end{aligned}
    \label{eq:wh_trans_lor_fields}
\end{align}
where, for convenience, we have split the Lorentzian source into two parts $s_1,s_2$, where the source $s_1(t,\varphi)$ is defined to vanish for $t \geq t_0$ and the source $s_2(t,\varphi)$ is defined to vanish for $t \leq t_0$. As before, we also express the regulated, sourceless Euclidean fields as
\begin{align}
    \begin{aligned}
        \Phi_a & = \sum_k e^{ik\varphi} \int_{\mathcal{W}^<}\dd{\omega} e^{-\omega \tau} f^\lambda V_{aj}\tilde{s}_j \,,     \\
        \Phi_b & = \sum_{k} e^{ik\varphi} \int_{\mathcal{W}^>}\dd{\omega} e^{-\omega \tau}f^{\lambda'} W_{bj}\tilde{s}_j \,.
    \end{aligned}
\end{align}
Near the matching points, the Lorentzian fields behave as (closing the contours appropriately):
\begin{align}
    \begin{aligned}
        \Phi_1(t \sim -T)  & = \sum_{k}e^{ik\varphi}\left[ \int_{\mathcal{W}^<}\dd{\omega}e^{-i\omega t}f^{\lambda}(\delta_{1j}+ V_{1j})\tilde{s}_j + \int_{\mathcal{W}^>}\dd{\omega}e^{-i\omega t} f^\lambda W_{1j}\tilde{s}_j \right],    \\
        \Phi_1(t \sim t_0) & = \sum_{k}e^{ik\varphi}\left[ \int_{\mathcal{W}^<}\dd{\omega}e^{-i\omega t}f^{\lambda}V_{1j}\tilde{s}_j + \int_{\mathcal{W}^>}\dd{\omega}e^{-i\omega t} f^{\lambda} (\delta_{1j}+W_{1j})\tilde{s}_j \right],   \\
        \Phi_2(t \sim t_0) & = \sum_{k}e^{ik\varphi}\left[ \int_{\mathcal{W}^<}\dd{\omega}e^{-i\omega t}f^{\lambda'}(\delta_{2j}+V_{2j})\tilde{s}_j + \int_{\mathcal{W}^>}\dd{\omega}e^{-i\omega t} f^{\lambda'} W_{2j}\tilde{s}_j \right], \\
        \Phi_2(t \sim T)   & = \sum_{k}e^{ik\varphi}\left[ \int_{\mathcal{W}^<}\dd{\omega}e^{-i\omega t}f^{\lambda'}V_{2j}\tilde{s}_j + \int_{\mathcal{W}^>}\dd{\omega}e^{-i\omega t} f^{\lambda'} (\delta_{2j}+W_{2j})\tilde{s}_j \right]. \\
    \end{aligned}
    \label{eq:trans:nearmatching}
\end{align}
\paragraph{Applying matching conditions} 
The matching (\ref{eq:wh_trans_mathching_cond}) conditions at $t=\pm T$ give:
\begin{equation}\label{eq:eucl_match_res_wh_trans}
    W_{1j} =0 \,, \qquad V_{2j} = 0 \,.
\end{equation}
The matching condition (\ref{eq:wh_trans_mathching_cond}) at $t=t_0$ in principle give a matching of sums over the residus of the \emph{different} poles of $f^\lambda$ and $f^{\lambda'}$. However, working to first order in $\delta \lambda = \lambda' - \lambda$, we can write:
\begin{equation}\label{eq:f_lambda_prim}
    f^{\lambda'} = f^\lambda + \frac{R^2}{r_+}\frac{4i\omega}{\lambda} \frac{\kappa^2_\lambda}{(1-\kappa^2_\lambda)^2}\qty(f^+ - f^-)\delta \lambda + O(\delta \lambda^2) \,.
\end{equation}
Thus, at this order, it follows that the normal mode poles of $f^\lambda$ and $f^{\lambda'}$ are the same (i.e. the solutions to $\kappa_\lambda^2=1$). We can denote $\omega_{nk}^+$ to be the positive normal mode solutions of (\ref{eq:wh_normal_modes}), where $n = 1,2,\dots$ labels successively larger solutions and $k \in \mathbb{Z}$. Similarly, we denote $\omega_{nk}^-$ to be the negative solutions of (\ref{eq:wh_normal_modes}). Note that the $\mathcal{O}((\delta \lambda)^n)$ term in this expansion goes as $\sim (\omega\, \delta \lambda)^n$, so that the expansion breaks down for sufficiently large $\omega$.

We define the following residues:
\begin{align}\label{eq:residues_trans}
    \begin{aligned}
        \oint_{\omega^\pm_{nk}}\dd{\omega} \frac{1}{1-\kappa^2_{\lambda}}     & =  \pm 2\pi \frac{1}{q_\lambda(\omega^\pm_{nk},k)} \,,                                              \\
        \oint_{\omega^\pm_{nk}}\dd{\omega} \frac{1}{(1-\kappa^2_{\lambda})^2} & =  \pm 2\pi \frac{A_\lambda(\omega^\pm_{nk},k)}{q_\lambda(\omega^\pm_{nk},k)} \,,
    \end{aligned}
\end{align}
where the circular integral around $\omega^+_{nk}$ (resp. $\omega^-_{nk}$) is defined to have negative (resp. positive) direction to conform with our definition of $\mathcal{W}^\gtrless$. The quantity $q_\lambda(\omega,k)$ is given by:
\begin{align}
    q_\lambda(\omega,k) & = \frac{R^2}{r_+} \left[ 4 \log \qty( \frac{\lambda}{2}) + \psi\qty(1+ \frac{i}{2}(r_+ \overline{\omega} -\overline{k})) + \psi\qty(1- \frac{i}{2}(r_+ \overline{\omega} -\overline{k})) -2 \psi(1-i r_+ \overline{\omega})\right. \nonumber \\
                        & \qquad \qquad \left. + \psi\qty(1+ \frac{i}{2}(r_+ \overline{\omega} +\overline{k})) + \psi\qty(1- \frac{i}{2}(r_+ \overline{\omega} +\overline{k})) - 2 \psi (1+i r_+ \overline{\omega}) \right] .
\end{align}
The quantity $A_\lambda(\omega,k)$ has a similar (but lengthier) expression in terms of polygamma functions.

Using the residues (\ref{eq:residues_trans}), the basis functions (to first order in $\delta \lambda$) (\ref{eq:deftildef}) and (\ref{eq:f_lambda_prim}), together with the Euclidean matching results (\ref{eq:eucl_match_res_wh_trans}), the behaviour of the fields (\ref{eq:trans:nearmatching}) near $t\sim t_0$ can be rewritten as:
\begin{align}
    \begin{aligned}
        \Phi_1(t \sim t_0) & = 2\pi \sum_{k} e^{ik\varphi}\left[ -\sum_{n}e^{-i\omega^-_{nk}t} \frac{g^-(\omega^-_{nk},k,r)}{q_\lambda(\omega^-_{nk},k)} V_{1j}(\omega^-_{nk})\tilde{s}_j(\omega^-_{nk},k)\right.                                                                                                                            \\
                           & \left. \qquad \qquad + \sum_{n}e^{-i\omega^+_{nk}t} \frac{g^-(\omega^+_{nk},k,r)}{q_\lambda(\omega^+_{nk},k)} \delta_{1j}\tilde{s}_j(\omega^+_{nk},k) \right],                                                                                                                                                  \\
        \Phi_2(t \sim t_0) & = 2\pi \sum_{k}e^{ik\varphi} \left[ - \sum_{n}e^{-i\omega^-_{nk}t} \frac{g^-(\omega^-_{nk},k,r)}{q_\lambda(\omega^-_{nk},k)} \qty(1+ \frac{R^2}{r_+}\frac{4i\omega^-_{nk}}{\lambda} A_\lambda(\omega^-_{nk},k)\delta \lambda)\delta_{2j}\tilde{s}_j(\omega^-_{nk},k) \right. \\
                           & \left. \qquad \qquad +\sum_{n}e^{-i\omega^+_{nk}t} \frac{g^-(\omega^+_{nk},k,r)}{q_\lambda(\omega^+_{nk},k)} \qty(1+ \frac{R^2}{r_+}\frac{4i\omega^+_{nk}}{\lambda} A_\lambda(\omega^+_{nk},k)\delta \lambda)W_{2j}(\omega^+_{nk})\tilde{s}_j(\omega^+_{nk},k)\right],
    \end{aligned}
\end{align}
where we used the definition (\ref{eq:gdef}) for $g^-$. Using these expressions, the matching condition (\ref{eq:wh_trans_mathching_cond}) can now be performed separately for each of the (linearly independent) residue terms in the sum. This gives us the remaining contour-correcting functions (to first order in $\delta\lambda$):
\begin{align}
    \begin{aligned}
        V_{1j}(\omega,k) & = \qty(1+ \frac{R^2}{r_+}\frac{4i\omega}{\lambda} A_\lambda(\omega,k)\delta \lambda)\delta_{2j} \,, \quad \omega < 0 \,, \\
        W_{2j}(\omega,k) & = \qty(1- \frac{R^2}{r_+}\frac{4i\omega}{\lambda} A_\lambda(\omega,k)\delta \lambda)\delta_{1j} \,, \quad \omega > 0 \,. \\
    \end{aligned}
\end{align}

\paragraph{Extracting the correlator} Without loss of generality, we take one operator insertion to be at time $t < t_0$; then, as we slide the other operator insertion between $t' \in[-T,T]$, we obtain for the Feynman correlator:
\begin{align}
    iG^F_{\lambda \to \lambda'}(x,x') & = \theta(t_0-t')iG^{11}(x,x') + \theta(t'-t_0)iG^{12}(x,x') \nonumber                                                                                                                                                             \\
                                      & = \frac{1}{2\pi^2 i} \sum_{k}e^{ik(\varphi - \varphi')} \left[ \theta(t_0-t')\int_{\mathcal{F}}\dd{\omega}e^{-i\omega(t-t')}\alpha \beta^\lambda \right. \nonumber                                                                \\
                                      & \left. \qquad \qquad \qquad \qquad \qquad + \theta(t'-t_0)\int_{\mathcal{W}^<}\dd{\omega}e^{-i\omega(t-t')}\alpha \beta^\lambda \qty(1+ \frac{R^2}{r_+}\frac{4i\omega}{\lambda} A_\lambda\delta \lambda)\right].
\end{align}
We can recast this result as
\begin{align}
    iG_{\lambda \to \lambda'}^F(x,x') & = iG_{\lambda \to \lambda}^F(x,x') + \frac{\theta(t'-t_0)}{2\pi^2 i} \sum_{k}e^{ik(\varphi - \varphi')} \int_{\mathcal{W}^<}\dd{\omega}e^{-i\omega(t-t')}\alpha \beta^\lambda \qty(\frac{R^2}{r_+}\frac{4i\omega}{\lambda}A_\lambda\delta\lambda) \nonumber \\
\label{eq:feynproptrans}                                      & = iG_{\lambda \to \lambda}^F(x,x') - \frac{4R^2\theta(t'-t_0)}{\pi r_+}\frac{\delta\lambda}{\lambda} \sum_{n,k}e^{ik(\varphi - \varphi')}\eval{\qty(e^{-i\omega(t-t')}\alpha(\beta^+ - \beta^-)\frac{\omega A_\lambda}{q_\lambda})}_{\omega = \omega^-_{nk}}  \,,
\end{align}
where $iG_{\lambda \to \lambda}^F(x,x')$ is the Feynman propagator of a wormhole that retains the same parameter $\lambda$ throughout the whole Lorentzian (and Euclidean) evolution, i.e. the propagator as follows from (\ref{eq: wh retarded}) and section \ref{sec:SvRWH}. Note that this expression is in principle only valid where the expansion (\ref{eq:f_lambda_prim}) is valid, so (as mentioned above after (\ref{eq:f_lambda_prim})), it cannot be trusted at arbitrary high frequencies. We will discuss the position-space propagator following from (\ref{eq:feynproptrans}) in section \ref{sec:WHtransposspace}.

\section{Retarded Wormhole Correlator via Hybrid WKB}\label{sec:AdS3WKB}
In this section we aim to apply the hybrid WKB technique, developed by Bena-Heidmann-Monten-Warner (BHMW) in \cite{Bena:2019azk}, to the wormhole toy model.\footnote{This method was used in \cite{Bena:2020yii} in an asymptotically flat setting; we also used this method in our previous paper \cite{Dimitrov:2020txx} to calculate correlators for the extremal BTZ wormhole.} Our goal is to obtain the free scalar retarded correlator for the non-rotating wormhole and compare this WKB result to the real-time holography result we obtained above in (\ref{eq: wh retarded}). We will write this retarded propagator as:
\begin{align}
    iG^R(\Delta t, \Delta \varphi) = {\theta(\Delta t) \over 2\pi^2 i } \sum_k \int_{\mathbb{R}+i\epsilon} \dd{\omega} e^{-i\omega \Delta t+ik \Delta \varphi} \mathcal{R}(\omega,k) \,,
    \label{eq:GfuncRinitial}
\end{align}
with $\Delta t = t-t'$ and $\Delta\varphi = \varphi-\varphi'$.
The hybrid WKB technique will then give us $\mathcal{R}(\omega,k)$.

The BHMW WKB approximation works in a number of steps. First, we apply the traditional WKB method to find an approximate solution to the wave equation.
In order to proceed with finding the WKB approximation to the propagator, one needs to analyze the asymptotic structure of the approximate solution around the boundary. This is generally a hard task as the WKB solution is not written in a basis that is conducive to asymptotic analysis. Thus, in the second step of the method, one finds a new potential that asymptotically approaches the original potential and for which the wave equation can be solved exactly.
Then, the WKB solutions can be matched to the asymptotic solutions, which finally allows one to efficiently analyze the asymptotic structure of the approximate solutions to the wave equation and ultimately extract the propagator.

\subsection{Calculation of the WKB Propagator}\label{sec:AdS3WKB:nonrot}
We will now apply the BHMW WKB method step-by-step to a free scalar in the (non-rotating) wormhole background with metric (\ref{eq:WH3metric}).

\paragraph{Setting up the radial equation and potential}
We wish to split the radial part of the scalar as:
\begin{equation}
    \Phi_r(r) = h(r) \phi(r) \,,
\end{equation}
and we perform a coordinate transformation $ r \rightarrow x(r)$ to write the radial wave equation (\ref{eq:scalar3Dradial}) for $\phi$ in the Schr{\"o}dinger form:
\begin{equation}\label{eq:WKBSchr}
    \frac{\partial^2}{\partial x^2} \phi(x)- V(x) \phi(x)=0 \,.
\end{equation}
The absence of a term proportional to $\partial_x\phi$ leads to the condition:
\begin{equation}\label{eq:WH3feq}
    h(r) = \frac{1}{\left(r^2X(r)Y(r)\left(\frac{\partial x(r)}{\partial r}\right)^2\right)^{1/4}} \,.
\end{equation}
We are still free to choose $x(r)$, which will affect the form of the potential $V(x)$.
To align with the analysis in BHMW \cite{Bena:2019azk}, we choose $x(r)$ such that the potential approaches a particular constant in the UV:
\begin{equation} \label{eq:WKBVasympt}
    \lim_{x\rightarrow \infty} V(x) = 1+m^2 R^2 \equiv \mu^2 \,.
\end{equation}
where we have defined the dimensionless parameter $\mu$.
This can be achieved by fixing the function $x(r)$ such that:
\be \label{eq:WKBfr} h(r) = \frac{R}{r} \,.\ee

\begin{figure}[t!]\centering
    \includegraphics[width=\textwidth]{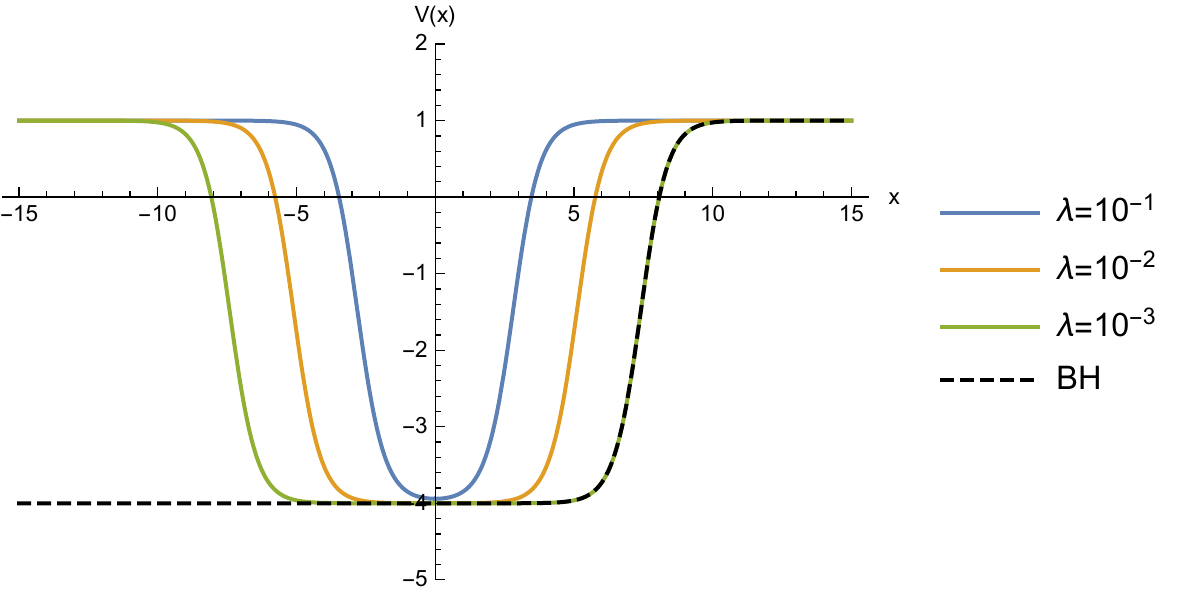}
    \caption{The potential $V(x)$ plotted as a function of $x$ with for $\mu=1,k=1,\omega=2,r_\lambda=1$, and $\lambda=\{10^{-1},10^{-2},10^{-3}\}$. As $\lambda$ decreases, the width of the potential becomes larger and larger. The black hole potential is also depicted, shifted to line up with the $\lambda=10^{-3}$ graph for $x>0$.}
    \label{fig:potential}
\end{figure}

A choice of $x(r)$ that satisfies (\ref{eq:WH3feq}) and (\ref{eq:WKBfr}) is:
\begin{equation}\label{eq:xcoord}
    x(r) = \log \left(\sqrt{\frac{r^2-r_\lambda^2}{r_\lambda^2}}+\sqrt{\frac{r^2-r_+^2}{r_\lambda^2}}\right) - \log{ \lambda} \,,
\end{equation}
Note that $x(r_+)=0+\mathcal{O}(\lambda^2)$.\footnote{We could have also chosen e.g. $x(r_t)=0+O(\lambda^2)$; this would introduce an extra constant term $-\log 1/2(1+\sqrt{5})$ in (\ref{eq:xcoord}). Such a constant does not change the derivation of the end result (\ref{eq:thetaintermed}).}
The $x$ coordinate spans the whole wormhole and ranges from $-\infty$ to $\infty$.
The relation between $x(r)$ and $r$ can be inverted to find
\begin{equation}
    r^2=r_\lambda^2 \left(1+\frac{\lambda^2}{4(1-\lambda^2)^2} (e^{-x}+e^{x}(1-\lambda^2))^2   \right) = r_+^2\left(1 + \lambda^2\sinh^2 x   \right) + O(\lambda^4)\,.
\end{equation}
The potential now takes a relatively simple form in terms of the radial coordinate $x$:
\begin{equation}
    V(r(x)) = \mu^2  + \frac{R^2}{r^2}\left(k^2 -m^2 r_\lambda^2-R^2\omega^2\right) -\frac{r_\lambda^2}{r^4}\left( k^2 R^2 + r_+^2\right) \,.
\end{equation}
Figure \ref{fig:potential} is a plot of the potential for different values of $\lambda$.
The potential approaches $\mu^2$ in the UV and at the throat position\footnote{We saw above that at the throat position $r=r_+$, we actually have $x=O(\lambda^2)$ instead of $x=0$; however (\ref{eq:throatexpans}) will still be the value of the potential at the throat position to $O(\lambda^2)$.} $x=0$ we find:
\begin{equation}\label{eq:throatexpans}
    V(x=0) = -\frac{\omega^2R^4}{r_+^2}+O\left(\lambda ^2\right) \,,
\end{equation}

The following condition needs to be met in order for the WKB approximation to be valid
\begin{equation}\label{eq:wkbapprox}
    \left|\left(V(x)\right)^{-\frac{3}{2}}\frac{\partial V(x)}{\partial x}\right| \ll 1 \,.
\end{equation}
This quantity should be evaluated away from the turning points, defined by $V(x_\text{turn})=0$, where the left side diverges.
Expanding the left hand side of (\ref{eq:wkbapprox}) for small $\lambda$, we find:
\begin{equation}\label{eq:wkbapproxnonrot}
    \left|(V(x))^{-\frac{3}{2}}\frac{\partial V(x)}{\partial x}\right| = \lambda^2 \left|\frac{r_+}{R^6\omega^3}(k^2R^2+r_+^2(1+\mu^2)+\omega^2R^4)\sinh 2x \right| + \mathcal{O}(\lambda^4) \,.
\end{equation}
The prefactor of $\lambda^2$ ensures that \eqref{eq:wkbapprox} is satisfied around the throat position (as long as the frequency is sufficiently high (e.g. $\omega R \gg \lambda$).

\paragraph{Asymptotic solutions}
Applying the same procedure outlined above to the non-rotating BTZ black hole metric (\ref{eq:BTZmetric}) gives as black hole $x$ coordinate:
\be \label{eq:xcoordBH} x^\text{BH}(r)=\log{\sqrt{\frac{r^2}{r_+^2}-1}},\ee
and potential:
\begin{equation}
    V_\text{asymp}(x) = \frac{e^{2 x}}{\left(1+e^{2 x}\right)^2}\left(1+\frac{k^2R^2}{r_+^2}-\frac{\omega^2R^4 \left(1+e^{-2 x}\right)}{r_+^2}+\mu^2 \left(1+e^{2 x}\right)\right) \,.
\end{equation}
Now, the wormhole potential is well approximated by:
\begin{equation}\label{eq:Vapprox}
    V(x) = \theta(x)V_\text{asymp}(x+\log(\lambda/2)) + (x\rightarrow-x) +\mathcal{O}(\lambda^2)\,,
\end{equation}
where the shift in argument by $\log(\lambda/2)$ comes from comparing the wormhole $x^{WH}$ coordinate (\ref{eq:xcoord}) and the black hole $x^{BH}$ coordinate (\ref{eq:xcoordBH}) at the matching radius $r=r_t=r_+(1+\lambda^2/8)$ (see (\ref{eq:rstarnonrotWH})); i.e. we have $x^{WH}=\mathcal{O}(\lambda^2)$ and $x^{BH} = \log(\lambda/2)$. Physically, this shift in \eqref{eq:Vapprox} for the $x$ coordinate arises from the wormhole of length $L_\lambda$ opening up.
The expression \eqref{eq:Vapprox} thus nicely encapsulates that for small $\lambda$, we get two copies of the BTZ potential connected by a wormhole of length $L_\lambda$.

The exact solutions $\phi_\text{asymp}(x)$ are the well-known solutions to the BTZ wave equation (\ref{eq:BTZ_modes}), which we repeat here in the $x$ coordinate:
\begin{equation}\label{eq:phiBTZ}\begin{aligned}
        \phi_\text{asymp}(x) & = \sum_{\pm} c_\pm \left(1+e^{2 x}\right)^{\frac{1}{2}-\frac{i k R}{2r_+}} e^{\pm\frac{ i \omega R^2 x}{r_+}}                                           \\
                             & _2F_1\left(\frac{1-\mu}{2}-i R\frac{k\mp \omega R}{2r_+},\frac{1+\mu }{2}-i R\frac{k\mp \omega R}{2r_+};1\pm\frac{i \omega R^2}{r_+};-e^{2 x}\right)\,.
    \end{aligned}\end{equation}
%
%
%
%
This general solution (\ref{eq:phiBTZ}) has the structure:
\begin{equation}
    \phi_\text{asymp}(x)= c_1 \phi_\text{asymp}^\text{grow}(x)+c_2 \phi_\text{asymp}^\text{dec}(x)\,,
\end{equation}
where in the UV we have defined the two linearly independent solutions as those that behave at large $x$ as:\footnote{Note that the full radial function $\Phi_r$ in (\ref{eq:scalar3Dsep}) is still given by $\Phi_r(r) = \phi(x)R/r$.}
\begin{equation}\label{eq:WKBstructasymptphi}\begin{aligned}
        \phi_\text{asymp}^\text{grow}(x) & = e^{\mu x}\left(1+\dots\right)      \\
        \phi_\text{asymp}^\text{dec}(x)  & =  e^{-\mu x}\left(1+\dots\right)\,,
    \end{aligned}\end{equation}
where the omitted terms are subleading in $x$ at large $x$.
When $\mu$ is an integer, there are also pieces proportional to $x$ in the UV expansion; we will assume $\mu$ not to be an integer and analytically continue the final result to integer $\mu$ (as we are interested in $\mu=1$).
Note that both $\phi_\text{asymp}^\text{grow}$ and $\phi_\text{asymp}^\text{dec}$ are real functions of $x$ (for real $r_+,k,\omega,\mu,R$).

\paragraph{Calculation of the WKB propagator}
Now we have all ingredients to evaluate the propagator that follows from the hybrid WKB technique.
According to \cite{Bena:2019azk} the propagator is given by
\begin{equation}
    \mathcal{R}_\text{WKB} = \left(\mathcal{A}+\frac{\sqrt{3}}{2}\right)e^{-2I_+}-\frac{\phi_\text{asymp}^\text{grow}(x_+)}{\phi_\text{asymp}^\text{dec}(x_+)} \,.
\end{equation}
where
\begin{equation}
    I_+ = -\mu x_+ + \int_{x_+}^\infty \left(\sqrt{|V(z)|}-\mu\right)dz \,,
\end{equation}
and the turning points $x_{\pm}$ are defined such that $V(x_\pm)=0$.
The expression for $\mathcal{A}$ depends on the number of turning points.
For a single turning point, we have
\begin{equation}
    \mathcal{A} = \textrm{sign}(\omega) \frac{i}{2} \,.
\end{equation}
For two turning points $x_-$ and $x_+$, we have
\begin{equation}\begin{aligned}
        \mathcal{A} & = \frac{1}{2} \tan(\Theta) \,,         \\
        \Theta      & = \int_{x_-}^{x_+}\sqrt{|V(z)|}dz \,.
    \end{aligned}\end{equation}
This will be the formula we will need to evaluate to find the wormhole propagator.
The general expression for $\mathcal{A}$ for more than two turning points can be found in an appendix of \cite{Bena:2019azk}.

It is useful to first apply the WKB approximation to the (non-rotating) BTZ black hole itself.
The BTZ potential has only one turning point, so one finds
\begin{equation} \label{eq:WKBBTZprop}
    \mathcal{R}^\text{BH}_\text{WKB} \approx \left(\frac{i}{2}\textrm{sign}(\omega)+\frac{\sqrt{3}}{2}\right)e^{-2I_+}-\frac{\phi_\text{asymp}^\text{grow}(x_+)}{\phi_\text{asymp}^\text{dec}(x_+)} \,.
\end{equation}
As noted earlier, the exact solutions $\phi_\text{asymp}^\text{grow,dec}$ are real and so is $I_+$.
This allows us to write
\begin{equation}\begin{aligned}
        \textrm{Re}\left(\mathcal{R}^\text{BH}_\text{WKB}\right) & \approx \frac{\sqrt{3}}{2}e^{-2I_+}-\frac{\phi_\text{asymp}^\text{grow}(x_+)}{\phi_\text{asymp}^\text{dec}(x_+)} \,, \\
        \textrm{Im}\left(\mathcal{R}^\text{BH}_\text{WKB}\right) & \approx \frac{1}{2}\textrm{sign}(\omega) e^{-2I_+}\,.
    \end{aligned}\end{equation}

The wormhole potential is essentially two copies of the black hole potential, leading to two turning points, so $\mathcal{A}$ will differ with respect to the BTZ propagator (\ref{eq:WKBBTZprop}).
However, due to \eqref{eq:Vapprox} we can see that $I_+^\text{wormhole}=I_+^\text{BH}+\mathcal{O}(\lambda^2)$, and thus we can write
\begin{equation}\begin{aligned}\label{eq:RWKB}
        \mathcal{R}_\text{WKB}^\text{WH} & \approx \textrm{Re}\left(\mathcal{R}^\text{BH}\right)+2 \,\textrm{sign}(\omega) \, \mathcal{A} \, \textrm{Im}\left(\mathcal{R}^\text{BH}\right) \,,
    \end{aligned}\end{equation}
where $\approx$ now denotes both inaccuracy due to the WKB approximation and corrections of order $\lambda^2$.
Now, we need to evaluate $\mathcal{A}$ or equivalently $\Theta$:
\begin{align}
    \nonumber\Theta          & = \int_{x_-}^{x_+}\sqrt{|V_\text{asymp}(z)|}dz + \mathcal{O}(\lambda^2)                                                                                                                                                                                                                             \\
    \nonumber                & = \left|\log\left[e^{-\frac{1}{2} (\pi  \mu)} \left(\frac{16\omega^2R^2}{\lambda ^2}\right)^{\omega R^2/r_+}\left(-\frac{r_+^2 \left(\mu^2+1\right)+2 i r_+ \mu \omega R^2+k^2R^2-\omega^2R^4}{r_+^2 \left(\mu^2+1\right)-2 i r_+ \mu \omega R^2+k^2R^2-\omega^2R^4}\right)^{i \mu/2}\right.\right. \\
    \nonumber                & \left(-2 \omega^2 \left(k^2R^2-r_+^2 \left(\mu^2-1\right)\right)+\left(\frac{r_+^2}{R^2} \left(\mu^2+1\right)+k^2\right)^2+\omega^4R^4\right)^{\frac{\sqrt{r_+^2+k^2R^2}-\omega R^2}{2 r_+}}                                                                                                        \\
    \label{eq:thetaintermed} & \left.\left.\left(\omega \left(2 \sqrt{r_+^2+k^2R^2}+\omega R^2\right)+\frac{r_+^2}{R^2} \left(\mu^2+1\right)+k^2\right)^{-\frac{\sqrt{r_+^2+k^2R^2}}{r_+}}\right]\right| + \mathcal{O}(\lambda^2) \,.
\end{align}

This expression simplifies in the ``low-frequency'' limit (\ref{eq:SvRWHlowfreqregime}), $\omega\sim L_\lambda^{-1}$ (where wormhole throat length $L_\lambda$ is defined in (\ref{eq:WH3defthroatlength})), and in the ``high-frequency'' limit (\ref{eq:SvRWHhighfreqregime}), $\omega\gg r_+/R^2$. In both cases, the expression that we get is (for $\mu=1$, and subtracting integer multiples of $\pi$):
\be \label{eq:thetaapprox} \Theta = \left| \omega\, L_\lambda \right| \,.\ee
We discarded terms of $O(\lambda^2)$ as well as terms of $O(\lambda^0\omega)$ as they will always be subleading.
We can understand this expression for $\Theta$ as follows: when $L_\lambda$ becomes very large ($\lambda$ is exponentially small), the wormhole throat contribution dominates the integral in (\ref{eq:thetaintermed}). Since the potential is very flat for (almost) the entire wormhole, as can be seen from  (\ref{eq:Vapprox}) and figure \ref{fig:potential}, we can approximate the integral by:
\begin{equation}\begin{aligned}
        \Theta & \approx \sqrt{|V(0)|} \, L_\lambda + \mathcal{O}(\lambda^0)   \\
               & =\left|\omega \, L_\lambda\right| + \mathcal{O}(\lambda^0)\,,
    \end{aligned}\end{equation}
where we used the value of $V(0)$ in \eqref{eq:throatexpans}. The result is indeed in agreement with the direct evaluation (\ref{eq:thetaapprox}).

\subsection{Hybrid WKB vs. Real Time Holography}\label{sec:rel_between_freq_space_corrs}
For the non-rotating BTZ black hole, we calculated the black hole retarded scalar correlator using real-time holography in section \ref{sec:SvRBTZ} as:
\begin{align}
    \mathcal{R}^{\text{BH}} = \alpha \beta^- \,, 
\end{align}
Using the fact that $(\alpha \beta^\pm)^* = \alpha \beta^\mp$, we note that:
\begin{align}
    \textrm{Re}\qty(\mathcal{R}^{\text{BH}}) = {1\over 2} (\alpha \beta^+ + \alpha \beta^-) \,,  \qquad \textrm{Im}\qty(\mathcal{R}^{\text{BH}}) = {i\over 2} (\alpha \beta^+ - \alpha \beta^-) \,. 
\end{align}
For the corresponding non-rotating Solodukhin wormhole, in section \ref{sec:SvRWH} we calculated:
\be \label{eq:WHcorrfullintermed} \mathcal{R}^{\text{WH}} =  \frac{1}{1-\kappa^2}\alpha \beta^+ -\frac{\kappa^2}{1-\kappa^2} \alpha \beta^- = \textrm{Re}\qty(\mathcal{R}^{\text{BH}}) - i\, \frac{1+\kappa^2}{1-\kappa^2} \textrm{Im}\qty(\mathcal{R}^{\text{BH}}) \,,  \ee
where $\kappa$ is given by (\ref{eq:deftildef}) (and (\ref{eq:normalization_bh_modes})). Noting that $|\kappa|=1$, we can define the angle $\eta$ through:
\be \kappa \equiv i e^{-i\eta} \,, \ee
so that (\ref{eq:WHcorrfullintermed}) becomes:
\be \label{eq:WHcorrfull} \mathcal{R}^{\text{WH}} =  \textrm{Re}\qty(\mathcal{R}^{\text{BH}}) +\tan \eta\, \textrm{Im}\qty(\mathcal{R}^{\text{BH}}) \,,  \ee

As an aside, we note that, curiously, there is a simple way to go from the black hole to wormhole correlator in the large frequency limit: one simply does an ``analytic continuation'' of the inverse temperature $\beta$ to the travel time $\Delta t$, where $\Delta t= 2 L_\lambda$ in the wormhole. Then $e^{-\beta\omega} \rightarrow e^{-i\Delta t\omega}$, so the ``information losing'' factor becomes an oscillatory factor which shows the recovery of information in a time $\Delta t$. 

The real-time holography result (\ref{eq:WHcorrfull}) is to be compared to the full WKB answer for the propagator (where we set $\mu=1$), which we derived in the previous section:
\begin{equation}\begin{aligned}\label{eq:RWKBfull}
        \mathcal{R}_\text{WKB}^\text{WH} & \approx \textrm{Re}\left(\mathcal{R}^\text{BH}\right)+ \left(\textrm{sign}(\omega)\tan \Theta\right) \, \textrm{Im}\left(\mathcal{R}^\text{BH}\right) \,,
    \end{aligned}\end{equation}
where $\Theta$ is given by (\ref{eq:thetaintermed}).

    \paragraph{Comparison}
Comparing (\ref{eq:RWKBfull} to (\ref{eq:WHcorrfull}), we see that the WKB approximation does a good job of capturing the propagator when $\Theta\approx \eta$ (for $\omega>0$).

\begin{figure}[ht]
    \centering
    \includegraphics[width=0.7\textwidth]{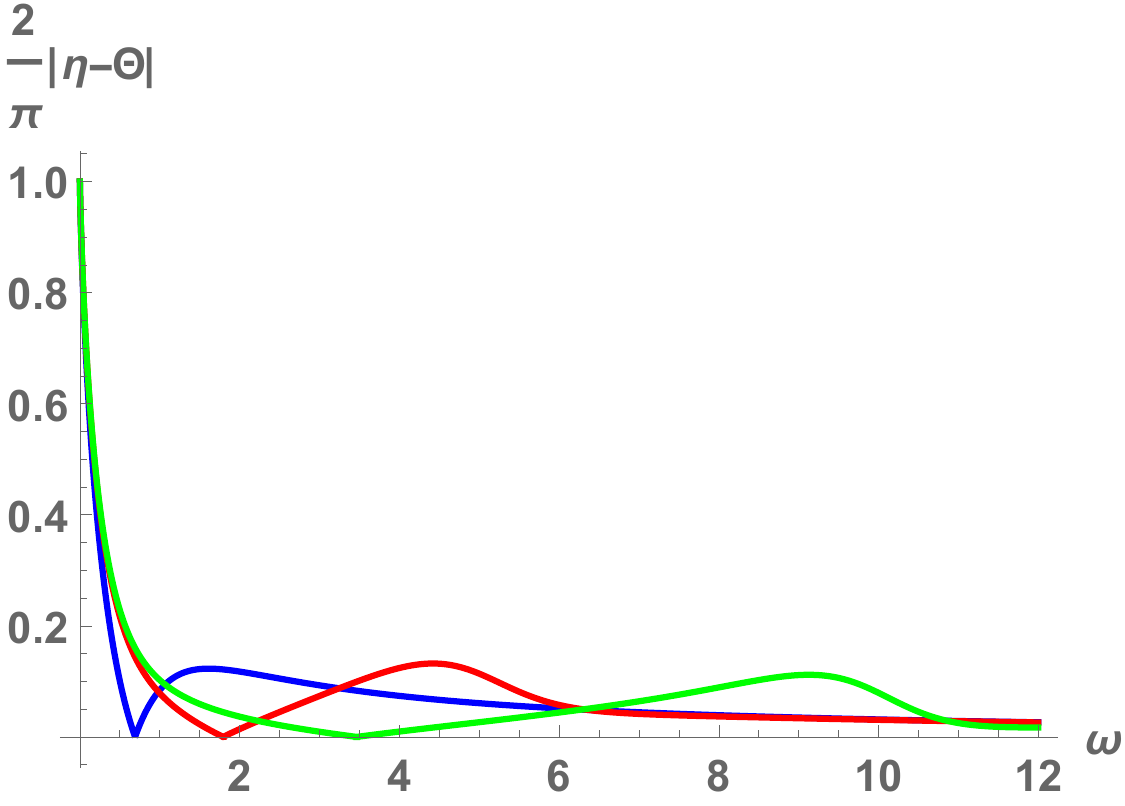}
    \caption{A plot of $(2/\pi)\left|(\eta-\Theta)\mod \pi\right|$ as a function of $\omega$. We take $R=r_+=1$, and $\lambda=10^{-20}$ ($L_\lambda\approx 94.88$). The blue, red, and green lines respectively correspond to $k=0,5,10$. The mismatch starts out maximally at $\pi/2$ (see (\ref{eq:etavsomegalow})) but asymptotes (quickly) to zero (see (\ref{eq:etavsomegahigh})).}
    \label{fig:etavstheta}
\end{figure}

In the low-frequency regime (\ref{eq:SvRWHlowfreqregime}), we have:
\be \label{eq:etavsomegalow} \eta \approx \omega L_\lambda + \frac{\pi}{2} \,,  \qquad \Theta \approx \omega L_\lambda  \,, \ee
so that the mismatch is maximal at low frequencies.\footnote{Note that we must actually take the entire WKB approximation with a grain of salt for small $\omega$, as mentioned above around (\ref{eq:wkbapproxnonrot}). Small frequencies should be treated separately, see \cite{Bena:2019azk}.} On the other hand, in the high-frequency regime (\ref{eq:SvRWHhighfreqregime}), expanding upon (\ref{eq:thetaapprox}):
\be \label{eq:etavsomegahigh} \eta \approx \omega L_\lambda + \frac12 \left(\frac{R^2}{r_+}\omega\right)^{-1} - \frac{5}{864} \left(\frac{R^2}{r_+}\omega\right)^{-3}  +\mathcal{O}(\omega^{-5}) \,,  \qquad  \Theta \approx \omega L_\lambda + \frac23  \left(\frac{R^2}{r_+}\omega\right)^{-3} + \mathcal{O}(\omega^{-7}) \,. \ee
Thus, we see that the the propagators start differing in the high-frequency regime at $\mathcal{O}(\omega^{-1})$. We give an example of the difference between $\eta$ and $\Theta$ in figure \ref{fig:etavstheta}; for large wormhole throat lengths $L_\lambda$, we see that the (large) mismatch is only at very low frequencies so that $\Theta$ asymptotes to $\eta$ rather quickly. Thus, we can conclude that the BHMW hybrid WKB approximation captures the wormhole propagator quantitatively very well.


\section{Position-Space Correlators}\label{sec:posspaceprop}
In this section, we will integrate explicitly the frequency-space propagators found in section \ref{sec:SvR_calculations} to obtain the position-space propagators for both the (non-rotating) BTZ black hole and the corresponding Solodukhin wormhole. We will also present the position-space correlator and interpretation thereof for the ``wormhole transition'' correlator of section \ref{sec:wormhole_transition}.

We repeat the general relation (\ref{eq:GfuncRinitial}) between the frequency space retarded propagator $\mathcal{R}$ and the position-space one, valid for all of the correlators we are considering:
\begin{align}
    iG^R(\Delta t, \Delta \varphi) = {\theta(\Delta t) \over 2\pi^2 i } \sum_k \int_{\mathbb{R}+i\epsilon} \dd{\omega} e^{-i\omega \Delta t+ik \Delta \varphi} \mathcal{R}(\omega,k) \,.
    \label{eq:retarded_general_f-la}
\end{align}
Note that compared to (\ref{eq:BHretardedinfuncLLRR}) or (\ref{eq: wh retarded}), we have set $\Delta t = t-t'$ and $\Delta\varphi = \varphi-\varphi'$.

\subsection{Black Hole Correlator}
\label{sec:bh_pos_space_correlators}
To calculate the position-space retarded propagator for the BTZ black hole, we need to consider the contributions from $G^{LL}$ and $G^{RR}$ separately. We refer to appendix A of \cite{Botta-Cantcheff:2018brv} for details of this calculation; here, we review this derivation briefly.

Starting with $G^{LL}$ in (\ref{eq:bh_correlators}), we take $t>t'$ and thus close the $\omega$ integral in the lower half plane. The functions $\alpha$ and $\beta^+$ have no poles here, while we pick up the residues of the poles $\omega_{nk\pm}^-$ given in (\ref{eq:BTZQNMs}) of $\beta^-$. Note that $n_\omega = (e^{\beta\omega}-1)^{-1}$ also has (purely imaginary) poles at $\omega = 2\pi \beta^{-1} i n$ for $n\in \mathbb{Z}$ (depicted by circles on figure \ref{fig:complexcontours}). As discussed in \cite{Botta-Cantcheff:2018brv}, in principle one needs to take the residues of these poles into account as well, but since they end up giving a zero contribution in the final result, we will omit them here. Then we get:
\be \label{eq:GLLintermed} i G^{LL}(\Delta t, \Delta \varphi) = -\frac{2}{\pi i} \frac{r_+}{R^2} \sum_{k\in\mathbb{Z}} \sum_{n=1}^\infty \sum_{\pm} e^{ik (\Delta \varphi \mp \frac{\Delta t}{R})} e^{-2\frac{r_+}{R^2}n\Delta t }\alpha(\omega_{nk\pm}^-) \frac{e^{\beta \omega^-_{nk\pm}}}{e^{\beta\omega^-_{nk\pm}}-1} \,. \ee
To evaluate the $k$ sum, we use the Poisson resummation formula:\footnote{For a periodic function $\tilde f(x)=\tilde f(x+2\pi n)$, we can write the two expansions:
\be \tilde f(x) = \sum_{n\in\mathbb{Z}} C_n e^{i n x}, \qquad \tilde f(x) = \frac{1}{2\pi}\int dk f(k) e^{ikx} \,,  \ee
and one finds the relation between the two expansions:
\be f(k) =  \int dx \tilde f(x) e^{-ik x} =  \sum_n C_n \delta(n-k) \,, \ee
so that, using $\sum_k \delta(k-n) = \sum_m e^{2\pi mn i}$:
\be \sum_k f(k) =  \sum_n C_n \sum_m e^{2\pi m n i} =  \sum_m \tilde f(2\pi m) =  \sum_m \int dk' f(k') e^{2\pi k m i} \,. \ee
}
\be \sum_k f(k) = \sum_m \int dk f(k) e^{2\pi k m i} \,, \ee
to rewrite the sum over $k$ in (\ref{eq:GLLintermed}) as an integral over $k$ and a ``sum over images'' $m$:
\be \label{eq:GLLintermed2} i G^{LL}(\Delta t, \Delta \varphi) = -\frac{2}{\pi i} \frac{r_+}{R^2} \sum_{m\in\mathbb{Z}} \sum_{n=1}^\infty \sum_{\pm} \int dk\, e^{ik (\Delta \varphi \mp \frac{\Delta t}{R}+2\pi m)} e^{-2\frac{r_+}{R^2}n\Delta t }\alpha(\omega_{nk\pm}^-) \frac{e^{\pm 2\pi \frac{R}{r_+} k}}{e^{\pm 2\pi \frac{R}{r_+} k}-1} \,, \ee
where we used $\beta = 2\pi R^2/r_+$.
To regulate this integral as $k\rightarrow \pm \infty$, we must send $\Delta t \rightarrow \Delta t(1-i \epsilon)$ with $\epsilon>0$; this regulates both ends of the integral for both the terms with $\pm$ in (\ref{eq:GLLintermed2}). Then, for $\Delta\varphi\mp\Delta t/R<0$, we can close this $k$ integral in the lower half plane. Only the factors $(e^{\pm 2\pi R/r_+ k}-1)^{-1}$ have poles (when $k=- i\, l\, r_+/R$ for $l\in \mathbb{Z}$), and so we pick up the residus:
\be i G^{LL}(\Delta t, \Delta \varphi) = \frac{2}{\pi } \frac{r_+^2}{R^3} \sum_{m\in\mathbb{Z}} \sum_{n=1}^\infty \sum_{l=1}^\infty \sum_{\pm}\pm e^{l \frac{r_+}{R} (\Delta \varphi \mp \frac{\Delta t}{R}(1-i\epsilon)+2\pi m)} e^{-2\frac{r_+}{R^2}n\Delta t(1-i\epsilon) }n(n\pm l)  \,. \ee
We can explicitly perform the sum over the two terms with $\pm$ as well as the infinite sums over $n,l$, resulting in:
\be \label{eq:GLLfinal} i G^{LL}(\Delta t, \Delta \varphi) = \frac{1}{2\pi}\frac{r_+^2}{ R^3} \sum_{m\in\mathbb{Z}} \frac{1}{ \left( \cosh\left[ \frac{r_+}{R^2}\Delta t(1-i\epsilon)\right]  - \cosh\left[ \frac{r_+}{R}(\Delta \varphi+2m\pi)\right]\right)^2 } \,, \ee
which agrees with \cite{Botta-Cantcheff:2018brv}. Even though (\ref{eq:GLLfinal}) was derived for $\Delta\varphi\mp\Delta t/R<0$, it can be analytically continued outside of this region.

\begin{figure}[ht]\centering
 \includegraphics[width=0.49\textwidth]{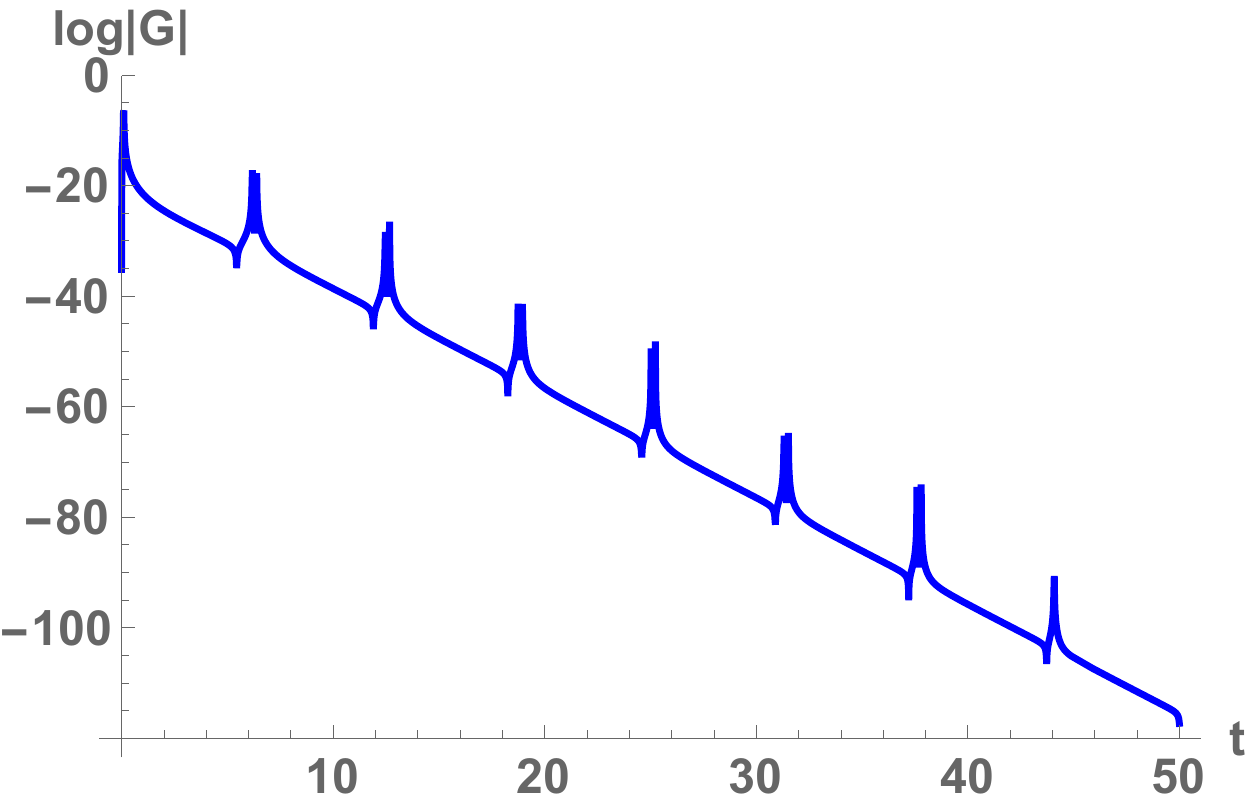} \includegraphics[width=0.49\textwidth]{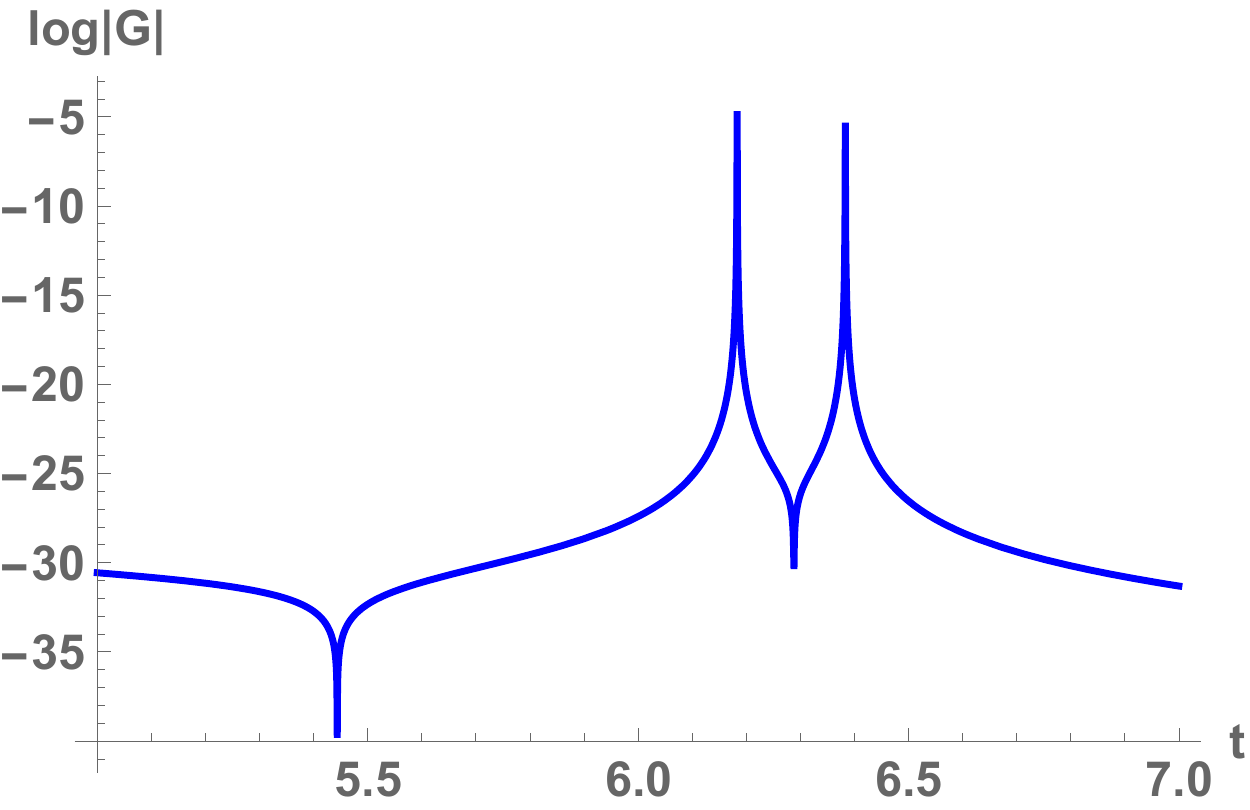}
 \caption{The black hole correlator $G^{R}_\text{BH}(\Delta t,\Delta\varphi=0.1)$ as given in (\ref{eq:BHcorrpsfinal}). We take $R=r_+=1$ and $\epsilon=10^{-10}$. The correlator decays exponentially for $t>0$; note that each subsequent (small) peak at $t>0$ come from the contribution of an ``image'' $m>0$. The ``double-peak'' structure (apparent from the close-up on the right) is typical for $\Delta\varphi\neq 0$.}
 \label{fig:BHcorr}
\end{figure}

To calculate $G^{RR}$, we again start with (\ref{eq:bh_correlators}) and proceed analogously. The only difference is that to regulate the resulting $k$ integral, we will need to send $\Delta t\rightarrow \Delta t(1+i\epsilon)$.\footnote{The origin of this difference in regulator is that in $G^{RR}$ in (\ref{eq:bh_correlators}), $\beta^-$ is multiplied by $n$ instead of $(n+1)$ as it is in $G^{LL}$.} Thus, we indeed find $i G^{RR} = [i G^{LL}]^*$. The retarded propagator is then given by (\ref{eq:BHretardedinfuncLLRR}), i.e.:
\begin{align}
\nonumber i G_{\text{BH}}^R(\Delta t, \Delta \varphi) = \frac{\theta(\Delta t)}{2\pi}\frac{r_+^2}{ R^3}&\sum_{m\in\mathbb{Z}} \left[ \frac{1}{ \left( \cosh\left[ \frac{r_+}{R^2}\Delta t(1-i\epsilon)\right]  - \cosh\left[ \frac{r_+}{R}(\Delta \varphi+2m\pi)\right]\right)^2 } \right.\\
\label{eq:BHcorrpsfinal} & \left. - \frac{1}{ \left( \cosh\left[ \frac{r_+}{R^2}\Delta t(1+i\epsilon)\right]  - \cosh\left[ \frac{r_+}{R}(\Delta \varphi+2m\pi)\right]\right)^2 } \right].
\end{align}
A plot of this correlator is given in figure \ref{fig:BHcorr}.

\subsection{Three Ways to the Wormhole Correlator}\label{sec:WHPSprop}
For the wormhole position-space propagator  (\ref{eq:WHcorrfull}), we will first illustrate its basic properties using a quick and dirty, ``naive'' calculation. We will show the expressions for the actual wormhole propagator and attempt a numerical evaluation of this. Finally, we will simplify the wormhole by altering its matching condition to be frequency-dependent, which will allow us to integrate the propagator analytically.

\subsubsection{Naive high-frequency calculation}\label{sec:naiveWHcorr}
In the high-frequency limit (\ref{eq:etavsomegahigh}), $\eta\approx \omega L_\lambda$ thus the correlator (\ref{eq:WHcorrfull}) simplifies to:
\begin{align}
   \mathcal{R}^\text{WH} & \approx  \textrm{Re}\qty(\mathcal{R}^{\text{BH}}) +\tan(\omega L_\lambda) \textrm{Im}\qty(\mathcal{R}^{\text{BH}}) \,, 
    \label{eq:WHBHcorrhighfreq}
\end{align}
Note that the poles of $\tan\omega L_\lambda$ indeed correspond to the normal modes (\ref{eq:SvRWHhighfreqNMs}) in this regime.\footnote{
Note that if one tries to calculate the position-space retarded propagator using (\ref{eq:retarded_general_f-la}) and (\ref{eq:WHBHcorrhighfreq}), in addition to the contributions from residus around the (approximated) normal modes of the wormhole coming from the poles of $\tan(\omega L_\lambda)$, there are also contributions from the residus of poles of $\alpha \beta^-$ (i.e. the quasinormal modes of the original BTZ black hole). This is an artifact of the approximation used to obtain (\ref{eq:WHBHcorrhighfreq}), which is only valid in a particular regime for \emph{real} $\omega$. The exact result (\ref{eq:WHcorrfull}) does not have any poles off the real axis; (\ref{eq:WHBHcorrhighfreq}) is not valid in the entire complex plane.}

A naive way to evaluate the wormhole propagator approximately would then be to use the distributional identity:
\be \tan x = \sum_{n=1}^\infty 2(-1)^{n-1} \sin (2nx) \,, \ee
and its Fourier transform:
\be \mathcal{F}(t) \equiv \frac{1}{2\pi}\int d\omega e^{-i\omega t} \tan \omega L_\lambda = i \sum_{n=1}^\infty (-1)^{n-1} \left(\delta(t+2 n L_\lambda) - \delta(t-2n L_\lambda)\right) \,, \ee
Then, using the convolution theorem for Fourier transforms on the high-frequency retarded propagator (\ref{eq:WHBHcorrhighfreq}):
\begin{align}
 G^R_\text{WH}(\Delta t, \Delta \varphi) &\approx   G^0(\Delta t,\Delta \varphi) + (G^1*\mathcal{F})(\Delta t,\Delta \varphi) \,, \\
 G^0(\Delta x) & \equiv -\frac{\theta(\Delta t)}{2\pi^2 } \sum_k e^{ik\Delta \varphi}\int d\omega e^{-i\omega \Delta t} \text{Re}(\mathcal{R}^\text{BH}) \,, \\
  G^1(\Delta x) & \equiv -\frac{\theta(\Delta t)}{2\pi^2 } \sum_k e^{ik\Delta \varphi}\int d\omega e^{-i\omega \Delta t} \text{Im}(\mathcal{R}^\text{BH}) \,, 
 \end{align}
so that $G^R_\text{BH}(\Delta t,\Delta \varphi) = G^0(\Delta t,\Delta \varphi) + iG^1(\Delta t,\Delta \varphi)$, and the convolution gives:
\begin{align} \nonumber (G^1*\mathcal{F})(\Delta t,\Delta \varphi) &=    i \sum_{n=1}^\infty (-1)^{n-1} \int dt' G^1(\Delta t-t', \Delta \varphi) \left(\delta(t'+2 n L_\lambda) - \delta(t'-2n L_\lambda)\right)\\
&=  i \sum_{n=1}^\infty (-1)^{n-1} \left(G^1(\Delta t +2n L_\lambda) - G^1(\Delta t-2n L_\lambda)\right) \,, 
\end{align}
so that we have the relation:
\be \label{eq:naiveWHGR} G^R_\text{WH}(\Delta t, \Delta \varphi) \approx G^0(\Delta t,\Delta \varphi) + i \sum_{n\in \mathbb{Z}} (-1)^{n}\text{sgn}(n) G^1(\Delta t-2n L_\lambda) \,, \ee
where we define the $n=0$ term of the sum to be $G^1(\Delta t)$ --- although note that the Fourier transform of the $\tan$, as naively calculated above, does not contain this $n=0$ term.

From the expression (\ref{eq:naiveWHGR}), we can clearly see the appearance of ``echoes'' in the propagator, i.e. recurrences at periodic intervals $2L_\lambda$ of the correlator. Thus, at early times $0<\Delta t\ll L_\lambda$, the propagator will mimic the BTZ one (\ref{eq:BHcorrpsfinal}) and decay exponentially; only at a time $\Delta t\sim L_\lambda$ will the propagator start to differ qualitatively from the black hole one due to the appearance of the first echo. This is the most important property of the wormhole propagator that we wish to highlight; we will confirm this qualitative echo behavior of (\ref{eq:naiveWHGR}) below in a (numerical) calculation of the full propagator (\ref{eq:WHcorrfull}), and also (analytically) in a frequency-dependent wormhole alteration.

\subsubsection{Full propagator}\label{sec:WHfullprop}
The derivation of (\ref{eq:naiveWHGR}) was rather imprecise; for example, we did not take care of the correct contour for the $\omega$ integration in (\ref{eq:retarded_general_f-la}), and we resorted to the approximation of the full propagator (\ref{eq:WHcorrfull}) by the high-frequency regime expression (\ref{eq:WHBHcorrhighfreq}).

\begin{figure}[ht]\centering
 \includegraphics[width=0.7\textwidth]{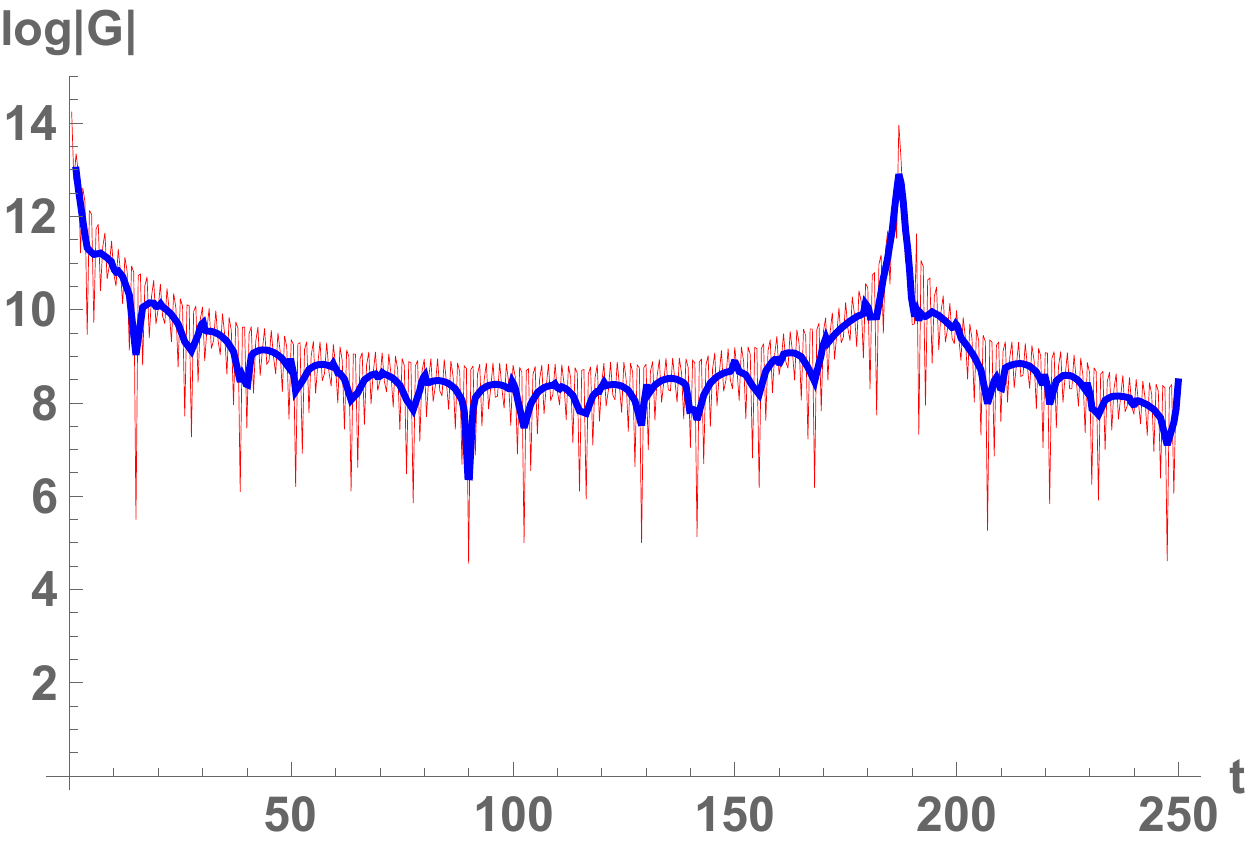}
 \caption{The wormhole correlator $G^{R}_\text{WH}(\Delta t,\Delta\varphi=0.1)$ as given in (\ref{eq:GRWHexpr}). We take $R=r_+=1$, $\lambda=2\cdot 10^{-20}$ and $\epsilon=10^{-1}$, and evaluate each of the four sums in (\ref{eq:GRWHexpr}) up to $n_{max}=500,k_{max}=50$. The red line is the computed value and the blue line is a smoothed out average. We can clearly see the first recurrence or ``echo'', which occur at $\Delta t\approx 2n L_\lambda\approx 186.98n$.}
 \label{fig:WHcorr}
\end{figure}

Calculating the full retarded propagator (\ref{eq:WHcorrfull}) is difficult as we do not have an analytic expression for the propagator's poles. As discussed in section \ref{sec:rel_between_freq_space_corrs}, these poles are only located on the real axis (i.e. they are normal modes) and are given at the location where $\kappa = \pm 1$ with $\kappa$ given in (\ref{eq:deftildef}) and (\ref{eq:normalization_bh_modes}). Note that $\kappa$ satisfies the following properties:
\be \label{eq:kappaproperties} |\kappa| = 1 \,,  \qquad \left[\kappa(\omega,k)\right]^* = \kappa(-\omega,k) \,, \qquad \kappa(\omega,k) = \kappa(\omega, -k) \,. \ee
For a fixed $k\in\mathbb{Z}$, we can denote the positive normal modes as $\omega_{nk}^+$ with $n=1,2,\cdots$ and  $n=1$ corresponding to the first (strictly) positive normal mode. Similarly, $\omega_{nk}^-$ are the negative normal modes. From the properties (\ref{eq:kappaproperties}), it follows that $\omega_{nk}^- = -\omega_{nk}^+$  and also $\omega^\pm_{n(-k)} = \omega^\pm_{nk}$, for any $n,k$. Therefore, we can choose to express everything in terms of $k\geq 0$ and $\omega_{nk}^+$.
Evaluating the $\omega$ integral by closing in the lower half plane then gives:
\begin{align}
\nonumber i G_\text{WH}^R(\Delta t, \Delta \varphi)  &= \frac{\theta(\Delta t)}{2\pi^2 i}\left[ \sum_{k=0}^\infty \sum_{n=1}^{\infty} e^{-i \omega_{nk}^+ \Delta t + i k \Delta\varphi} \mathcal{G}(k,\omega_{kn}^+) -\sum_{k=0}^\infty \sum_{n=1}^{\infty} e^{+i \omega_{nk}^+ \Delta t + i k \Delta\varphi} \mathcal{G}(k,\omega_{kn}^+) \right.\\
\label{eq:GRWHexpr} & \left. + \sum_{k=1}^\infty \sum_{n=1}^{\infty} e^{-i \omega_{nk}^+\Delta  t - i k \Delta \varphi} \mathcal{G}(k,\omega_{kn}^+) -\sum_{k=1}^\infty \sum_{n=1}^{\infty} e^{+i \omega_{nk}^+ \Delta t - i k \Delta \varphi} \mathcal{G}(k,\omega_{kn}^+) \right],\\
 \mathcal{G}(k,\omega) & \equiv \frac{i \pi^2}{2r_+}\frac{(k^2-R^2\omega^2) \sum_\pm \pm \coth \left(\pi \frac{R}{2r_+}(k\pm R\omega)\right) }{ \sum_\pm\left[-2\psi\left(\pm i\frac{R^2}{r_+}\omega\right)+ \psi\left( i\frac{R}{2r_+}(k\pm R\omega)\right) +\psi\left( i\frac{R}{2r_+}(-k\pm R\omega)\right)\right] } \,, 
 \end{align}
 where we used that $\mathcal{G}(k,\omega) = \mathcal{G}(-k,\omega)= -\mathcal{G}(k,-\omega)$. We split the $k,n$ sums (each) into positive and negative frequency parts. Then, to evaluate these four sums directly, we put regulators by changing:
 \be e^{\mp i \omega_{nk}^+ \Delta t} \rightarrow e^{\mp i \omega_{nk}^+ \Delta t(1\mp i\epsilon)} \,,  \qquad e^{\pm i k \Delta\varphi} \rightarrow e^{\pm i k \Delta\varphi(1\pm i\epsilon)} \,, \ee
 which gives the correct regulation of all four sums for $\Delta t,\Delta \varphi>0$. Then, we can numerically evaluate the four sums in (\ref{eq:GRWHexpr}) to a given cut-off value $n_{max},k_{max}$, see figure \ref{fig:WHcorr}. Although the result clearly suffers from numerical noise (brought on by incomplete sums),\footnote{For example, note that we do not see images appearing, nor the ``double-peak'' structure, as was obtained for the black hole correlator (see figure \ref{fig:BHcorr}).} we can already clearly see the recurrences or ``echoes'' in the correlator occurring at $\Delta t\approx 2n L_\lambda$.
 
\subsubsection{Low- and high-frequency approximation}\label{sec:propfreqaltWH}
The main obstacle to being able to explicitly find the wormhole correlator, and in particular evaluate the infinite sums in (\ref{eq:GRWHexpr}), is that we do not have an analytic expression for the normal modes $\omega^\pm_{nk}$. These normal modes are solutions to the transcendental equation $\kappa^2 = 1$ with $\kappa$ given by (\ref{eq:deftildef}).

However, in the low- and high-frequency limits (see (\ref{eq:SvRWHlowfreqregime}) and (\ref{eq:SvRWHhighfreqregime})), $\kappa$ simplifies considerably and we can find the normal modes explicitly. Therefore, let us consider replacing $\kappa$ entirely by either its low- or high-frequency limit:
\be \label{eq:tildekappadef} \tilde \kappa^{\text{(low)}} =\exp\left(- i L_\lambda \omega\right), \qquad \tilde{\kappa}^{\text{(high)}} =\exp\left(- i L_\lambda \omega+ i\frac{\pi}{2}\right), \ee
Physically, we can interpret replacing $\kappa$ by one of the $\tilde\kappa$'s as introducing a frequency-dependent boundary condition at the middle of the wormhole throat (replacing (\ref{eq:rtmatchingcond})).

\begin{figure}[ht]\centering
\begin{subfigure}{0.49\textwidth}
 \includegraphics[width=\textwidth]{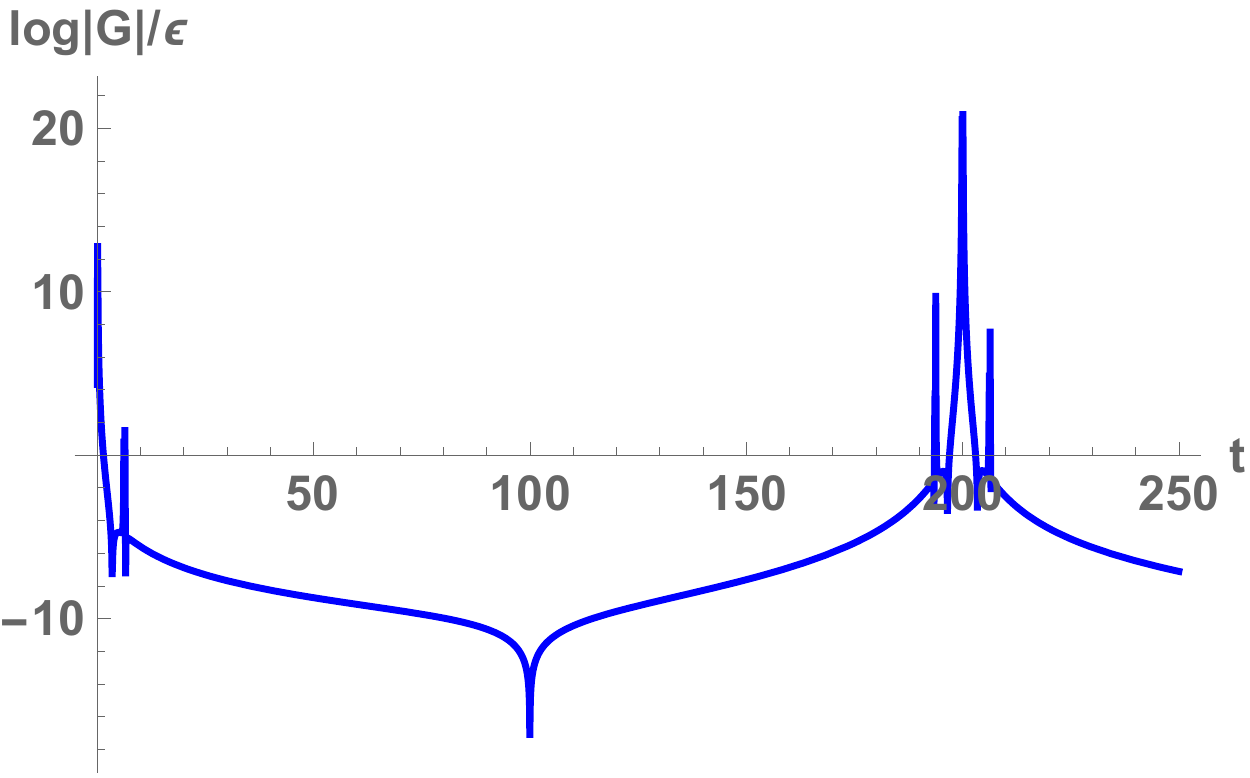}
 \caption{The wormhole correlator}
 \end{subfigure}
\begin{subfigure}{0.49\textwidth}
 \includegraphics[width=\textwidth]{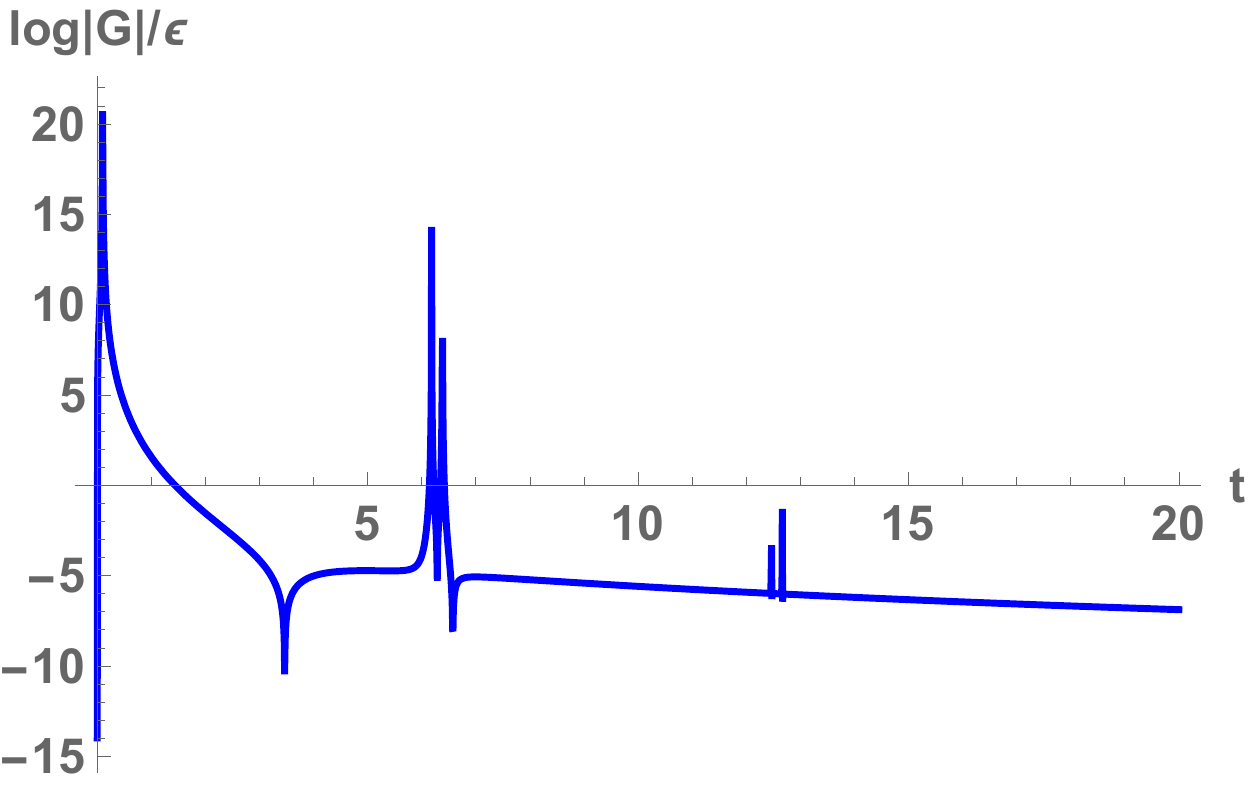}
 \caption{The wormhole correlator around $\Delta t=0$}
 \end{subfigure}
 \begin{subfigure}{0.49\textwidth}
 \includegraphics[width=\textwidth]{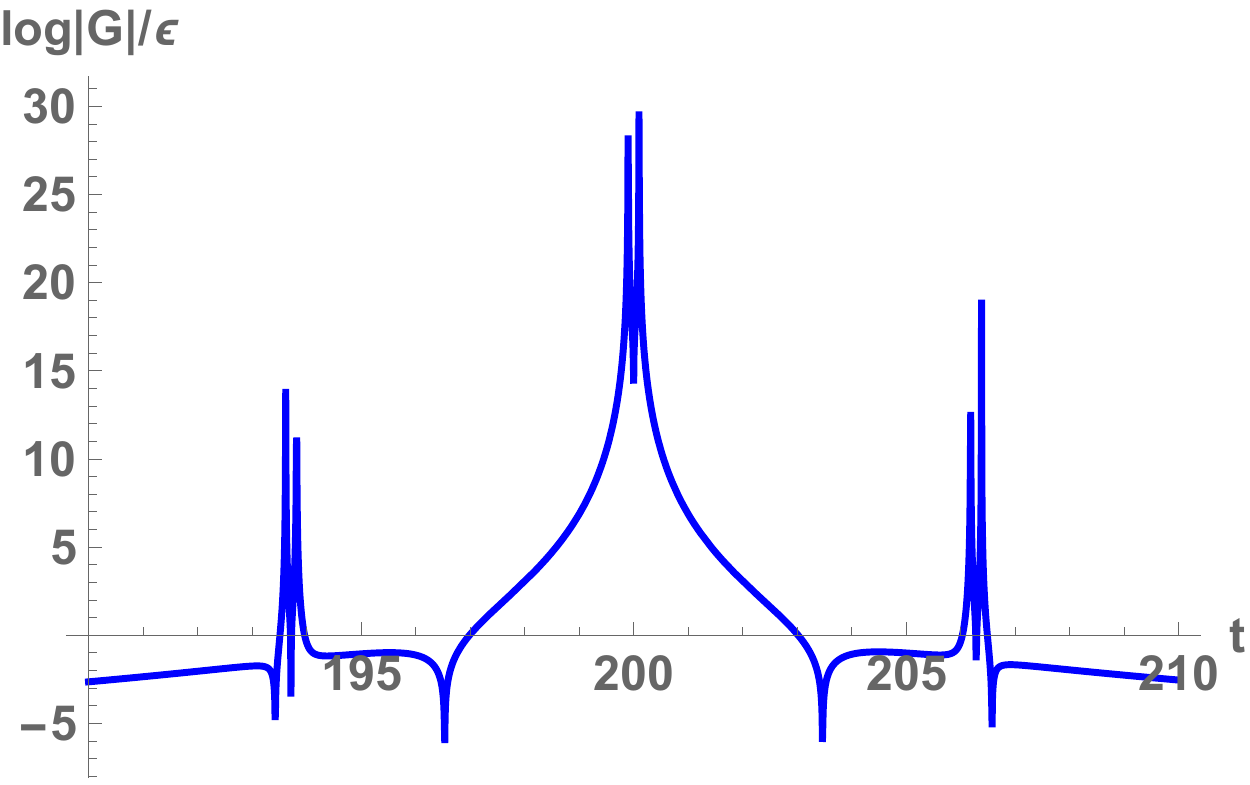}
  \caption{The wormhole correlator around $\Delta t=2L_\lambda$}
 \end{subfigure}
 \begin{subfigure}{0.49\textwidth}
 \includegraphics[width=\textwidth]{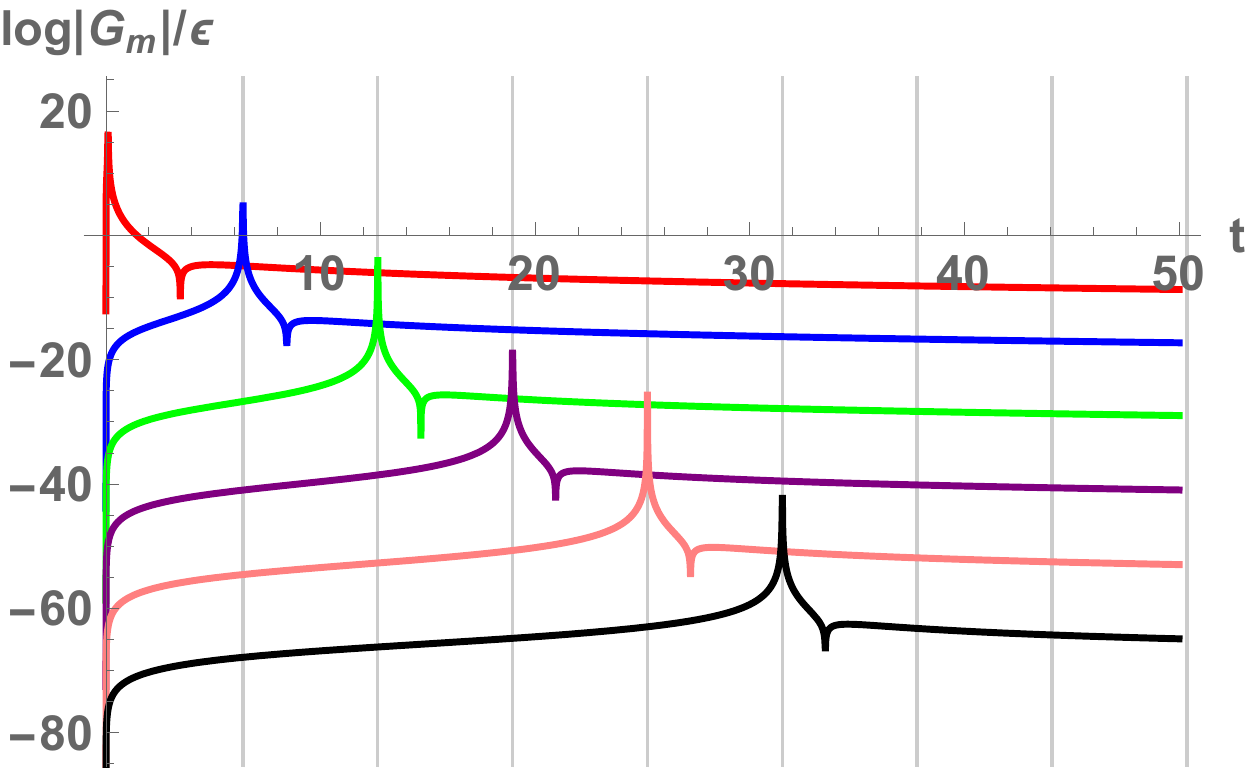}
   \caption{The $m=0,\cdots, 5$ terms in (\ref{eq:WHcorrfreqmod}) separately}
   \label{fig:WHcorrfreqmod:mterms}
 \end{subfigure}
 \caption{The altered wormhole correlator $G^R_{\widetilde{\text{WH}}_\text{(low)}}(\Delta t,\Delta \varphi=0.1)$ as given in (\ref{eq:WHcorrfreqmod}). We take $R=r_+=1$, $\epsilon= 10^{-10}$, and $L_\lambda = 100$. We see the clear onset of the first echo at $\Delta t=2L_\lambda$.  }
 \label{fig:WHcorrfreqmod}
\end{figure}

Replacing $\kappa$ by either one of the $\tilde\kappa$'s in the wormhole modes $f^\lambda$ given in (\ref{eq:deftildef}), it is immediately clear that the normal modes of this altered wormhole satisfy $\tilde \kappa^2 = 1$ and are thus now given by:
\be \label{eq:modWHNMs} \tilde\omega_n^{\text{(low)}} = n \frac{\pi}{L_\lambda}, \qquad \tilde\omega_n^{\text{(high)}} = \left(n+\frac12\right)\frac{\pi}{L_\lambda}, \ee
with $n\in\mathbb{Z}$. We will choose to work with the low-frequency limit in the rest of this section (as opposed to the high-frequency limit that we chose in section \ref{sec:naiveWHcorr}); the high-frequency case proceeds entirely analogously and moreover gives very similar end-results (such as figure \ref{fig:WHcorrfreqmod}).

We consider the retarded propagator (\ref{eq: wh retarded}) and close the $\omega$ integral in the lower half plane:
\be i G^R_{\widetilde{\text{WH}}_\text{(low)}}(\Delta t, \Delta \varphi) =\frac{R^2 \theta(\Delta t)}{8L_\lambda^3r_+^2} \sum_{k\in\mathbb{Z}}\sum_{n\in\mathbb{Z}}\sum_\pm e^{ik\Delta \varphi- i \frac{n \pi }{L_\lambda} \Delta t}(n^2\pi^2R^2-k^2 L_\lambda^2) \coth\left(\frac{n\pi^2 R^2}{2r_+ L_\lambda} \pm  \frac{\pi R}{2r_+}k\right)   .\ee
We use the same Poisson resummation trick as we in the calculation of the black hole propagator, which converts the sum over $k$ into an integral and a sum over images, introducing a factor $e^{i 2\pi k m}$, with $m\in\mathbb{Z}$:
\begin{align}
 \nonumber i G^R_{\widetilde{\text{WH}}_\text{(low)}}(\Delta t, \Delta \varphi) =\frac{R^2 \theta(\Delta t)}{8L_\lambda^3r_+^2}\sum_{m\in\mathbb{Z}}\sum_{n\in\mathbb{Z}}\sum_\pm\int dk\, & e^{ik(\Delta \varphi+2\pi m)- i \frac{n \pi }{L_\lambda} \Delta t }(n^2\pi^2R^2-k^2 L_\lambda^2)\\
 & \times \coth\left(\frac{n\pi^2 R^2}{2r_+ L_\lambda} \pm  \frac{\pi R}{2r_+}k\right).
 \end{align}
For $\Delta \varphi>0$, we can close the $k$ contour in the upper half plane; note that the integrand is already well-behaved as $k\rightarrow \pm\infty$ so there is no need for introducing an extra regulator (as there was in the black hole case). The $\coth$ factors have poles when:
\be k = \pm \frac{n\pi R}{L_\lambda} + 2i l \frac{r_+}{R}, \ee
for $l\in\mathbb{Z}$, so we obtain:
\begin{align}\label{eq:modWHcorrafterkint}  i G^R_{\widetilde{\text{WH}}_\text{(low)}}(\Delta t, \Delta \varphi)  =&-\frac{2\theta(\Delta t)}{ R L_\lambda^2} \sum_{m\in\mathbb{Z}}\sum_{n\in\mathbb{Z}}\sum_{l=1}^\infty \sum_\pm l (  n \pi R^2 \pm i l L_\lambda r_+)  \\
\nonumber & \times \exp\left( -2l\frac{r_+}{R}(\Delta \varphi + 2m\pi) + i n \pi \left(  \pm\frac{R}{L_\lambda}(\Delta \varphi+2m\pi) - \frac{t}{L_\lambda}\right) \right) .
\end{align}
We can now perform the sums over $l$ and $n$ explicitly; just as for the empty $AdS$ correlator (see (\ref{eq:corremptyAdS})), we need to regulate the sum over negative $n$ by taking $\Delta t\rightarrow \Delta t(1+i\epsilon)$ and the sum over positive $n$ by taking $\Delta t\rightarrow \Delta t(1-i\epsilon)$. Using the shorthand notation:
\be \tilde{t}_\epsilon \equiv \frac{\pi }{L_\lambda}\Delta t(1+i\epsilon), \qquad \tilde{\varphi}_m \equiv \frac{\pi R}{L_\lambda}\left(\Delta \varphi + 2\pi m\right),\ee
we can express the final result as:
\begin{align} 
 \label{eq:WHcorrfreqmod} i G^R_{\widetilde{\text{WH}}_\text{(low)}}(\Delta t, \Delta \varphi) &= \frac{ \theta(\Delta t)}{2 R  L_\lambda^2}\sum_{m\in\mathbb{Z}}\left( \tilde{\mathcal{G}}(\tilde{t}_\epsilon, \tilde{\varphi}_m) - \tilde{\mathcal{G}}(\tilde{t}_{-\epsilon}, \tilde{\varphi}_m) \right),\\
\nonumber \tilde{\mathcal{G}}(\tilde{t}_\epsilon, \tilde{\varphi}_m) &\equiv \frac{\pi R^2\left(\cos \tilde{t}_\epsilon \cos \tilde{\varphi}_m-1\right)+L_\lambda r_+ \sin \tilde{\varphi}_m \coth \frac{r_+L_\lambda \tilde{\varphi}_m }{\pi R^2} \left(\cos \tilde{\varphi}_m-\cos \tilde{t}_\epsilon\right)}{\sinh^2\frac{r_+L_\lambda \tilde{\varphi}_m }{\pi R^2} \left(\cos \tilde{t}_\epsilon-\cos \tilde{\varphi}_m\right)^2}.
\end{align}
From the final correlator expression (\ref{eq:WHcorrfreqmod}), it is immediately clear that:
\be G^R_{\widetilde{\text{WH}}_\text{(low)}}(\Delta t+2n L_\lambda,\Delta \varphi) = G^R_{\widetilde{\text{WH}}_\text{(low)}}(\Delta t, \Delta \varphi),\ee
for any $n\in\mathbb{Z}$;, so that this correlator has exact, periodic echoes (as anticipated above in section \ref{sec:naiveWHcorr}).
We plot this correlator in figure \ref{fig:WHcorrfreqmod}; we can clearly see the the first echo at $\Delta t=2L_\lambda$. Note that there is also a ``dip'' in the correlator at $\Delta t=L_\lambda$. 
It is interesting to note that, as opposed to the black hole correlator (see figure \ref{fig:BHcorr}), the contribution from the images $m\geq 2$ is exponentially suppressed compared to the $m=0$ term as we can see figure \ref{fig:WHcorrfreqmod:mterms} (i.e. the red line for $m=0$ is above the $m>2$ peaks).
Another difference with the black hole correlator is that this correlator does not die off (exponentially) as fast as the black hole one; note additionally that the $L_\lambda\rightarrow\infty$ limit does not reproduce the black hole correlator. These differences with the black hole correlator are artifacts of extrapolating the low-frequency limit of the original wormhole to all frequencies, which (as mentioned above) can be interpreted as a frequency-dependent modification of the wormhole boundary condition. In particular, the normal modes $\tilde{\omega}_{n}^\text{(low)}$ in (\ref{eq:modWHNMs}) differ significantly from the actual normal modes of the original wormhole at high frequencies (\ref{eq:SvRWHhighfreqNMs}); of course, the original wormhole has a correlator that (by construction) must have the correct exponential fall-off for $0<\Delta t\ll L_\lambda$.\footnote{Note that the result for the correlator is very similar if we had selected instead the high-frequency limit $\kappa^\text{(high)}$ in (\ref{eq:tildekappadef}) as basis for our calculation. This is somewhat curious, since in this case it is of course at \emph{low} frequencies where this approximation and extrapolation diverges from the actual wormhole results.}

\subsection{Wormhole Transition Feynman Correlator}\label{sec:WHtransposspace}
We would like to calculate the position-space correlator for the ``off-diagonal'' Feynman correlator $\langle \lambda |\mathcal{O}\mathcal{O}| \lambda' \rangle$ in (\ref{eq:feynproptrans}) that we discussed in section \ref{sec:wormhole_transition}. The calculation of this correlator suffers from the same problems we encountered in section \ref{sec:WHfullprop}, namely that we cannot explicitly perform the relevant infinite sums over normal modes as we do not have an analytic expression for these modes. Instead, we will calculate the off-diagonal correlator given by (\ref{eq:feynproptrans}) in the low-frequency approximation introduced above in section \ref{sec:propfreqaltWH}, which we will be able to find analytically.\footnote{Just as for the diagonal correlator discussed in section \ref{sec:propfreqaltWH}, the calculation for the off-diagonal correlator using instead the high-frequency approximation $\tilde\kappa^\text{(high)}$ is completely analogous, and gives qualitatively the same result as the low-frequency approximation described here.}

Recall that the low-frequency approximation of section \ref{sec:propfreqaltWH} is obtained by replacing $\kappa$ in the scalar wave solutions (\ref{eq:deftildef}) by $\tilde\kappa^\text{(low)}$ given in (\ref{eq:tildekappadef}). The normal modes are then given by $\tilde{\omega}_n^\text{(low)}$ in (\ref{eq:modWHNMs}), which are the solutions to $(\tilde\kappa^\text{(low)})^2 = 1$.
It is then a straightforward exercise to calculate the residus of $(1-(\tilde\kappa^\text{(low)})^2)^{-1}$ and $(1-(\tilde\kappa^\text{(low)})^2)^{-2}$ at these normal modes, to find the modified expressions that replace $q,A$ in (\ref{eq:residues_trans}). We find the simple expressions:
\be \tilde q = 2L_\lambda, \qquad \tilde A = 1.\ee
The off-diagonal Feynman correlator then follows from (\ref{eq:feynproptrans}), after replacing the normal modes $\omega_{nk}^\pm$ by their modified versions $\tilde\omega_{n}^\text{(low)}$ and replacing the functions $A,q$ by $\tilde A, \tilde q$. Note that we can also replace $\delta\lambda$ by $\delta L_\lambda$ using the following relation:
\be \label{eq:deltalambdaforL} \delta L_\lambda = - 2 \frac{R^2}{r_+} \frac{\delta\lambda}{\lambda}.\ee
The actual calculation of the Feynman correlator proceeds precisely analogously as in section \ref{sec:propfreqaltWH} for the diagonal (retarded) correlator, although we note that the sum over $n$ is only over the \emph{negative} poles ($n<0$) as indicated in (\ref{eq:feynproptrans}). As usual, we trade the sum over $k$ for an integral over $k$ and sum over images $m$; after performing the integrals over $\omega$ and $k$, we obtain (using $\Delta t = t-t'$,  $\Delta\varphi = \varphi-\varphi'$ and assuming $t'>t_0>t$): 
\begin{align} \nonumber i G^F_{\widetilde{\text{WH}}_\text{(low)},\lambda\rightarrow\lambda'}&(\Delta t,\Delta \varphi)  =-\frac{2}{ R L_\lambda^2} \sum_{m\in\mathbb{Z}}\sum_{n<0}\sum_{l=1}^\infty \sum_\pm  l (  n \pi R^2 \pm i l L_\lambda r_+)\times\left(1 - 2 n \pi i  \frac{\delta L_\lambda}{L_\lambda}    \right)\\
&\times \exp\left( -2l\frac{r_+}{R}(\Delta\varphi + 2m\pi) + i n \pi \left(  \pm\frac{R}{L_\lambda}(\Delta\varphi+2m\pi) - \frac{\Delta t}{L_\lambda}\right) \right).
\end{align}
This is the analogue of (\ref{eq:modWHcorrafterkint}); note that compared to (\ref{eq:modWHcorrafterkint}), the only difference in the expression to be summed over is in the last factor on the first line. We can then again perform the $l,n$ sums explicitly, obtaining:
\begin{align} 
\label{eq:WHmodoffdiagcorr} &i G^F_{\widetilde{\text{WH}}_\text{(low)},\lambda\rightarrow\lambda'}(\Delta t,\Delta \varphi)) = \frac{ 1}{2 R  L_\lambda^2}\sum_{m\in\mathbb{Z}} \left(\tilde{\mathcal{G}}(\tilde{t}_\epsilon, \tilde{\varphi}_m)+ \frac{\pi \delta L_\lambda}{L_\lambda} \Delta\tilde{\mathcal{G}}(\tilde{t}_\epsilon, \tilde{\varphi}_m) \right),\\
 \nonumber &\Delta \tilde{\mathcal{G}}(\tilde{t}_\epsilon, \tilde{\varphi}_m) \equiv \sin\tilde{t}_\epsilon \frac{\pi R^2 \left(-3 +2\cos\tilde{\varphi}_m\cos\tilde{t}_\epsilon + \cos 2\tilde{\varphi}_m\right)+ 2 L_\lambda r_+ \coth \frac{r_+L_\lambda \tilde{\varphi}_m }{\pi R^2} \sin\tilde{\varphi}_m \left(\cos\tilde{\varphi}_m-\cos\tilde{t}_\epsilon\right)}{\sinh^2\frac{r_+L_\lambda \tilde{\varphi}_m }{\pi R^2} \left(\cos \tilde{t}_\epsilon-\cos \tilde{\varphi}_m\right)^3} ,
\end{align}
and $\tilde{\mathcal{G}}(\tilde{t}_\epsilon, \tilde{\varphi}_m)$ is the expression from the original (diagonal) correlator we found in (\ref{eq:WHcorrfreqmod}).

\begin{figure}[ht]\centering
\begin{subfigure}{0.49\textwidth}
 \includegraphics[width=\textwidth]{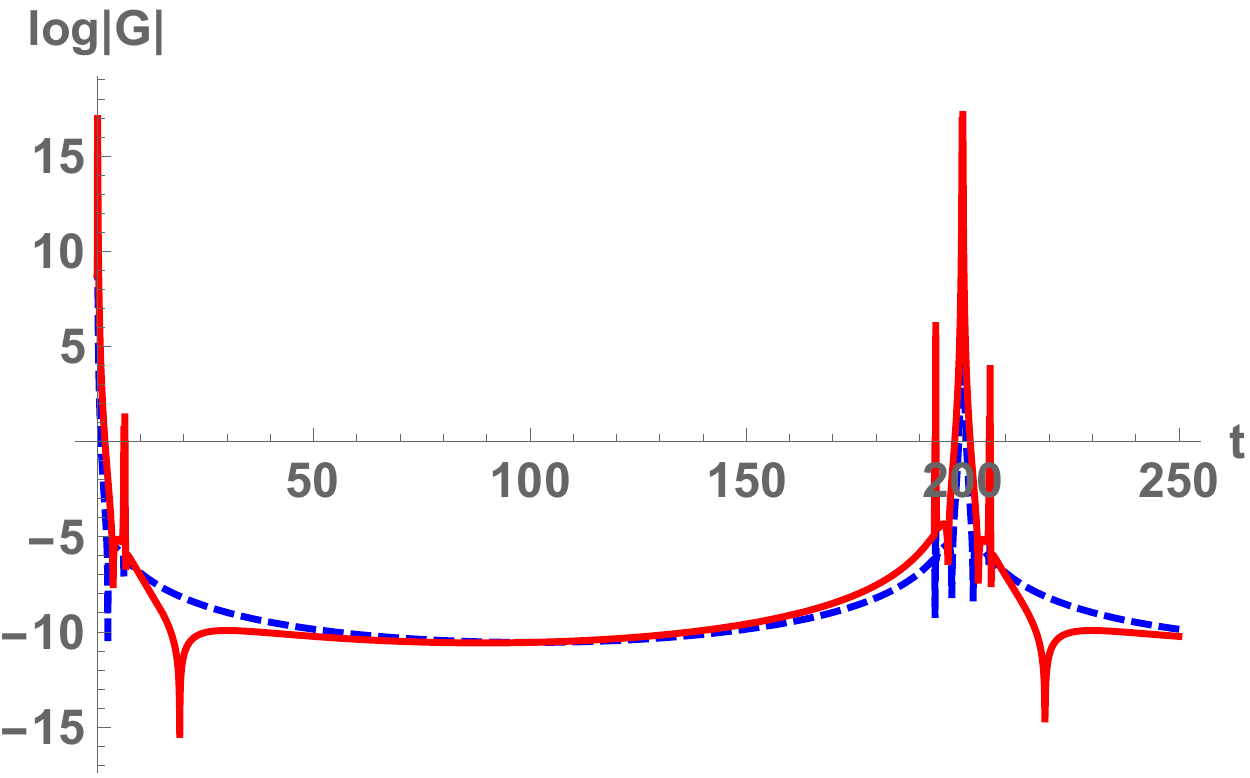}
 \end{subfigure}
\begin{subfigure}{0.49\textwidth}
 \includegraphics[width=\textwidth]{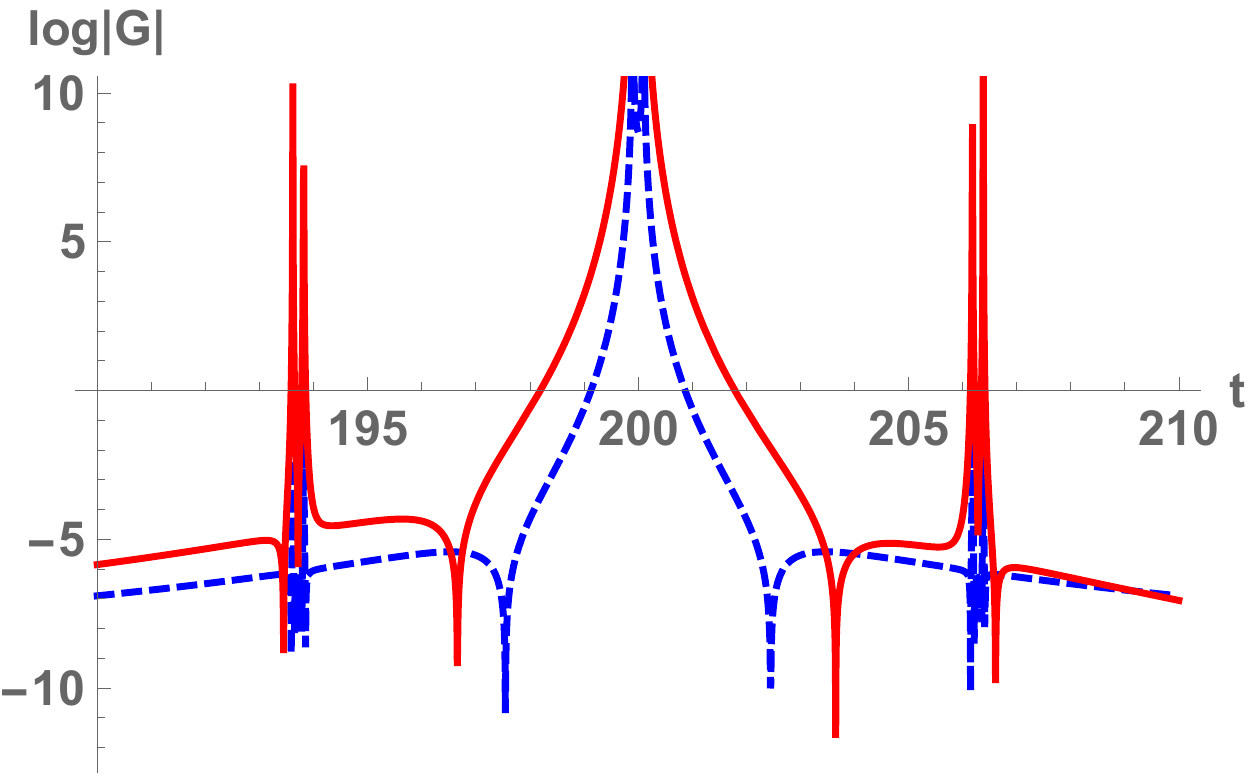}
 \end{subfigure}
 \caption{The altered, off-diagonal wormhole correlator $G^F_{\widetilde{\text{WH}}_\text{(low)},\lambda\rightarrow\lambda'}(\Delta t,\Delta \varphi=0.1)$ (in red) from (\ref{eq:WHmodoffdiagcorr}) together with the diagonal correlator $G^F_{\widetilde{\text{WH}}_\text{(low)},\lambda\rightarrow\lambda}(\Delta t,\Delta \varphi=0.1)$ (in blue, dashed) (which is (\ref{eq:WHmodoffdiagcorr}) with $\delta L_\lambda=0$). We take $R=r_+=1$, $\epsilon= 10^{-10}$, $L_\lambda = 100$ and $\delta L_\lambda = 5$.}
 \label{fig:WHmodoffdiag}
\end{figure}

We plot the correlator (\ref{eq:WHmodoffdiagcorr}) and compare it to the diagonal correlator (when $\delta L_\lambda = 0$) in figure \ref{fig:WHmodoffdiag}.
We can clearly see that the off-diagonal correlator has a wider echo peak than the diagonal correlator, suggesting that sharp features of the (diagonal) correlators (like the echoes at $\Delta t\sim 2nL_\lambda$) become less pronounced in off-diagonal correlators.
Note that in obtaining the off-diagonal correlator (\ref{eq:WHmodoffdiagcorr}), we have integrated over all modes $\omega$, including those that do not satisfy $\omega\ll (\delta L_\lambda)^{-1}$; for such high-frequency modes, it can easily be seen that the expansion (\ref{eq:f_lambda_prim}) (with $\kappa$ replaced by $\tilde\kappa^\text{(low)}$) breaks down.
Therefore, the obtained correlator (and in particular any spurious short-time behaviour) should be taken with a grain of salt.

\section*{Acknowledgments}
We would like to thank Bert Vercnocke for collaborations on earlier versions of this work.
We thank Nikolay Bobev, Geoffrey Comp\`ere, Pierre Heidmann, Tom Lemmens, Christian Maes, and Ruben Monten for useful discussions. VD would especially like to thank Natalia Pinzani Fokeeva for taking a lot of time to explain subtleties regarding \cite{Skenderis:2008dg,Skenderis:2008dh}, and Balt van Rees for discussing the same papers at the pre-Strings Workshop 2019 in Leuven.
The  work  of  VD is supported by a  doctoral  fellowship  from  the  Research  Foundation  -  Flanders (FWO).
The work of DRM is supported by ERC Advanced Grant 787320 - QBH Structure, ERC Starting Grant 679278- Emergent-BH, and FWO Research Project G.0926.17N.
  The work of VSM was supported by a doctoral fellowship from the FWO.
   This work is also partially supported  by the KU Leuven C1 grant ZKD1118 C16/16/005.


\appendix
\section{Schwinger-Keldysh Formalism in Field Theory}
\label{sec:appendix_SK}
In this appendix, we briefly review the Schwinger-Keldysh formalism \cite{Schwinger:1960qe,Bakshi:1962dv,Bakshi:1963bn,Keldysh:1964ud} for $d$-dimensional QFTs in Lorentzian signature. This formalism was developed for cases where the in state and the out state are not \textit{both} eigenstates of the Hamiltonian and as such it is particularly useful for studying non-equilibrium phenomena. It provides a natural definition of the various two-point correlation functions in Lorentzian signature (Feynman, anti-Feynman, retarded, advanced, etc.) in a pure or a thermal state, which is relevant for the discussions in this paper.
\begin{figure}[ht]
    \centering
    \includegraphics{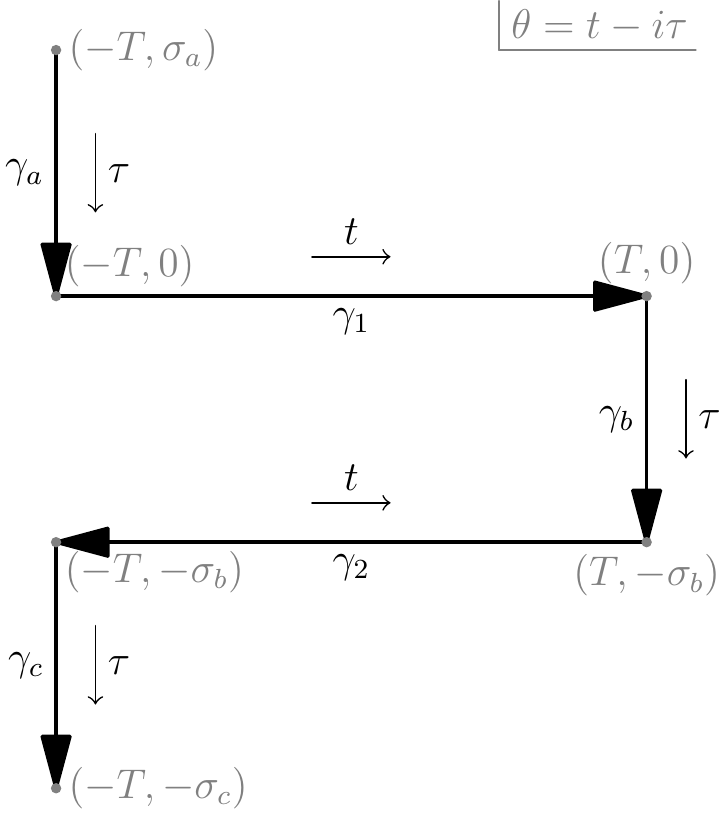}
    \caption{General Keldysh contour given by (\ref{eq:general_SK_contour})}
    \label{fig:generalSKcont}
\end{figure}
We begin by defining a Keldysh contour, $\gamma$, in the complex time plane $\theta = t-i\tau$. Without loss of generality, we can take $\gamma$ to consist of three Euclidean segments $\gamma_I$, where $I = \{a,b,c\}$, and two Lorentzian segments $\gamma_i$, where $i = \{1,2\}$, whose complex times run as (see figure \ref{fig:generalSKcont})\footnote{This type of Schwinger-Keldysh contour is known as an in-in contour, since the Euclidean pieces that prepare the state are both glued at Lorentzian time $t=-T$. For particle scattering problems it is customary to instead work with an in-out contour whose Euclidean pieces are glued at times $t=-T$ and $t=T$, since both the in and out states are eigenstates of the Hamiltonian. The downside of this other configuration is that it can only produce the Feynman correlator explicitly. We will work with such an in-out contour when we calculate the Feynman correlator for a time dependent geometry dual to a wormhole transitioning state in section \ref{sec:wormhole_transition}.}
\begin{align}
    \begin{aligned}
        \gamma_a   & : \quad \theta \in \qty[-T+i\sigma_a,-T] \,,             \\
        \gamma_{b} & : \quad \theta \in \qty[T,T-i\sigma_{b}] \,,             \\
        \gamma_c   & : \quad \theta \in \qty[-T-i\sigma_{b},-T-i\sigma_c] \,,
    \end{aligned} \qquad
    \begin{aligned}
        \gamma_1   & : \quad \theta \in \qty[-T,T] \,,                         \\
        \gamma_{2} & : \quad \theta \in \qty[-T-i\sigma_{b},T-i\sigma_{b}] \,.
    \end{aligned}
    \label{eq:general_SK_contour}
\end{align}
The Euclidean segments prepare the state $\rho$. When $\rho$ is the vacuum or some other pure state, we set $\sigma_{b} \rightarrow 0$ and $\sigma_a=\sigma_c \rightarrow \infty$. When $\rho$ is a thermal state of temperature $\beta$ we set $\sigma_a = \sigma_{b} = \sigma_c = \beta/2$ and identify the end points of the contour. The partition function, evaluated in the state $\rho$, in the presence of a source  $s$ coupled to some operator $\mathcal{O}$, is given by
\begin{align}
    Z^{\text{QFT}}_\rho [s] = \int_\rho \mathcal{D}\chi \ \exp\qty(i\int_\gamma \dd{\theta} \int \dd{\vec{x}} \qty[ \mathcal{L}_{\text{QFT}}[\chi, \partial_\mu \chi] + s(\theta,\vec{x})\mathcal{O}(\theta,\vec{x}) ] ),
    \label{eq:cft_partition_function}
\end{align}
where $\chi$ collectively describes the QFT degrees of freedom. Omitting the spatial $\vec{x}$ integral in what follows, the source term in the partition function reads
\begin{align}
    \int_\gamma \dd{\theta} s(\theta)\mathcal{O}(\theta) & = \int^0_{-\sigma_a} \dd{\tau}s(-T-i\tau) \mathcal{O}(-T-i\tau)+ i\int^T_{-T} \dd{t} s(t)\mathcal{O}(t)                                                           \nonumber               \\
                                                         & \ \ \ \ + \int^{\sigma_{b}}_{0} \dd{\tau}s(T-i\tau) \mathcal{O}(T-i\tau) + i \int^{-T}_T \dd{t} s(t-i\sigma_{b})\mathcal{O}(t-i\sigma_{b})                       \nonumber                \\
                                                         & \ \ \ \ + \int^{\sigma_c}_{\sigma_{b}} \dd{\tau}s(-T-i\tau) \mathcal{O}(-T-i\tau)                                                                                \nonumber                \\
                                                         & \equiv \int^0_{-\sigma_a} \dd{\tau}s_a(\tau) \mathcal{O}_a(\tau)+ i\int^T_{-T} \dd{t} s_1(t)\mathcal{O}_1(t) + \int^{\sigma_{b}}_{0} \dd{\tau}s_{b}(\tau) \mathcal{O}_{b}(\tau) \nonumber \\
                                                         & \ \ \ \ - i \int^{T}_{-T} \dd{t} s_{2}(t)\mathcal{O}_{2}(t) + \int^{\sigma_c}_{\sigma_{b}} \dd{\tau}s_c(\tau) \mathcal{O}_c(\tau) \,,
\end{align}
where we have defined
\begin{align}
    \begin{aligned}
        s_a(\tau)   & = s(-T-i\tau) \,, & \tau       & \in \qty[-\sigma_a,0] \,,       \\
        s_{b}(\tau) & = s(T-i\tau) \,,  & \quad \tau & \in \qty[0,\sigma_{b}]   \,,    \\
        s_c(\tau)   & = s(-T-i\tau) \,, & \quad \tau & \in \qty[\sigma_b,\sigma_c] \,,
    \end{aligned} \qquad
    \begin{aligned}
        s_1(t)   & = s(t) \,,               & t & \in[-T,T] \,, \\
        s_{2}(t) & = s(t - i\sigma_{b}) \,, & t & \in[-T,T] \,,
    \end{aligned}
    \label{eq:sources_on_different_segments}
\end{align}
and analogously for $\mathcal{O}_{i}$ and $\mathcal{O}_I$. The vev of $\mathcal{O}_{i}$, in the state $\rho$, and in the presence of the source $s$, is given by
\begin{align}
    \ev{\mathcal{O}_{i}(x)}_{\rho, s} = {(-1)^{\delta_{i2}} \over i} { \delta \ln Z^{\text{CFT}}_\rho [s] \over \delta s_{i}(x)} \,,
\end{align}
where $x=(t,\vec{x})$. The Keldysh propagator, in the state $\rho$, is
\begin{align}
    i G^{ij}_{\rho}(x,x') = \eval{{(-1)^{\delta_{ij}-\delta_{i2}+1} \over i} { \delta \ev{\mathcal{O}_i(x)}_{\rho, s} \over \delta s_j(x')} }_{s=0} = \ev{T_\gamma \mathcal{O}_i(x)  \mathcal{O}_j(x')}_{\rho} \,,
    \label{eq:generic_SK_ordered_correlator}
\end{align}
where $T_\gamma$ stands for time ordering along the Keldysh contour. In our conventions the Feynman ($iG^F_\rho$), Wightman ($iG^>_\rho$), anti-Wightman ($iG^<_\rho$), and anti-Feynman ($iG^{\overline{F}}_\rho$) correlators are defined as
\begin{align}
    \begin{aligned}
        \ev{T_{\gamma} \mathcal{O}_1(x)\mathcal{O}_1(x')}_{\rho}   & \equiv \ev{T \mathcal{O}(x)\mathcal{O}(x')}_{\rho} = iG^F_\rho(x,x')        \,,                  \\
        \ev{T_{\gamma} \mathcal{O}_1(x)\mathcal{O}_{2}(x')}_{\rho} & \equiv \ev{\mathcal{O}(x')\mathcal{O}(x)}_{\rho} = iG^<_\rho(x,x')             \,,               \\
        \ev{T_\gamma \mathcal{O}_{2}(x)\mathcal{O}_1(x')}_{\rho}   & \equiv \ev{\mathcal{O}(x)\mathcal{O}(x')}_{\rho} = iG^>_\rho(x,x')                \,,            \\
        \ev{T_{\gamma} \mathcal{O}_{2}(x)\mathcal{O}_2(x')}_{\rho} & \equiv \ev{\overline{T} \mathcal{O}(x)\mathcal{O}(x')}_{\rho} = iG^{\overline{F}}_\rho(x,x') \,,
    \end{aligned}
\end{align}
where the equivalences stem from the fact that any operator that appears on the backwards piece, $\gamma_{2}$, by construction has larger Keldysh time than any operator appearing on the forwards contour, $\gamma_1$, and we have dropped the subscripts $(1,2)$ from the operators. It is easy to see that
\begin{align}
    \begin{aligned}
        iG^>(x,x')   & = \theta(t-t') iG^F(x,x') + \theta(t'-t) iG^{\overline{F}}(x,x') \,, \\
        iG^{<}(x,x') & = \theta(t'-t) iG^F(x,x') + \theta(t-t') iG^{\overline{F}}(x,x') \,. \\
    \end{aligned}
    \label{eq:wightman_correlators_in_terms_of_Feynmans}
\end{align}
Out of the components of the Keldysh propagator one can build the retarded ($iG^R_\rho$) and advanced ($iG^A_\rho$) propagators
\begin{align}
    \begin{aligned}
        iG^R_\rho(x,x') & = iG^F_\rho(x,x') - iG^<_\rho(x,x') = \theta(t-t')\qty[ iG^F_\rho(x,x') - iG^{\overline{F}}_\rho(x,x') ] \,,  \\
        iG^A_\rho(x,x') & = iG^F_\rho(x,x') - iG^>_\rho(x,x') = -\theta(t'-t)\qty[ iG^F_\rho(x,x') - iG^{\overline{F}}_\rho(x,x') ] \,.
    \end{aligned}
    \label{eq:retarded_and_advenced_correlators_in_terms_of_Feynmans}
\end{align}
Of particular physical interest is the retarded correlator; this appears as the causal linear response of the system due to external perturbations. We can see this when calculating the variation of the one-point function on $\gamma_1$, in state $\rho$ and in the presence of the \textit{same} external source $s_1 = s_{2}$ on $\gamma_1$ and $\gamma_2$:
\begin{align}\label{eq:GRphysical}
    \delta \ev{\mathcal{O}_1(x')}_{\rho,s} & = \int \dd{x} \qty[ \delta s_1(x) {\delta \ev{\mathcal{O}_1(x')}_{\rho,s} \over \delta s_1(x)} +  \delta s_{2}(x) {\delta \ev{\mathcal{O}_1(x')}_{\rho,s} \over \delta s_{1'}(x)} ] \nonumber \\
                                           & = i\int \dd{x} \delta s_1(x) \qty[\ev{T\mathcal{O}(x)\mathcal{O}(x')}_{\rho} -  \ev{\mathcal{O}(x)\mathcal{O}(x')}_{\rho}]                                                    \nonumber       \\
                                           & =-\int\dd{x} \delta s_1(x) G^R_\rho(x,x') \,.
\end{align}
A good consistency check is that the correlators we obtain should obey the identities:
\begin{align}
    \begin{aligned}
        \qty[iG^>(x,x')]^* & = iG^<(x,x') \,, & \qty[iG^F(x,x')]^*         & = iG^{\overline{F}}(x,x')  \,,  \\
        iG^>(x',x)         & = iG^<(x,x') \,, & iG^{F, \overline{F}}(x',x) & = iG^{F,\overline{F}}(x,x') \,.
    \end{aligned}
    \label{eq:consistency_correlators}
\end{align}

\section{SvR Real-Time Holography}
\label{sec:appendix_SvR}
In this appendix, we briefly review the parts of the Skenderis-van Rees (SvR) formalism for real-time $\text{AdS}_{d+1}/\text{CFT}_d$ holography \cite{Skenderis:2008dh,Skenderis:2008dg} which are most relevant for our discussion in section \ref{sec:SvR_calculations}, see \cite{Skenderis:2008dg} for a much more complete discussion and derivation. The SvR formalism can be seen as the holographic dual of the Schwinger-Keldysh formalism in field theory, which we briefly review in appendix \ref{sec:appendix_SK}.

We will focus on a scalar field $\Phi(\theta,\vec{x},r)$ with mass $m$ dual to the operator $\mathcal{O}(\theta, \vec{x})$ of conformal dimension $\Delta$, which satisfies:
\begin{align}
    R^2 m^2 = \Delta(\Delta -2) \,,
\end{align}
using $R$ as the AdS scale. The bulk dual to a field theory state $\rho$ can be prepared by allowing $\Phi$ to fluctuate on top of a mixed-signature spacetime $\mathcal{M}$ with fixed background metric $g^{E,L}_{\mu \nu}$, which is Euclidean or Lorentzian depending on the segment. The bulk partition function, as a functional of the boundary value of the field\footnote{Here, we do not yet specify the precise meaning of $\eval{\Phi}_{\partial\mathcal{M}}$; the correct interpretation of this equation can be found in (\ref{eq:appsourceasphi}).} $\eval{\Phi}_{\partial\mathcal{M}} (\theta,\vec{x})\equiv s(\theta, \vec{x})$, is given by
\begin{align}
    Z^{\text{bulk}}_{g_{\mu \nu}}[s] = \int_{\eval{\Phi}_{\partial\mathcal{M}} = s} \mathcal{D}\Phi  \ e^{iS_{\text{bulk}}[\Phi]} \approx e^{iS^{\text{on-shell}}_{\text{bulk}}\qty[s]} \,.
    \label{eq:bulk_partition_function}
\end{align}
The bulk action along the contour $\gamma$ is
\begin{align}
    iS_{\text{bulk}}[\Phi] = \int_\gamma \dd{\theta} L\qty(\theta) & = \int^0_{-\sigma_a} \dd{\tau} L(-T -i\tau) + i\int^T_{-T} \dd{t} L(t) + \int^{\sigma_{b}}_{0} \dd{\tau} L(T -i\tau)          \nonumber   \\
                                                                   & \ \ \ \ + i\int^{-T}_{T} \dd{t} L(t-i\sigma_b) + \int^{\sigma_c}_{\sigma_{b}} \dd{\tau} L(-T -i\tau)                          \nonumber   \\
                                                                   & \equiv -\int^0_{-\sigma_a} \dd{\tau} L_a^E(\tau) + i\int^T_{-T} \dd{t} L_1^L(t) - \int^{\sigma_{b}}_{0} \dd{\tau} L_{b}^E(\tau) \nonumber \\
                                                                   & \ \ \ \ - i\int^{T}_{-T} \dd{t} L_{2}^L(t) - \int^{\sigma_c}_{\sigma_{b}} \dd{\tau} L_c^E(\tau) \,,
\end{align}
where the Lorentzian and Euclidean Lagrangians are given by
\begin{align}
    \begin{aligned}
        L^{L}_i\qty(t)  & = {1 \over 2} \int \dd{\vec{x}}\dd{r} \sqrt{-g^L}\qty[-g^{L,\mu\nu} \partial_\mu \Phi_i(t,\vec{x},r) \partial_\nu \Phi_i(t,\vec{x},r) - m^2 \Phi_i^2(t,\vec{x},r)]  \,,      \\
        L^E_I\qty(\tau) & = {1 \over 2} \int \dd{\vec{x}}\dd{r} \sqrt{g^E}\qty[g^{E,\mu\nu}\partial_\mu \Phi_I(\tau,\vec{x},r) \partial_\nu \Phi_I(\tau,\vec{x},r) + m^2 \Phi_I^2(\tau,\vec{x},r)] \,,
    \end{aligned}
\end{align}
and the fields $\Phi_i$ and $\Phi_I$ are defined analogously to (\ref{eq:sources_on_different_segments}). Further, in (\ref{eq:bulk_partition_function}) we have approximated the path integral using the saddle point approximation around the mixed-signature classical solution $\Phi^{\text{cl}}(\theta,\vec{x},r)$. We build this by first solving for the two Lorentzian ($\Phi^{\text{cl}}_{i}(t,\vec{x},r)$) and the three Euclidean pieces ($\Phi^{\text{cl}}_{I}(\tau,\vec{x},r)$) separately, which satisfy the respective equations of motion
\begin{align}
    \begin{aligned}
        {1 \over \sqrt{-g^L} }\partial_\mu \qty(\sqrt{-g^L} g^{L,\mu\nu} \partial_\nu \Phi^{\text{cl}}_i ) - m^2 \Phi^{\text{cl}}_i & =0 \,, \\
        {1 \over \sqrt{g^E} }\partial_\mu \qty(\sqrt{g^E} g^{E,\mu\nu} \partial_\nu \Phi^{\text{cl}}_I ) - m^2 \Phi^{\text{cl}}_I   & =0 \,.
    \end{aligned}
    \label{eq:eom}
\end{align}
Then, the properly extremized mixed-signature classical solution is such that, at the spatial hypersurfaces where Euclidean and Lorentzian segments meet, the piecewise fields obey the matching conditions\footnote{In equation (4.1.15) of \cite{Skenderis:2008dg}, the matching conditions have different signs. The difference stems from the fact that they parametrize the time on $\gamma_{2}$ to run in the opposite direction. In our convention the minus appearing in front of the action on $\gamma_{2}$ cancels with the minus coming from the fact that the derivation $-\partial_\tau$ on $\gamma_{b}$ runs in the opposite direction compared to the derivation $i\partial_t$ on $\gamma_{2}$, thus (\ref{eq:unversal_matching_conditions}) applies universally at all meeting hypersurfaces. }
\begin{align}
    \Phi^{\text{cl}}_i  = \Phi^{\text{cl}}_I \,,  \qquad  -i\partial_t \Phi^{\text{cl}}_i  = \partial_\tau \Phi^{\text{cl}}_I \,.
    \label{eq:unversal_matching_conditions}
\end{align}
The classical solution puts the bulk action on-shell: $iS_{\text{bulk}}\qty[\Phi^{\text{cl}}]=iS^{\text{on-shell}}_{\text{bulk}}\qty[s]$, and renders it a functional of the boundary value of the field $\eval{\Phi}_{\partial\mathcal{M}} (\theta,\vec{x})\equiv s(\theta, \vec{x})$. We assume we can solve the Lorentzian and Euclidean equations of motion by separation of variables:
\begin{align}
    \Phi_i^{\text{cl}} & = \sum_{l, \vec{m}}\int d\omega e^{-i \omega t} Y_{l\vec{m}}(\vec{x}) f_i(\omega,l,\vec{m},r) \,, & \Phi_I^{\text{cl}} & =  \sum_{l, \vec{m}}\int d\omega e^{-\omega \tau} Y_{l\vec{m}}(\vec{x}) f_I(\omega,l,\vec{m},r) \,.
    \label{eq:modes_general_d}
\end{align}
The radial parts $f_i,f_I$ then obey the same second-order ODE,\footnote{A Euclidean mode will satisfy exactly the same radial equation as a Lorentzian one, since e.g. $e^{i \omega t} g^{L, tt} \partial^2_t e^{-i \omega t}  = e^{\omega \tau}  g^{E, \tau \tau} \partial^2_\tau e^{-\omega \tau}$ (similar relations hold involving off-diagonal time-space components of the metric).  As such, there is no need to distinguish between the Lorentzian and Euclidean radial functions.} which needs to be supplied with two boundary conditions. If the geometry in question has only one boundary (e.g. empty AdS), then one boundary condition comes from demanding regularity in the IR (e.g. at the origin of AdS) and another from demanding that the field asymptotes to the source in the UV (e.g. at the boundary of AdS). If the geometry has two boundaries (e.g. the the eternal black hole, then we have two UV boundary conditions. To implement UV boundary condition(s), we expand the fields in powers of $r$ near the boundary\footnote{We use coordinates where the boundary is at $r\rightarrow \infty$ and looks like:
\be ds^2|_{r\rightarrow\infty} = \frac{r^2}{R^2}\left( -d\theta^2 + d\vec{x}^2\right) + \frac{R^2}{r^2}dr^2 \,, \ee
where $d\theta^2 = dt^2$ or $d\theta^2 = -d\tau^2$, depending on if we are in Lorentzian or Euclidean signature.
}
\begin{align}
    \Phi^{\text{cl}}_{i,I}(x,r) = \qty({a \over r})^{d-\Delta}\phi_{i,I}^{(d-\Delta)}(x)+ \, \dots \, +\qty({a \over r})^{\Delta}\qty(\phi_{i,I}^{(\Delta)}(x)+\psi_{i,I}^{(\Delta)}(x)\log{a^2 \over r^2})+\dots \,,
    \label{eq:asymp_exp}
\end{align}
where $x=(t,\vec{x})$ or $x=(\tau, \vec{x})$ depending on whether we are on an Euclidean or a Lorentzian segment, and $a$ is some distance scale of the geometry in question. The first set of dots in (\ref{eq:asymp_exp}) contains terms between orders $d-\Delta+2$ and $\Delta-2$,  while the second set of dots contains terms of orders beyond $\Delta$, which all might have $\log$ pieces to them. These $\log$ pieces only appear when $d$ is even \textit{and} when $\Delta \in \mathbb{Z}$, and are related to the conformal anomaly in the CFT. From these expansions, we read off the sources as:
\begin{align} \label{eq:appsourceasphi}
    s_{i,I}(x) =\lim_{r \rightarrow \infty} \qty(r \over a)^{d-\Delta} \Phi_{i,I}(x,r)  \equiv \phi_{i,I}^{(d-\Delta)}(x) \,.
\end{align}
We can also define the Fourier transformed sources $\tilde{s}_{i,I}$ through:
\begin{align}
    s_i(x) & = \sum_{l,\vec{m}} \int d\omega e^{-i \omega t} Y_{l\vec{m}}(\vec{x}) \tilde{s}_i(\omega,l,\vec{m})f(\omega,l,\vec{m},r) \,,  \\
    s_I(x) & = \sum_{l,\vec{m}}\int d\omega e^{-\omega \tau} Y_{l\vec{m}}(\vec{x})  \tilde{s}_I(\omega,l,\vec{m})f(\omega,l,\vec{m},r) \,,
\end{align}
which can alternatively be seen as fixing the normalization of the radial functions $f$. Having obtained the piecewise bulk fields $\Phi_{i,I}$ that solve the equations of motion with the correct boundary conditions and obey the matching conditions, we build up the mixed-signature bulk field $\Phi$ using the analogous expression to (\ref{eq:sources_on_different_segments}) and express the on-shell action as
\begin{align}
    iS^{\text{on-shell}}_{\text{bulk}}[s] = \lim_{r \rightarrow \infty}{i\over 2} \qty(r\over a)^\Delta \int_\gamma \dd{\theta} \int \dd{\vec{x}} s(\theta,\vec{x}) r \partial_r \Phi(\theta,\vec{x},r).
\end{align}
It is clear that all terms beyond order $\Delta$ in (\ref{eq:asymp_exp}) decay when the limit is taken, while all terms between orders $d-\Delta$ and $\Delta-2$ diverge. We can cancel these divergences with appropriate counterterms; this procedure is called holographic renormalization \cite{Skenderis:2002wp}. The final, renormalized result is:
\begin{align}
    iS^{\text{on-shell}}_{\text{bulk}}[s] = i(2\Delta -d)\int_\gamma \dd{\theta} \int \dd{\vec{x}} s(\theta,\vec{x}) \phi^{(\Delta)}(\theta,\vec{x}) \,.
\end{align}
With this setup one can extend the Gubser-Klebanov-Polyakov-Witten dictionary \cite{Witten:1998qj,Gubser:1998bc}, originally developed only for Euclidean AdS/CFT, to these mixed-signature bulks and boundaries. The dictionary states that the boundary value of the field, $\eval{\Phi}_{\partial\mathcal{M}}$, acts as a source for the operator $\mathcal{O}$. Further the boundary partition function in the state $\rho$ is equivalent to the bulk partition function on the background $g_{\mu\nu}$, which in the large $N$ limit can be approximated by the classical on-shell action
\begin{align}
    Z^{\text{CFT}}_\rho [s] = Z^{\text{bulk}}_{g_{\mu \nu}}[s] \approx e^{iS^{\text{on-shell}}_{\text{bulk}}\qty[s]} \,.
    \label{eq:Zbulk=ZCFT}
\end{align}
Thus, in the large $N$ limit, we can compute correlation functions holographically as
\begin{align}
    \begin{aligned}
        \ev{\mathcal{O}_{i}(x)}_{\rho, s} & ={ (-1)^{\delta_{i2}} \over i} { \delta iS^{\text{on-shell}}_{\text{bulk}}[s] \over \delta s_{i}(x)} = (2\Delta -d) \phi^{(\Delta)}_{i}(x)                                                                                                           \,,   \\
        i G^{ij}_{\rho}(x,x')             & = \eval{{(-1)^{\delta_{ij}-\delta_{i2}+1} \over i} { \delta \ev{\mathcal{O}_i(x)}_{\rho, s} \over \delta s_j(x')} }_{s=0} = {(-1)^{\delta_{ij}-\delta_{i2}+1}(2\Delta -d) \over i} \eval{ { \delta \phi^{(\Delta)}_i(x)  \over \delta s_j(x')} }_{s=0} \,.
    \end{aligned}
    \label{eq:Keldysh_propagator_from_bulk}
\end{align}

\subsection{Correlators of Empty AdS}\label{sec:app:emptyAdS}
In this section, we briefly review the real-time holography calculation that obtains the two-point function of a minimally coupled scalar field in empty $\text{AdS}_3$. This calculation can  be found in section 4.1 of \cite{Skenderis:2008dg}.

\begin{figure}[ht]
    \centering
    \includegraphics{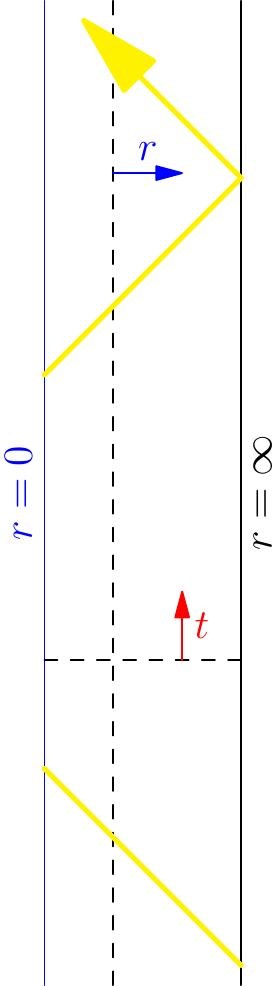}
    \caption{Empty $\text{AdS}_3$ with a radial light ray (depicted in yellow). The light ray appears discontinuous on the diagram because we have suppressed the angle $\varphi$: after reaching the center of AdS the light ray continues propagating in the antipodal part of the spacetime, reflects from the antipodal point on the boundary and re-emerges at $r=0$ after time $\pi R$. }
    \label{fig:empty_AdS}
\end{figure}

The metric of global (Lorentzian) $\text{AdS}_3$ is:
\begin{align}
    \dd{s}^2 = - \qty(1+{r^2 \over R^2}) \dd{t}^2 +\qty(1+{r^2 \over R^2})^{-1} \dd{r}^2 + r^2 \dd{\varphi}^2 \,,  \qquad
    \begin{cases}
        \begin{aligned}
            r       & \in [0,\infty)  \,,      \\
            t       & \in (-\infty,\infty) \,, \\
            \varphi & \in [0,2\pi) \,.
        \end{aligned}
    \end{cases}
    \label{eq:ads_global_metric}
\end{align}
The Euclidean metric is obtained by taking $t\rightarrow i\tau$. A conformal diagram of empty $\text{AdS}_3$ is given in figure \ref{fig:empty_AdS}.

\paragraph{Schwinger-Keldysh contour} We take the following Schwinger-Keldysh consisting of two Euclidean ($\gamma_I$, $I = a,b$) and two Lorentzian ($\gamma_i$, $i=1,2$) segments, whose Keldysh time $\theta = t - i\tau$ runs as
\begin{align}
    \begin{aligned}
        \gamma_a   & : \quad \theta \in \qty[-T+i\infty,-T] \,, \\
        \gamma_{b} & : \quad \theta \in \qty[T,T-i\infty] \,,
    \end{aligned} \qquad
    \begin{aligned}
        \gamma_{1} & : \quad \theta \in \qty[-T,T] \,, \\
        \gamma_{2} & : \quad \theta \in \qty[-T,T] \,.
    \end{aligned}
    \label{eq:ads_SK_contour}
\end{align}
The bulk dual to this contour is made out of two Lorentzian pieces and one Euclidean piece, which is split in two at $t=0$. The two Euclidean sub-pieces are then glued to the Lorentzian pieces as depicted in figure \ref{fig:pure_state_SK_contours}.  Note that the boundary contours of empty AdS and the Solodukhin wormhole are the same; however, the blue line in the bulk contour has a different interpretation: for the wormhole, it is the center of the wormhole throat, whereas for empty AdS it is the center of AdS at $r=0$. These gluings dictate the matching conditions that the piecewise bulk fields obey\footnote{The matching condition between the two Lorentzian fields is the $\sigma_b \rightarrow 0$ limit of the matching conditions of the general contour (\ref{eq:general_SK_contour}) in appendix \ref{sec:appendix_SK}.}
\begin{align} \label{eq:AdSmatchingconds}
    \begin{aligned}
        \Phi_a(0,\varphi,r) & = \Phi_1(-T,\varphi,r) \,, & \partial_{\tau} \Phi_a(0,\varphi,r) & = -i \partial_{t}\Phi_1(-T,\phi,r)  \,,   \\
        \Phi_1(T,\varphi,r) & = \Phi_2(T,\varphi,r) \,,  & \partial_{t} \Phi_1(T,\varphi,r)    & =  \partial_{t}\Phi_2(T,\varphi,r)  \,,   \\
        \Phi_b(0,\varphi,r) & = \Phi_2(-T,\varphi,r) \,, & \partial_{\tau}\Phi_b(0,\varphi,r)  & = -i\partial_{t} \Phi_2(-T,\varphi,r) \,.
    \end{aligned}
\end{align}
\paragraph{Scalar wave solutions}
We can use separation of variables as in (\ref{eq:scalar3Dsep}) to solve the equations of motion for the minimally coupled scalar on the different pieces of the Schwinger-Keldysh contour:
\begin{align}
    \Phi_i= e^{-i \omega t} e^{ik\varphi} \tilde{s}_i(\omega,k) f(\omega,k,r) \,,  \qquad \Phi_I = e^{-\omega \tau} e^{ik\varphi} \tilde{s}_I(\omega,k) f(\omega,k,r) \,,
\end{align}
with $i=\{1,2\}$ for the Lorentzian segments and $I=\{a,b\}$ for the Euclidean segments. The solutions to the wave equation (\ref{eq:scalar3Dradial}) for the AdS metric (\ref{eq:ads_global_metric}) are
\begin{align}
    f^{\pm}(\omega, k, r) = \qty(r \over R)^{\pm k} \qty(1+{r^2 \over R^2})^{R \omega \over 2} \, _2F_1\qty({1\over 2}(R\omega \pm k), 1+{1\over 2}(R\omega \pm k);1\pm k;-{r^2 \over R^2} ) \,.
    \label{eq:bh_modes}
\end{align}
Demanding regularity in the IR (near the origin $r=0$) necessitates that the only allowed solution is the linear combination:
\begin{align}
    f(\omega,k,r) = \mathcal{N}_{\omega k}\qty[\theta(k)f^{+}(\omega,k,r)+\theta(-k)f^{-}(\omega,k,r)] = \mathcal{N}_{\omega k} f^{+}(\omega,\abs{k},r) \,,
\end{align}
where $\mathcal{N}_{\omega k}$ is a constant of normalization
\begin{align}
    \mathcal{N}_{\omega k} = {\Gamma\qty(1+{1\over 2}(R\omega + \abs{k})) \Gamma\qty(1-{1\over 2}(R\omega - \abs{k}))\over \Gamma(1+\abs{k})} \,,
\end{align}
which is chosen such that the leading behavior at the UV (near the boundary $r=\infty$) is normalized to unity. Then the asymptotic expansion of the modes near the boundary is:
\begin{align}
    f(\omega,k,r)    & = 1+{R^2 \over r^2} \alpha(\omega,k) \qty( \beta(\omega,k)+\log{R^2 \over r^2}) + \dots \label{eq:asymp_exp_ads}  \,,                                                                          \\
    \alpha(\omega,k) & = -{1\over 4}\qty(R^2\omega^2 - \abs{k}^2) \nonumber                                                                 \,,                                                                       \\
    \beta(\omega,k)  & = \psi\left(1-\frac{1}{2} (R \omega -\abs{k})\right)+\psi\left(1+\frac{1}{2} (R \omega +\left| k\right| )\right)+\frac{2(R\omega + \abs{k}) }{R^2 \omega^2 -\left| k\right|^2 } \,,  \nonumber
\end{align}
where $\psi(\cdot)=\Gamma'(\cdot)/\Gamma(\cdot)$ is the diagamma function and we have given $\beta(\omega,k)$ modulo constants which do not affect the poles of $\alpha \beta$. The poles of $f$ are given by
\begin{align}
    R\omega^\pm_{nk} = \pm (2n + \abs{k}) \,,  \qquad n = 1,2,\dots \,,  \quad  k \in \mathbb{Z} \,.
\end{align}
These (real) poles constitute the normal modes of empty AdS.

\paragraph{Construction of the piecewise bulk fields}
Our next task is to construct the piecewise bulk fields. A generic Lorentzian field is expanded as
\begin{align}
    \Phi_i & = \sum_k e^{i k\varphi} \int_{\mathcal{C}_i}\dd{\omega} e^{-i\omega t}f   \delta_{ij}\tilde{s}_j                                                                                                                                             \nonumber \\
           & = \sum_k e^{i k\varphi} \qty[ \int_{\mathcal{F}}\dd{\omega} e^{-i\omega t}f  \delta_{ij} \tilde{s}_j + \int_{\mathcal{W}^<}\dd{\omega} e^{-i\omega t}f V_{ij}  \tilde{s}_j + \int_{\mathcal{W}^>}\dd{\omega} e^{-i\omega t}f W_{ij} \tilde{s}_j] \,,
    \label{eq:ads_lorentzian_field_ansatz}
\end{align}
where in the first line $\mathcal{C}_i$ is an arbitrary contour that avoids the poles of $f$. In the second line we have parametrized the freedom in choosing the contour $\mathcal{C}_i$ by taking a Feynman contour and adding ``contour-correcting'' functions $V_{ij}$ and $W_{ij}$ to it. These contours are depicted in figure \ref{fig:complexcontours}. The Feynman contour corresponds to the non-normalizable modes while the Wightman contours pick up normalizable mode contributions. Note that our ansatz has the correct UV asymptotics since:
\begin{align}\label{eq:FTofsAdS}
    \lim_{r \rightarrow \infty} \Phi_i = \sum_k e^{ik\varphi} \int_{\mathbb{R}}\dd{\omega} e^{-i\omega t} \delta_{ij}\tilde{s}_j = s_i \,.
\end{align}
We do not allow sources on the Euclidean pieces as we are preparing empty AdS without excitations, so the Euclidean fields are composed of normalizable pieces alone:\footnote{Note that we have already assumed no sources on the Euclidean segments when we constrained the generic linear combinations of the sources $V_{ij}\tilde{s}_j$ and $W_{ij}\tilde{s}_j$ to run only along the Lorentzian sources.}
\begin{align}
    \Phi_I = \sum_k e^{i k\varphi}\qty[\int_{\mathcal{W}^<}\dd{\omega} e^{-\omega \tau} fV_{Ij} \tilde{s}_j + \int_{\mathcal{W}^>}\dd{\omega} e^{-\omega \tau}f W_{Ij} \tilde{s}_j] \,.
\end{align}
Regularity imposes a further restriction on the allowed Euclidean pieces. On $\gamma_a$, $\tau<0$ so regularity as $\omega\rightarrow+\infty$ demands that $W_{aj}= 0$. Similarly, regularity for $\tau>0$ on $\gamma_b$ as $\omega\rightarrow -\infty$ is enforced by $V_{bj} = 0$.

\paragraph{Applying matching conditions}
Now, we wish to apply the matching conditions (\ref{eq:AdSmatchingconds}). First of all, the Fourier transform of $\tilde{s}_i$ in (\ref{eq:ads_lorentzian_field_ansatz}) is given by (\ref{eq:FTofsAdS}).
Noting that the source must only have support for $-T<t'<T$, at $t\sim -T$, the factor $e^{-i \omega (t-t')} = e^{i \omega \abs{t-t'}}$ blows up in the lower half plane so that we must close the Feynman integral in (\ref{eq:ads_lorentzian_field_ansatz}) in the upper half plane. Similarly, when $t\sim T$, we close the $\mathcal{F}$ integral in the lower half plane. This results in the expressions:
\begin{align}
    \begin{aligned}
        \Phi_i(t\sim -T) & = \sum_k e^{i k\varphi} \qty[\int_{\mathcal{W}^<}\dd{\omega} e^{-i\omega t} f(\delta_{ij}+V_{ij}) \tilde{s}_j + \int_{\mathcal{W}^>}\dd{\omega} e^{-i\omega t}f W_{ij }\tilde{s}_j]  \,, \\
        \Phi_i(t\sim T)  & = \sum_k e^{i k\varphi} \qty[\int_{\mathcal{W}^<}\dd{\omega} e^{-i\omega t} f V_{ij} \tilde{s}_j + \int_{\mathcal{W}^>}\dd{\omega} e^{-i\omega t} f(\delta_{ij}+W_{ij}) \tilde{s}_j] \,.
    \end{aligned}
    \label{eq:ads_lorentzian_fields_at_early_and_late_times}
\end{align}
Now, we can apply the matching conditions (\ref{eq:AdSmatchingconds}); since each pole is independent we must apply them separately to the $\mathcal{W}^<$ and $\mathcal{W}^>$ integrals:
\begin{align}
     & \begin{aligned}
        \mathcal{W}^<:
         &  & V_{aj}e^{-i\omega T} & = \delta_{1j} + V_{1j} \,, \\
         &  & V_{1j}               & = V_{2j}            \,,    \\
         &  & 0                    & = \delta_{2j} + V_{2j} \,,
    \end{aligned} &
     & \begin{aligned}
        \mathcal{W}^>:
         &  & 0                    & = W_{1j}            \,,    \\
         &  & \delta_{1j} + W_{1j} & = \delta_{2j} + W_{2j} \,, \\
         &  & W_{bj}e^{-i\omega T} & = W_{2j} \,.
    \end{aligned}
    \label{eq:ads_matching_equs}
\end{align}
The solution to these matching conditions is given by:
\begin{align}
    \begin{aligned}
        V_{ij} & = \mqty(0 & -1                                                                 \\ 0 & -1)_{ij} \,, & W_{ij} &= \mqty(0 & 0 \\ 1 & -1)_{ij} \,, \\
        V_{aj} & = \mqty(1 & -1)_j e^{i\omega T} \,, & W_{bj} & = \mqty(1 & -1)_j e^{i\omega T} \,,
    \end{aligned}
\end{align}
so that the Euclidean and Lorentzian fields are now completely determined in terms of the sources $\tilde{s}_i$.

\paragraph{Extracting the correlators}
Finally, having solved the bulk fields entirely, we can extract the correlation functions from them. For this, we extract the $r_+^2 / r^2$ term in the expansion of the Lorentzian fields (see appendix \ref{sec:appendix_SvR}):
\begin{align}
    \phi^{(2)}_i = \sum_k e^{i k\varphi} \qty[ \int_{\mathcal{F}}\dd{\omega} e^{-i\omega t} (\alpha \beta) \delta_{ij} \tilde{s}_j  + \int_{\mathcal{W}^<}\dd{\omega} e^{-i\omega t} (\alpha \beta) V_{ij} \tilde{s}_j + \int_{\mathcal{W}^>}\dd{\omega} e^{-i\omega t} (\alpha \beta) W_{ij} \tilde{s}_j] \,.
\end{align}
Then, we use the Fourier transform $s(t,\varphi)$ of $\tilde{s}_j(\omega,k)$ defined in (\ref{eq:FTofsAdS}) and perform functional derivatives with respect to $s_j(t',\varphi')$ to obtain the correlators:
\begin{align}
    i G^{ij}_{\text{AdS}}(x,x') & = {2(-1)^{\delta_{ij}-\delta_{i2}+1} \over i} \eval{ { \delta \phi^{(2)}_i(x)  \over \delta s_j(x')} }_{s_i=0}                                                             \nonumber      \\
                                & = {(-1)^{\delta_{ij}-\delta_{i2}+1} \over 2 \pi^2 i} \sum_k e^{ik (\varphi-\varphi')}\left[ \int_{\mathcal{F}}\dd{\omega} e^{-i\omega (t-t')} \delta_{ij} \alpha \beta  \right. \nonumber \\
                                & \left. \qquad  + \int_{\mathcal{W}^<}\dd{\omega} e^{-i\omega (t-t')} V_{ij} \alpha\beta + \int_{\mathcal{W}^>}\dd{\omega} e^{-i\omega (t-t')} W_{ij} \alpha\beta\right].
\end{align}
For example, the retarded correlator is given by:
\begin{align}
    \nonumber iG^{R}_{\text{AdS}}(x,x') & = \theta(t-t')\qty[iG^{F}_{\text{AdS}}(x,x') - iG^{\overline{F}}_{\text{AdS}}(x,x')] = \theta(t-t')\qty[iG^{11}_{\text{AdS}}(x,x') - iG^{22}_{\text{AdS}}(x,x')] \\
    \nonumber                           & = {\theta(t-t') \over 2 \pi^2 i} \sum_k e^{ik (\varphi-\varphi')} \int_{\mathcal{F}-\overline{\mathcal{F}}}\dd{\omega} e^{-i\omega (t-t')} \alpha \beta          \\
    \nonumber                           & = {\theta(t-t') \over 2 \pi^2 i} \sum_k e^{ik (\varphi-\varphi')} \int_{\mathbb{R}+i\epsilon}\dd{\omega} e^{-i\omega (t-t')} \alpha \beta                        \\
                                        & = {\theta(t-t') \over 2 \pi^2 i} \sum_k e^{ik (\varphi-\varphi')} \int_{\mathcal{W}^>-\mathcal{W}^<}\dd{\omega} e^{-i\omega (t-t')} \alpha \beta \,,
\end{align}
where we have used the relation $\overline{\mathcal{F}} = -\mathcal{F}+\mathcal{W}^>+\mathcal{W}^<$, as well as $\mathcal{F} - \overline{\mathcal{F}} = \mathbb{R}+i\epsilon$ with $\epsilon$ a small positive number. In the last line we have employed the fact that the $\theta(t-t')$ function demands to close $\mathbb{R}+i\epsilon$ in the lower half plane, see figure \ref{fig:omega_plane_contours_AdS/WH} (where now the crosses should be considered to be poles of the empty AdS mode $f$). Finally, we compute this integral as a sum over the residues of $\beta$. We obtain (with $\Delta t=t-t'$ and $\Delta \varphi=\varphi-\varphi'$):
\begin{align}
    iG^{R}_{\text{AdS}}(x,x') & = {2\theta(\Delta t) \over \pi R} \qty[\sum_{n,k} e^{-i\omega^+_{nk} \Delta t + ik\Delta \varphi} \alpha(\omega^+_{nk}) - \sum_{n,k} e^{-i\omega^-_{nk} \Delta t + ik\Delta \varphi} \alpha(\omega^-_{nk})]     \nonumber \\
     \label{eq:corremptyAdS}                         & = -{\theta(\Delta t) \over 2 \pi R} \qty[ {1 \over \qty(\cos\qty(\Delta t(1 -i\epsilon) \over R) -\cos(\Delta \varphi))^2} -  {1 \over \qty(\cos\qty(\Delta t(1 +i\epsilon) \over R) -\cos(\Delta \varphi))^2} ] \,,
\end{align}
where the $i\epsilon$ prescription appears with a minus sign to regulate the $\omega^+_{nk}$ sum and with a plus sign to regulate the $\omega^-_{nk}$ sum.

\newpage
\bibliographystyle{toine}
\bibliography{bib}

\end{document}